\documentclass[showpacs,superscriptaddress,showkeys,nofootinbib,aps,prd,preprintnumbers,twocolumn]{revtex4-1}

\usepackage[utf8]{inputenc}
\usepackage{amsmath,amssymb}
\usepackage{natbib}
\usepackage{graphicx}
\usepackage{tabularx}
\usepackage{dcolumn}
\usepackage{bm}
\usepackage{tablefootnote}
\usepackage{color}
\definecolor{linkcolor}{rgb}{0.0,0.3,0.5}
\usepackage[breaklinks, plainpages=false, colorlinks=true, anchorcolor=linkcolor, linkcolor=linkcolor, citecolor=linkcolor, urlcolor=linkcolor]{hyperref}
\usepackage[normalem]{ulem}

\def\apj{\rm{ApJ}}
\def\apjl{\rm{ApJ}}
\def\apjs{\rm{ApJS}}
\def\aap{\rm{A\&A}}

\def\mnras{\rm{MNRAS}}
\def\prd{\rm{PRD}}
\def\prl{\rm{PRL}}
\def\prx{\rm{PRX}}
\def\nat{\rm{Nature}}
\def\cqg{\rm{CQG}}

\def\pasp{\rm{PASP}}
\def\araa{\rm{ARA\&A}}
\def\aapr{\rm{A\&A~Rev.}} 
\def\llr{\rm{LLR}}

\makeatletter
\let\oldtheequation\theequation
\renewcommand\tagform@[1]{\maketag@@@{\ignorespaces#1\unskip\@@italiccorr}}
\renewcommand\theequation{(\oldtheequation)}
\makeatother

\begin{document}

\title{Mining Gravitational-wave Catalogs To Understand Binary Stellar Evolution: \\ A New Hierarchical Bayesian Framework}

\author{Stephen~R.~Taylor}
\thanks{NANOGrav Senior Postdoctoral Fellow}
\email{srtaylor@caltech.edu}
\affiliation{TAPIR 350-17, California Institute of Technology, 1200 E California Boulevard, Pasadena, CA 91125, USA}
\affiliation{Jet Propulsion Laboratory, California Institute of Technology, 4800 Oak Grove Drive, Pasadena, CA 91109, USA}

\author{Davide~Gerosa}
\thanks{Einstein Fellow}
\email{dgerosa@caltech.edu}
\affiliation{TAPIR 350-17, California Institute of Technology, 1200 E California Boulevard, Pasadena, CA 91125, USA}

\date{\today}
\begin{abstract}
Catalogs of stellar-mass compact binary systems detected by ground-based gravitational-wave instruments (such as Advanced LIGO and Advanced Virgo) will offer insights into the demographics of progenitor systems and the physics guiding stellar evolution. Existing techniques approach this through phenomenological modeling, discrete model selection, or model mixtures. Instead, we explore a novel technique that mines gravitational-wave catalogs to directly infer posterior probability distributions of the hyper-parameters describing formation and evolutionary scenarios (e.g.\ progenitor metallicity, kick parameters, and common-envelope efficiency). We use a bank of compact-binary population synthesis simulations to train a Gaussian-process emulator that acts as a prior on observed parameter distributions (e.g.\ chirp mass, redshift, rate). This emulator slots into a hierarchical population inference framework to extract the underlying astrophysical origins of systems detected by Advanced LIGO and Advanced Virgo. Our method is fast, easily expanded with additional simulations, and can be adapted for training on arbitrary population synthesis codes, as well as different detectors like LISA.
\end{abstract}

                     
\maketitle

\section{Introduction}
\label{sec:intro}

Over the last few years, the Advanced LIGO and Advanced Virgo interferometers have detected gravitational-waves (GWs) emitted during the final inspiral and merger of binary black holes and neutron stars. Among the many fruits of these ongoing searches have been the first direct detection of GWs from binary black-hole (BH) systems \cite{2016PhRvL.116f1102A}; a growing catalog of BHs at various masses, distances, and component spin orientations \cite{2016PhRvX...6d1015A,2016PhRvL.116x1103A,2017PhRvL.118v1101A,2017ApJ...851L..35A,2017PhRvL.119n1101A}; and the first double neutron-star (NS) merger signal \cite{2017PhRvL.119p1101A}, with a plethora of associated multi-messenger electromagnetic follow-up analysis \cite{2017ApJ...848L..12A}. The expected detection rate of binary BHs and NSs could be tens per year with current detectors \cite{2016PhRvX...6d1015A}, and promise a data explosion for future third-generation ground-based interferometers \cite{2012CQGra..29l4013S}. As we move from the dawn of GW astronomy into its source-rich golden-age, we will be able to perform detailed reconstructions of the demographics of stellar populations, the formation history of compact binary systems, and the physical processes guiding stellar evolution. 

There are undoubtedly individual GW detections that can provide invaluable physical and astrophysical insight. For instance, the detection of GW150914 proved that GWs could be directly detected \cite{2016PhRvL.116f1102A} and that GW emission was consistent with GR \cite{2016PhRvD..94h4002Y,2016PhRvL.116v1101A}. Perhaps even more crucially from an astrophysical standpoint, it gave the first irrefutable proof that BHs indeed form binary systems able to merger within a Hubble time. Likewise, the detection and electromagnetic follow-up of GW170817 showed that NS mergers could explain the origin of short gamma-ray bursts \cite{2017ApJ...848L..12A}; gave insight into the equation of state of nuclear matter \cite{2018arXiv180408583D,2018arXiv180511581T}; constrained the speed of the graviton to less than one part in $10^{-15}$ \cite{2017ApJ...848L..13A}; and even permitted a measurement of the Hubble constant \cite{2017Natur.551...85A}.  There will continue to be such ``golden'' systems offering unique physical insights. For instance, detections with particularly favorable orientations in the future might show signs of spin precession \cite{1994PhRvD..49.6274A}. 
 
But even with the small number of GW detections so far, emphasis is already shifting to answering questions about the population properties of GW sources. As we move towards the large-statistics regime of GW astronomy, focus will shift from inferring \emph{parameters} of single sources (masses, spins, redshifts) to characterizing \emph{hyper-parameters} describing formation and evolutionary processes of BH and NS populations. 

There are many challenges to understanding the formation channels of GW-detected compact binary systems \cite{2014LRR....17....3P}. Binary stellar evolutionary codes (e.g. \cite{2002MNRAS.329..897H,2004MNRAS.350..407I,2008ApJS..174..223B,2015MNRAS.451.4086S,2018MNRAS.474.2959G,2018ApJ...854L...1B,2018arXiv180105433K,2017NatCo...814906S}) have become very detailed, but still suffer from large theoretical uncertainties. To name a few, these include $(i)$ the dependence of remnant compact object masses (and thus NS or BH identities) on stellar winds and metallicity; $(ii)$ the magnitude of kicks received by BHs and NSs at formation; and $(iii)$ the efficiency with which orbital energy can be transferred to a common envelope, thereby tightening a binary. Adding to these uncertainties in classical isolated binary evolution are details of other proposed scenarios involving dynamical interactions with other bodies \cite{2013LRR....16....4B}. There is thus much poorly known stellar astrophysics that catalogs of GW detections can be mined for. 

Several techniques have been developed to perform GW population inference, ranging from phenomenological parametrized modeling to discrete model selection, with mixture modeling as a blending of the former two. In phenomenological models, the distribution of component masses, spins, and redshifts are reconstructed through relatively simple parametrizations (e.g. \cite{2012PhRvD..85b3535T,2017arXiv171109226Z,2017ApJ...846...82Z,2018ApJ...854L...9F,2018arXiv180506442W,2018ApJ...856..173T,2018arXiv180610610R}). Any inference with these models will only be a broad sketch of the complicated process of compact binary formation. Detailed stellar population modeling allows binary stars to be tracked from known astrophysical assumptions all the way through to compact binary formation (or not, depending on conditions). But these are computationally expensive (making real-time simulation runs during Bayesian analysis unfeasible), and are typically performed in small batches for comparisons to observations. This approach has been very successful, showing e.g.\ that GW$150914$'s stellar progenitor had a metallicity of $\sim 5\%\, Z_\odot$ \cite{2016Natur.534..512B,2016ApJ...818L..22A,2016MNRAS.463L..31L}. More systematic approaches have also been taken, where Bayesian model selection is performed on grids of discrete population synthesis simulations, or where simulations are mixed together with weightings inferred from the data \citep{2015ApJ...810...58S,2017PhRvD..95l4046G,2017MNRAS.471.2801S,2017ApJ...846...82Z,2018PhRvD..97d3014W}. Finally, non-parametric methods have been developed to allow recovery of binary parameter distributions that is more agnostic than the parametrized-model approach \citep{2017MNRAS.465.3254M}. These methods recover the bin heights of parameter distribution histograms, typically with Gaussian Process (GP) priors linking the bins to enforce smoothness. 

In this paper we present a qualitatively new approach that fuses non-parametric modeling with population-synthesis simulations. In brief, we model histograms of GW parameter distributions with bin heights constrained by informative parametrized-priors built out of population synthesis simulations. This allows us to fully exploit catalogs of GW detections to directly infer the properties of progenitors and the evolutionary path undertaken. Our methods give predictions of rates and parameter distributions of compact-binary systems by interpolating between a set of population-synthesis simulations informed by the data. Crucially, the framework developed here remains agnostic of the specific population synthesis code to used. 

\begin{figure}
\begin{center}
\includegraphics[width=0.5\textwidth]{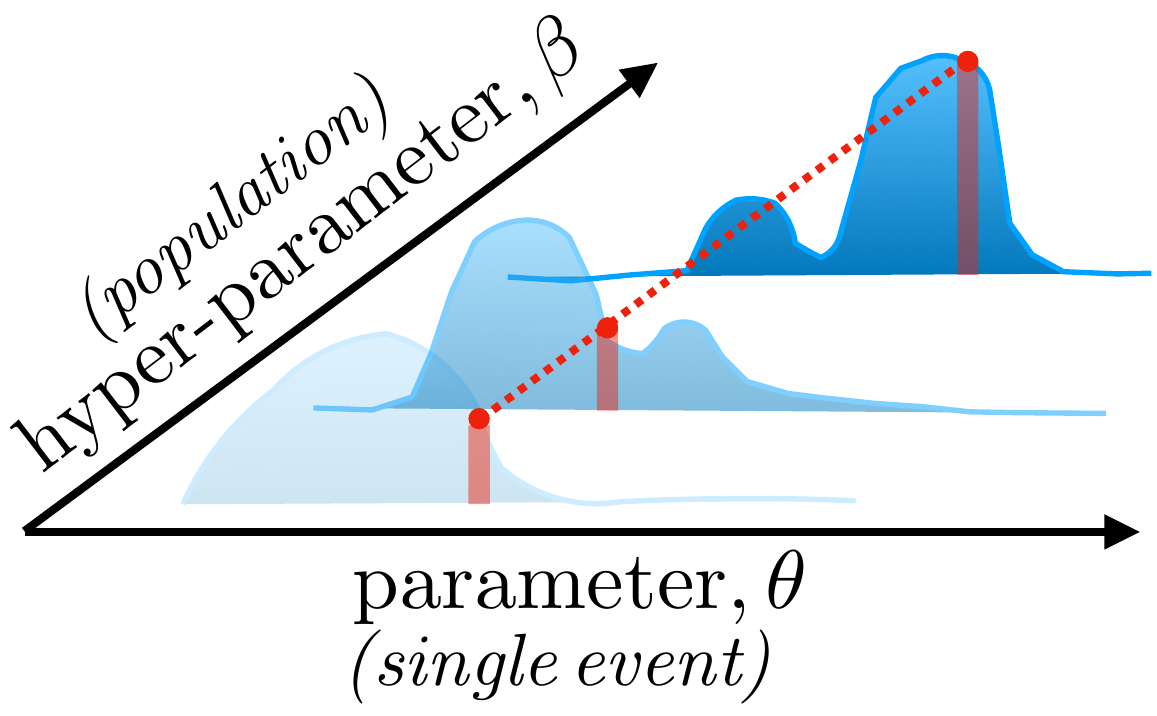}
\end{center}
\caption{A schematic representation of interpolating over parameter distributions( $\theta$, e.g.~masses, spins, redshift) as a function of population hyper-parameters ($\beta$, e.g. progenitor metallicity, common-envelope hardening efficiency, natal kicks, etc.). We carry out a restricted number of population synthesis simulations with different hyper-parameters, where each simulation produces compact binaries distributed over parameter space. These parameter distributions form the training data for our interpolant model. For each bin, pixel, or feature in the parameter distribution, we train a GP interpolant over the hyper-parameter space, allowing us to predict the distribution at any other hyper-parameter coordinate.}
\label{fig:hyper_gp_cartoon}
\end{figure}

We follow a multi-stage process (illustrated in \autoref{fig:hyper_gp_cartoon}), beginning with a design for the program of simulations across hyper-parameter space, compressing distributions of binary parameters to distill the most important features, and training a GP model to interpolate between the simulation hyper-parameter coordinates. These models are then fed to a hierarchical Bayesian pipeline to recover the joint posterior probability distribution of population hyper-parameters, while incorporating measurement uncertainties in each binary's parameters. GP emulation of computationally-expensive simulations has been used in cosmological matter power spectrum analysis \citep{2006ApJ...646L...1H,2007PhRvD..76h3503H}, pulsar-timing array GW constraints on supermassive binary BH dynamical environments \citep{2017PhRvL.118r1102T,2018ApJ...859...47A}, and has been suggested in principle for stellar-mass binary BH population inference \citep{2017IAUS..325...46B}. Here we fully develop this emulation approach, embedding it in a complete end-to-end statistical framework, starting from the simulation program design and following through to GW catalog analysis. 

This paper is laid out as follows. In \autoref{sec:stats} we describe how to choose locations in the hyper-parameter space where we should perform simulations, how to compress distributions of simulated binary parameters, and how we interpolate over these compressed distributions using GPs. We introduce our inference tools in \autoref{sec:inference}, including Bayesian GW parameter estimation, a scheme to convolve the intrinsic simulated binary distributions with detector selection effects, and a pipeline to perform hierarchical Bayesian inference on catalogs of GW detections. We show our results in \autoref{sec:results}, where our entire framework is tested on three case studies that successively increase in complexity and astrophysical realism. These include (i) a toy analytic model, (ii) an example with publicly-available population synthesis simulations, and  (iii) finally an example with our custom program of simulations. We provide our conclusions and a discussion of future prospects in \autoref{sec:conclusions}.   
\section{Statistical Framework}
\label{sec:stats}

In this Section we describe a statistical framework for choosing points in hyper-parameter space at which to generate simulated astrophysical populations (\autoref{sec:simdesign}), defining a data-driven basis for the distributions of population parameters (\autoref{datacompression}), and training an interpolation scheme to emulate these parameter distributions (\autoref{trainingemulator}). Our framework closely follows the steps outlined for cosmological matter power spectrum studies in Refs.\ \citep{2006ApJ...646L...1H,2007PhRvD..76h3503H}. 

\subsection{Simulation design} \label{sec:simdesign}

We need a careful strategy for determining the locations in hyper-parameter space at which to perform the simulations that will eventually be used to train our emulator. While the temptation is to choose an $N$-dimensional grid-design, this turns out to be highly sub-optimal. The hyper-parameter space dictating stellar-mass binary evolution is $\mathcal{O}(10)$ dimensions, and grid-based designs quickly explode in the number of required simulations. For example, if we choose a simple grid with $3$ nodes along each dimension, then in $2$-dimensions this is a reasonable choice, requiring $9$ simulations in total. However, expanding this to $10$ dimensions requires $3^{10}\sim 6\times10^4$ simulations, which is a computationally prohibitive step for current population-synthesis codes. The entire purpose of constructing an emulator is to avoid the need for high numbers of costly simulation runs. Furthermore, grid-based designs are poor at covering low-dimensional projections of the full hyper-parameter space. If the distribution of BH masses and spins is dominated by only three hyper-parameters (say progenitor metallicity, natal kicks, and common-envelope efficiency) out of the full $10$ dimensional space, then our above-mentioned grid-based design only assigns $3^3=27$ unique simulated combinations of these important hyper-parameters out of the total $\sim 6\times 10^4$ simulations. The opposite case is a purely random design, which however suffers from large regions of sparsely populated hyper-parameter space because random sampling maintains no record of where previous points have been placed.

One thus needs a simulation design that gives good coverage over all lower-dimensional projections of the hyper-parameter space, while simultaneously being sparse enough in the full space to make the program of simulations computationally tractable. A popular solution is  given by stratified sampling. If $M$ points are to be drawn, the hyper-parameter volume is first divided into $M$ equally-probable sub-strata, within which random sampling for each point is employed. Specifically, we use space-filling Latin hypercube designs \citep{mbc79}, where each sample is the only one permitted to occupy the axis-aligned hyperplane containing it. One must define how many samples are to be drawn at the outset of sampling, and the sampler keeps a record of the position of each past draw. A variant on this technique for integers in the range $[0,9]$ produces the popular puzzle Sudoku.

\begin{figure} 
\begin{center}
\includegraphics[width=0.45\textwidth]{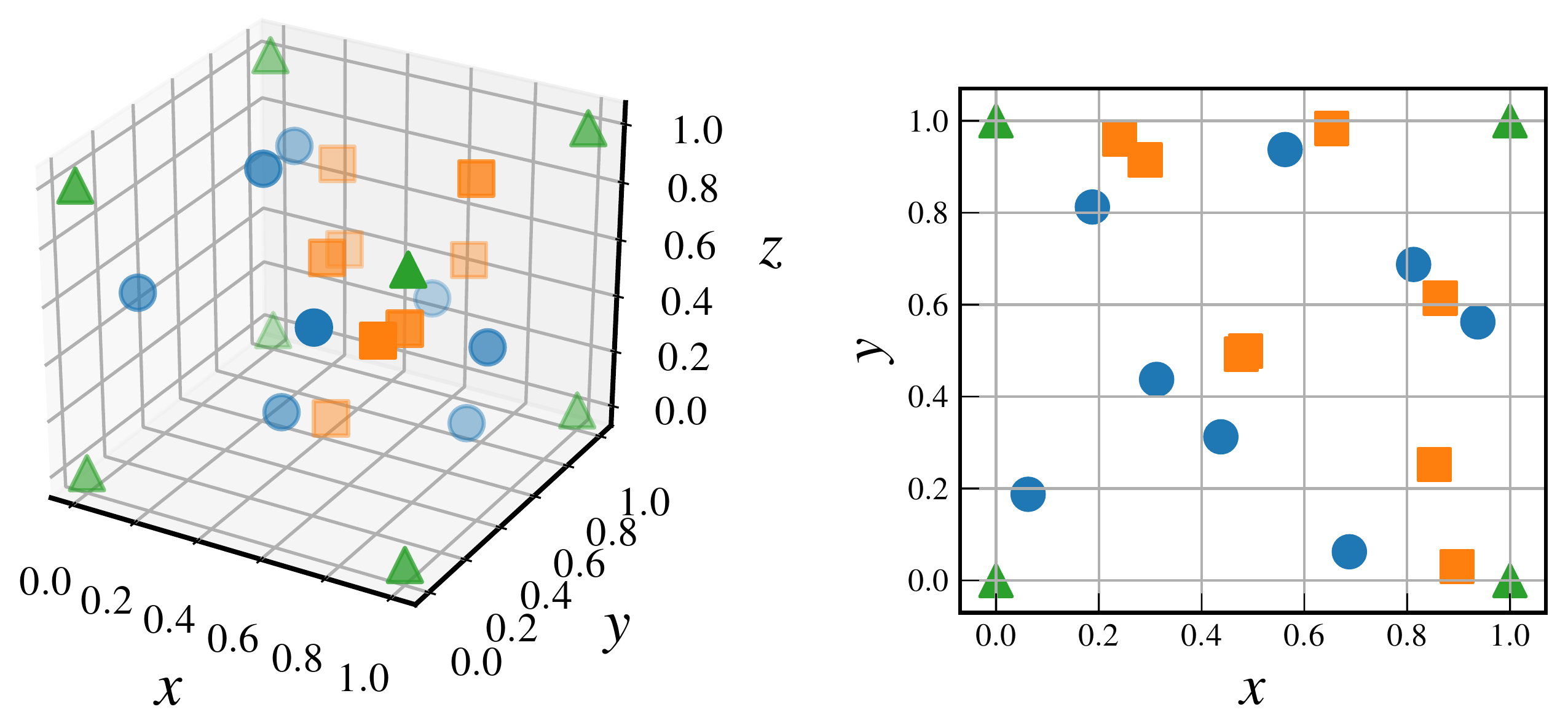}
\caption{Example of $\{x,y,z\}$ hyper-parameter locations assigned on an evenly-spaced grid (green triangles), randomly (orange squares), and with Latin hypercube sampling (blue circles), for $M=8$ training coordinates. A projection of these coordinates into the $\{x,y\}$ plane is shown on the right.}
\label{fig:sampling_demo}
\end{center}
\end{figure}

We use the \texttt{pyDOE} \cite{pyDOE} python module for all simulation designs in this paper. Various sampling options are available, but we choose to maximize the minimum separation between points in hyper-parameter space, while also centering them within the sampling intervals. We compute all simulation coordinates on the unit hypercube, then transform them to the physical hyper-parameter ranges of interest. Figure~\ref{fig:sampling_demo} shows a comparison of how $M=8$ training coordinates would be assigned in hyper-parameter space according to different simulation design schemes.

\subsection{Data compression}\label{datacompression}

Running population synthesis simulations will provide a catalog of systems, each one with associated measured parameters. In the case of  compact binaries, these parameters include component masses, spins, luminosity distance, perhaps eccentricity, etc. A natural way to summarize all this information is to produce histograms of the properties over the entire population; an interpolant could then be used to learn how the input simulation hyper-parameters affect the height of each histogram bin. Although there is nothing formally wrong with this strategy, it misses the opportunity to generate a data-driven basis on which to summarize the parameter distributions, rather than use naive binning. If we simply binned then we would need as many interpolants as bins, which might cause an unnecessary explosion of the computational cost. But if our training distributions lack pathological features, we can form a set of basis distributions that are smaller in number. 

To generate a data-driven basis for the simulated distributions of a binary property, we form a data matrix  $D$ of shape $N_\mathrm{bins}\times N_\mathrm{sims}$. Each column in this matrix corresponds to a single simulation, and contains the normalized bin heights in the histogram for the parameter (flattened over all parameter dimensions, if multi-dimensional histograms are considered), where we a-priori establish a common binning scheme across all simulations. We then use principal component analysis  (PCA) \citep{2007PhRvD..76h3503H} on the row-centered matrix to identify a new set of basis distributions: 
\begin{equation}
D = U\Sigma V^T,
\end{equation}
where the magnitude of the singular values along the diagonal of $\Sigma$ are used to assess the dimensionality of the new basis. We denote $N_\mathrm{basis}$ as the number of singular values above tolerance that form the restricted $\widetilde{\Sigma}$ diagonal matrix, while the column spaces of $U$ and $V$ are also restricted at $N_\mathrm{basis}$ to form $\widetilde{U}$ and $\widetilde{V}$. The columns of $\widetilde{U}\widetilde{\Sigma} / \sqrt{N_\mathrm{basis}}$ are principal components of the parameter distributions that form a natural basis, while columns of $\sqrt{N_\mathrm{basis}}\widetilde{V}$ correspond to the projection of the original data (bin heights) into the new basis. 
An interpolant can then be trained on the data in the new compressed basis, such that subsequent predictions are first made in lower dimension before being rotated back into the full-rank binning scheme. Any initial row-centering or scaling is also corrected after a prediction is rotated into full-rank. This data compression scheme identifies characteristic ``features'' in the parameter distributions. 

In the following, the choice of binning scheme (the range and size of bins) is explored case by case. We want to retain the dominant features in our parameter distributions that have sensitivity to hyper-parameters, but also want to avoid an interpolant learning Poisson fluctuations from low occupations bins. Also, for fixed $N_{basis}$, the compression fidelity may be lower in a finer binning scheme, where the bin heights may fluctuate significantly from Poisson noise.

\subsection{Training an emulator}\label{trainingemulator}

In regression analysis, or more specifically GW population inference, we need a model to fit to some data. We can assume a parametric form, but we can also be more flexible and let the model be data-driven. In the latter approach, we use the data to train an interpolant which connects the observations by e.g. straight lines ({linear interpolation}) or low-degree polynomials ({spline interpolation}). An even more powerful technique than straightforward linear or spline interpolation is GP regression, which treats noisy data as a single random draw from a multivariate Gaussian distribution with a mean vector and covariance function. By optimizing the parameters of a covariance function, and conditioning our predictions of the underlying function on the observations, we let the data tell us the nature of the underlying process rather than enforcing a strict parametric function.

In the rest of this section we define GPs and explain how they can be used as a powerful interpolation tool. There are many excellent treatments of this subject (for general theory see e.g.\ \citep{2006gpml.book.....R,m98,2015arXiv150502965E}; for ground-based GW applications see \citep{2015PhRvD..91l4062G,2016PhRvD..93f4001M,2013PhRvD..88h4061O}, and for recent applications to Pulsar Timing Arrays see \citep{2014PhRvD..90j4012V,2017PhRvL.118r1102T}), but here we only summarize the salient points that motivate our work.

\subsubsection{Gaussian processes}

The formal definition of a GP is a (possibly infinite) \textit{``collection of random variables, any finite number of which have a joint Gaussian distribution"} \citep{2006gpml.book.....R}. Instead of parametrizing the underlying function, we are placing a prior (in this case a Gaussian) on the space of possible functions characterized by a mean and covariance. The former is often set to zero and the latter describes how the $N$ points in our sample of the process are correlated \citep{m98}. Hence, if we model the underlying process, $f(\mathbf{x})$, as a GP from which our data $\mathbf{y} = \{y_1, y_2, \ldots, y_N\}$ are drawn, then formally we can write \citep{2006gpml.book.....R}:
\begin{align}
f(\mathbf{x})\, &{\sim}\, \mathcal{GP}(m(\mathbf{x}),k(\mathbf{x},\mathbf{x'})), \nonumber\\
\mathbf{y}\, &{\sim}\, \mathcal{N}(m(\mathbf{x}), k(\mathbf{x},\mathbf{x'})), 
\end{align}
where the covariance (or {kernel}) function is $k(\mathbf{x},\mathbf{x'}) = \langle (f(\mathbf{x})-m(\mathbf{x}))(f(\mathbf{x'})-m(\mathbf{x'})) \rangle$, and as mentioned above we set $m(\mathbf{x})=\mathbf{0}$.

\subsubsection{Predictions}
\begin{figure*}
\begin{center}
\includegraphics[width=\textwidth]{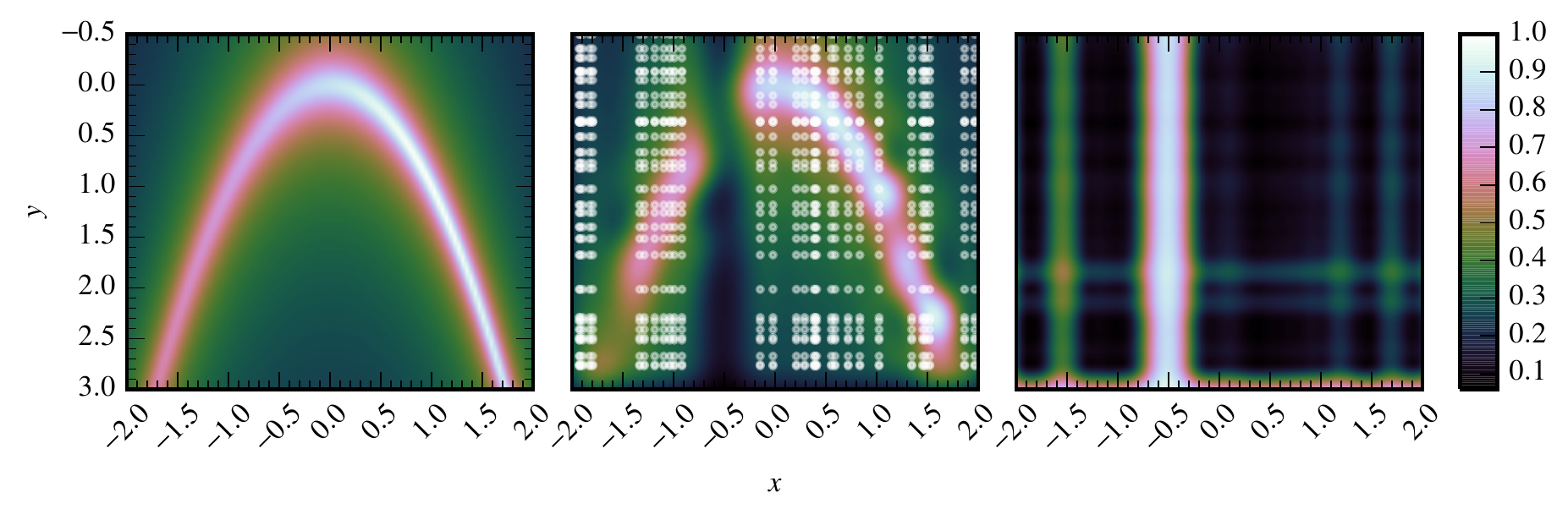}
\caption{Training a Gaussian process for prediction. In the left panel we show an inverted offset-Rosenbrock function. In the center panel we show the locations of our training data as white points, along with the GP predicted function values in the background. The right panel shows the uncertainty in the predicted function values of the center panel.}
\label{fig:gpr_example}
\end{center}
\end{figure*}
We need knowledge of the kernel to constrain the space of possible underlying functions. We train the GP by performing a limited sampling of the underlying process (which in our case are population-synthesis simulations), and condition further predictions on this training data. We account for possible measurement uncertainties on the training data, meaning that we are really measuring noisy values of the underlying process, i.e.
\begin{equation}
\mathbf{y} = f(\mathbf{x}) + \mathbf{n},
\end{equation}
where 
\begin{equation}
\mathbf{n}\sim\mathcal{N}(\mathbf{0},\sigma^2_n\delta(\mathbf{x}-\mathbf{x'}))\,.
\end{equation} 

If we have training data $\mathbf{y}$ measured at $\mathbf{x}$, and we want to predict function values at new points $\mathbf{x_*}$, then we first write the joint distribution of $\mathbf{y}$ and $\mathbf{y_*}$:
\begin{equation}
\begin{bmatrix} \mathbf{y} \\ \mathbf{y_*} \end{bmatrix} \sim \mathcal{N}\left( \mathbf{0}, \begin{bmatrix} K + \sigma_n^2I & K_*^T \\ K_* & K_{**} \end{bmatrix}\right),
\end{equation}
where $K$ is the matrix of kernel evaluations over the training data, $K_*$ is the matrix of kernel evaluations between the prediction points and the training data, and $K_{**}$ is the matrix of kernel evaluations over the prediction points.

The conditional distribution of $\mathbf{y_*}$ given $\mathbf{y}$ is \citep{2006gpml.book.....R}
\begin{equation} \label{eq:gppredict}
\mathbf{y_*}|\mathbf{y} \sim \mathcal{N}(\overline{\mathbf{y}}_*, \mathrm{cov}(\mathbf{y_*})),
\end{equation}
where,
\begin{align}
\overline{\mathbf{y}}_* &= K_*(K+\sigma_n^2I)^{-1}\mathbf{y}, \\
\mathrm{cov}(\mathbf{y_*}) &= K_{**} - K_*(K+\sigma_n^2I)^{-1}K_*^T.
\end{align}
Equation (\ref{eq:gppredict}) shows a key result --- namely that we have interpolated over our training data by conditioning predictions of new observations on their values. The mean of this conditional distribution  $\overline{\mathbf{y}}$ is our prediction, but equally important is the prediction uncertainty $\mathrm{cov}(\mathbf{y_*})$, which we can propagate through to subsequent inference.

\subsubsection{Kernel functions}

The choice of kernel function should be informed by some prior knowledge of the underlying process, but the only formal prerequisite is that it produce a positive-semidefinite covariance matrix. A common choice in the literature is the Squared Exponential (SE) kernel, whose popularity stems from the fact that it is stationary and infinitely differentiable. For training data whose input coordinates are multi-dimensional, this kernel function in a flat metric is:
\begin{equation}
k(x,x') = \sigma_k^2 \exp{ \left(- \frac{(x_i-x'_i)^2}{2\sigma_i^2} - \frac{(x_j-x'_j)^2}{2\sigma_j^2} - \cdots \right) },
\end{equation}
where each dimension of the input coordinate can have a separate variance $\{ \sigma_i^2, \sigma_j^2, \ldots \}$, and the kernel has an overall variance scaling $\sigma_k^2$. The variance of each dimension acts as a length parameter that dictates the degree with which distant observations can influence each other.

Throughout this paper we use \texttt{George} \citep{2015ITPAM..38..252A}, which is a powerful Python library for GP regression. As an example, we sample the following inverted offset-Rosenbrock function at $900$ random locations in $[x,y]$ space:
\begin{equation}
g(x,y) = \left[ (1-x)^2 + 100(y - x^2)^2 + 1 \right]^{-1/5}.
\end{equation}
This function is shown in the left panel of \autoref{fig:gpr_example}, while in the center panel we show the training data locations as white points and the predicted function values in the background. These function values have been predicted by training a GP with an SE kernel. The kernel hyper-parameters were not optimized, but merely set as $\{\sigma_k^2=1,\sigma_x^2=0.05,\sigma_y^2=0.05\}$. The prediction uncertainty is shown in the rightmost panel of \autoref{fig:gpr_example}, where we see that the predictive model accuracy is worst in the locations where there is a deficit of training data. This feature of GPs is particularly useful since it tells us where in parameter space we must take new samples (i.e. perform new populations synthesis simulations) so that we improve the accuracy of our model. Rather than assume a set of kernel hyper-parameters, we can optimize them; in this case the likelihood (or optimization function) is a Gaussian with an SE kernel, and the training data are treated as a draw from this Gaussian process. We can either map the posterior probability distribution of the kernel hyper-parameters (conditioned on the training data) or simply find the maximum a-posteriori values. In the following, we use MCMC techniques to sample the kernel hyper-parameter posterior distribution, and use the posterior samples to determine the maximum a-posteriori values.  
\section{Inference techniques}
\label{sec:inference}

In this Section we first outline Bayesian inference as a statistical framework allowing for robust detection and parameter estimation (\autoref{sec:bayes}). We then specify how it is applied to ground-based GW analysis, resulting in catalogs of measured compact-binary coalescences, each associated with a set of samples drawn from the posterior probability distribution of the event's physical parameters (\autoref{sec:gw_param_est}). Finally, we introduce a hierarchical Bayesian framework for inferring the evolutionary history and progenitor conditions of cataloged GW events, which uses the simulation-trained GP emulator as a parametrized prior (\autoref{sec:hierarchical}).

\subsection{Bayesian inference}
\label{sec:bayes}

Bayesian inference is a powerful statistical framework allowing models to be robustly tested against data, resulting in probability distributions of the model parameters that are conditioned on both prior expectations and new information \cite{1763RSPT...53..370B}. This framework employs Bayes' rule of conditional probabilities, such that the posterior probability of parameters $\Theta$ within a model $\mathcal{H}$, implied by data $\mathcal{D}$, is given by:
\begin{equation}
p(\Theta|\mathcal{D},\mathcal{H}) = \frac{{p}(\mathcal{D}|\Theta,\mathcal{H}){p}(\Theta|\mathcal{H})}{{p}(\mathcal{D}|\mathcal{H})},
\end{equation}
where ${p}(\mathcal{D}|\Theta,\mathcal{H})\equiv\mathcal{L}(\Theta)$ is the likelihood of the model parameters given the data, ${p}(\Theta|\mathcal{H})$ is the prior probability of the model parameters, and ${p}(\mathcal{D}|\mathcal{H})\equiv\mathcal{Z}$ is the fully-marginalized likelihood, or evidence. When inferring credible regions or upper limits for parameters within a single fixed model, the evidence acts as a constant and can be ignored. However it is an important feature for model selection, where the ratio between evidences under different models is known as the {Bayes factor}. When multiplied by an appropriate prior odds ratio, this becomes the {posterior odds ratio}, which is essentially the betting odds between the two models. 

In parameter estimation we are usually interested in the credible regions for a few parameters. Since Bayesian inference returns probability distributions, we can integrate over over all unwanted nuisance parameters while still incorporating their uncertainty into the measurement spread of parameters that we care about. This technique is known as marginalization. The high-dimensional parameter spaces of models is typically explored using numerical random sampling techniques like Markov Chain Monte Carlo, where the density of the chain samples in parameter space is proportional to the posterior probability density function. As such, all integrations can be trivially tackled through Monte Carlo techniques, e.g.:
\begin{equation} \label{eq:mcmc_marg}
\int\mathrm{d}x\, f(x)p(x|d,\mathcal{H}) \approx \frac{1}{N}\sum_{i=1}^N f(x_i),
\end{equation}
where $f(x)$ is an arbitrary function, and $p(x|d,\mathcal{H})$ is the posterior probability of $x$ given data $d$ under model $\mathcal{H}$, which we approximate with random samples $i\in[1,\ldots,N]$. We use \texttt{emcee} \cite{2013PASP..125..306F} for all sampling in the following.

\subsection{Gravitational-wave parameter estimation}
\label{sec:gw_param_est}

Bayesian inference needs a likelihood function to assess the fitness of the proposed model parameter choices against data, and a measure of the prior probability of these proposed parameters. For ground-based GW analysis, the data is the dimensionless strain computed from the raw interferometric output, which is composed of signal and noise processes. We treat the noise processes as Gaussian and stationary so that we can analytically marginalize over the noise strain, and consider only its power spectral density (PSD), which we assume to be known. For this, we use the Advanced LIGO noise PSD at design sensitivity \cite{LIGOcurve}, with a low frequency cutoff at $10$ Hz. The strain signal $h$ describing a compact-binary coalescence has $15$ parameters: $2$ sky-location, $1$ polarization angle, $1$ initial phase, $3$ components of an orbital angular-momentum vector, $2$ BH masses, and $2\times3$ components of the spin vectors. Appropriate sampling of this parameter space will return a set of independent draws from the posterior probability distribution of the signal model. We assume that a catalog of all detected GW events will eventually be issued in the form of sets of these posterior samples (see Refs.\ \cite{2018arXiv180408583D,2018arXiv180511579T} for initial steps in this direction).\footnote{While this work was being completed, the posterior samples were made available by the LIGO-Virgo Collaboration at \href{https://dcc.ligo.org/LIGO-T1800235/public}{dcc.ligo.org/LIGO-T1800235} and \citet{2017PhRvL.119y1103V} for the three events (GW$150914$, GW$151226$, LVT$151012$) in the Advanced LIGO detector's first observing run (O1).} 

In the following, we need a simple measure of the detection probability of a compact-binary system. We adopt a frequentist statistic for detection, corresponding to a threshold cut on the expected signal-to-noise ratio (SNR)
\begin{equation}
\rho^2 = 4
\int_0^\infty\mathrm{d}f\, \frac{\tilde{h}^*(f)\tilde{h}(f)}{S_n(f)},
\label{SNReq}
\end{equation}
where $S_n(f)$ is the one-sided noise PSD, and $\tilde{h}(f)$ is the Fourier-domain waveform. We employ the \textsc{IMRPhenomD} approximant \cite{2016PhRvD..93d4007K} and ignore spins in the SNR calculation, deferring its information content to future work (cf. \cite{2018arXiv180503046N} for possible biases). We access both the Advanced LIGO noise PSD and the waveform approximants through the \texttt{pyCBC} python package \citep{2014PhRvD..90h2004D,2016CQGra..33u5004U}.

A GW signal from a coalescing binary is described by the two polarizations
\begin{align}
h_+(t) &= A(t) \frac{1+\cos^2\iota}{2} \cos\Phi(t), 
\\
h_\times(t) &= A(t)  \cos\iota \sin\Phi(t),
\end{align}
where $\iota$ is the binary orbit inclination and all other dependencies are encoded in the signal amplitude $A(t)$ and phase $\Phi(t)$. The response of a (single) detector, 
\begin{equation}
h(t) = F_+ h_+(t) + F_\times h_\times(t), 
\end{equation}
is modulated by the antenna beam patterns $F_{+,\times}(\theta,\phi,\psi)$, where the three angles describe sky location and polarization content (e.g. Ref.\ \cite{2009LRR....12....2S}). One can then define the projection parameter  \cite{1993PhRvD..47.2198F,1996PhRvD..53.2878F,2015ApJ...806..263D} 
\begin{equation}
\omega = \sqrt{ \frac{(1+\cos^2\iota)^2}{4}F_+^2(\theta,\phi,\psi)+ \cos^2\iota F_\times^2(\theta,\phi,\psi)}\,
\end{equation}
and the phase offset 
\begin{equation}
\tan \Phi_0 = \frac{2 \cos\iota F_\times}{(1+\cos^2\iota) F_+}\,,
\end{equation}
such that
\begin{equation}
h(t) = A(t) \omega  \cos(\Phi(t)- \Phi_0)\,.
\end{equation}
The parameter $\omega$ encapsulate all the angular dependencies of the signal amplitude and satisfies $ \max_{\iota,\theta,\phi,\psi} \omega =1 $. From \autoref{SNReq} one thus obtains $\rho = w \rho_{\rm opt}$, where $\rho_{\rm opt}$ is the SNR for an optimally
oriented source. 

\begin{figure} 
\centering
\includegraphics[width=\columnwidth]{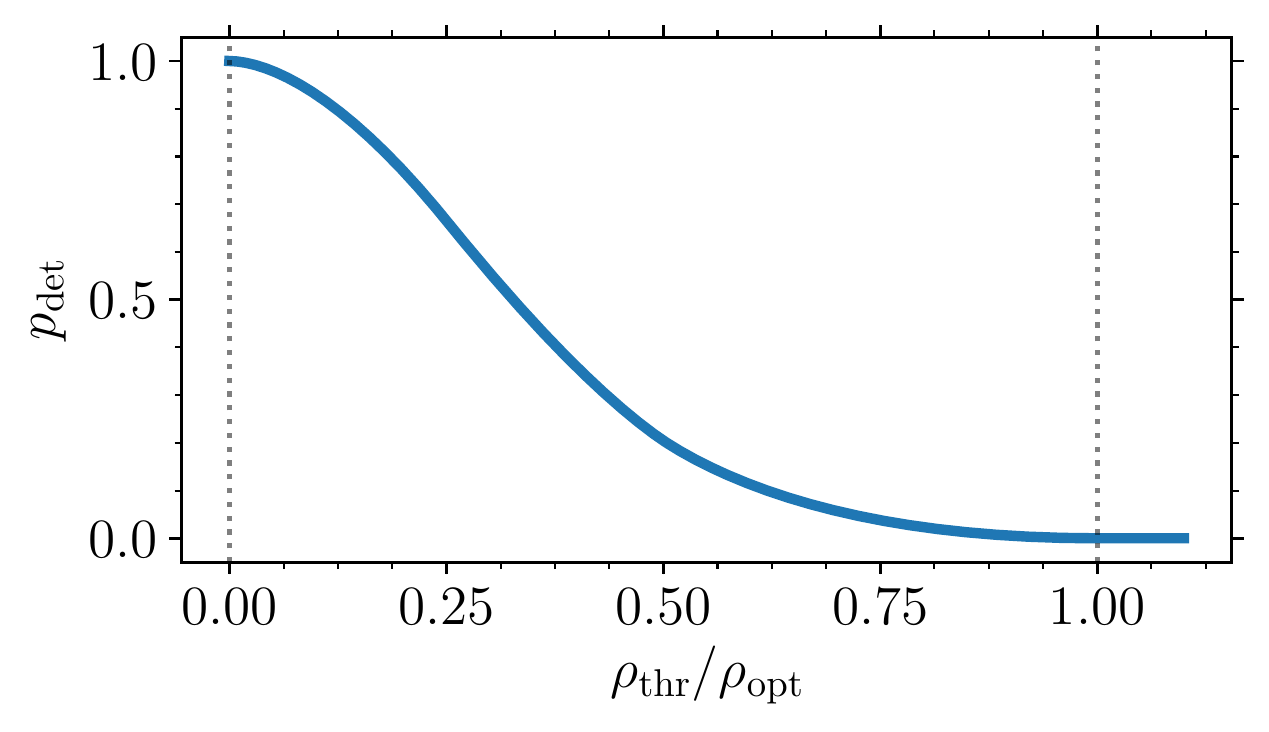}
\caption{Detection probability $p_{\rm det}$ as a function of the ratio between a given detection threshold $\rho_{\rm thr}$ and the SNR obtained assuming optimal orientation $\rho_{\rm opt}$. Here we work in the single-detector approximation and assume $\rho_{\rm thr}=8$.}
\label{fig:pomegafig}
\end{figure}

A population synthesis code would typically return a set of binary parameters like masses, spins and distance. The probability that those given binaries exceed a detection threshold $\rho_{\rm thr}$ is computed by averaging over sky location, polarization angle, and inclination. This is equivalent to evaluating the cumulative probability distribution $P(\omega)$ at the ratio between the threshold SNR and the optimal SNR, i.e. $p_{\rm det} = P(\rho_{\rm thr}/\rho_{\rm opt})$. The detectability function is shown in \autoref{fig:pomegafig}.
All of the binary realizations are detectable in the limit $\rho_{\rm opt}\to \infty$, i.e. $p_{\rm det}=1$. Conversely, none of the realizations are visible below detection threshold, i.e. $p_{\rm det}=0$ if $\rho_{\rm thr}=\rho_{\rm opt}$. For simplicity we use a single-detector SNR threshold  $\rho_{\rm thr}\geq8$, which has been found to act as a good proxy for more elaborate network analysis \cite{2010CQGra..27q3001A}. The function $P(\omega)$ is computed with a Monte Carlo as implemented in the python package \texttt{gwdet} \citep{davide_gerosa_2017_889966}.

\subsection{Hierarchical population inference}
\label{sec:hierarchical}

\subsubsection{Priors and hyper-parameters}

Choices of parameter priors may be motivated by underlying physical intuition (e.g. neutron star masses can not be greater than $\sim 4 M_\odot$) or fundamental constraints (e.g. masses should be positive, speeds can not exceed the speed of light, etc.). However, sometimes intuition or fundamental constraints do not lead us to a definitive prior, as in the case of the astrophysical distribution of compact object masses and spins. In some cases one might be able to make a reasonable guess at the form of the distribution (e.g. Gaussian), but the mean and width may be unknown. Or perhaps even the form itself is completely unknown, and only dictated by unknown properties of the progenitor system. In this case, we can extend our model to an additional level (hence \textit{hierarchical} inference) by using a parametrized prior. The parameters of these priors are the hyper-parameters, and they themselves will have hyper-priors.

\subsubsection{Likelihoods and posteriors}

Hierarchical inference is discussed in detail elsewhere (e.g. Refs.\ \cite{2010PhRvD..81h4029M,2010ApJ...725.2166H,2012PhRvD..86l4032A,2014ApJ...795...64F,2012PhRvD..85b3535T,2012PhRvD..86b3502T,2018arXiv180706726P}), but we summarize the salient points here. We make specific use of the formalism in \citet{mfg18} and \citet{selectioneffects}. The goal is to simultaneously infer the joint posterior probability distribution of the measured physical parameters of each event, as well as the hyper-parameters describing the statistical properties of the entire population. 

The joint probability of strain data from all GW signals $\{h_k\}$ (where $k\in[1,\ldots,N]$ indexes each event), and associated physical parameters describing each signal $\theta_k$ is
\begin{equation} \label{eq:hier_likelihood}
p(\{h_k\},\{\theta_k\}|\beta) =   p(\{h_k\}|\{\theta_k\})\,p(\{\theta_k\}|\beta),
\end{equation}
where $\beta$ are the population hyper-parameters. The GW signals will be produced at a certain rate in parameter space as a function of the hyper-parameters. We first consider a discrete representation of the physical parameter space (e.g. masses, redshits, etc) divided into bins, $l\in[1,\ldots,N_l]$. The data are then the number of events detected in a given bin in this parameter space $n_l$. Assuming non-overlapping statistically-independent signals (and thus bins)\footnote{This assumption is expected to fail for future 3rd-generation ground-based detectors, as well as the LISA space mission.}, the likelihood is the product of a Poisson process in each bin:
\begin{equation} \label{eq:poisson_discrete}
p(\{n_l\}|\beta) = \prod_{l=1}^{N_l}\frac{(r_l(\beta))^{n_l}e^{-r_l(\beta)}}{n_l!},
\end{equation}
where $r_l(\beta)$ is the expected rate of events in bin $l$ as function of hyper-parameters $\beta$. If we make the bins infinitesimally small, then each bin will either have $1$ or $0$ events. This gives the continuum limit
\begin{equation} \label{eq:poisson_continuous}
p(\{\theta_k\}|\beta) \propto e^{-N_\beta} \prod_{k=1}^N r(\theta_k|\beta),
\end{equation}
where $N_\beta = \iint dh\, d\theta\, p(h|\theta)\,r(\theta|\beta)$ is the expected total number of events for a population with hyper-parameters $\beta$, and $r(\theta|\beta) = N_\beta\, p(\theta|\beta)$ such that $\int d\theta\, p(\theta|\beta)=1$. The likelihood $p(h|\theta)$ is normalized over the data, so while the data integral is trivial here we will see soon why its explicit marginalization is useful. Plugging \autoref{eq:poisson_continuous} into \autoref{eq:hier_likelihood}, and again using the statistical independence of signals, gives
\begin{align} \label{eq:hier_likelihood_v2}
    p(\{h_k\},\{\theta_k\}|\beta) \propto e^{-N_\beta} & \prod_{k=1}^N p(h_k|\theta_k)\,r(\theta_k|\beta).
\end{align}
The measured data are usually thresholded using a detection statistic to decide which signals are robust events, and which are spurious or untrustworthy. Upon examining the data, we partition $N$ into ``observable" ($N_\mathrm{obs}$) and ``non-observable" ($N_\mathrm{nobs}$), so that \autoref{eq:hier_likelihood_v2} becomes
\begin{align} \label{eq:hier_likelihood_v3}
    p(&\{h_i\},\{\theta_i\},\{h_j\},\{\theta_j\}|\beta) \propto \nonumber\\
    &e^{-N_\beta} \left[\prod_{i=1}^{N_\mathrm{obs}} p(h_i|\theta_i)\,r(\theta_i|\beta)\right]\left[\prod_{j=1}^{N_\mathrm{nobs}} p(h_j|\theta_j)\,r(\theta_j|\beta)\right].
\end{align}
We now marginalize over the data and parameters of the non-observable events, and divide the probability by $N_\mathrm{nobs}!$ to mitigate over-counting through marginalization. We also marginalize over the number of non-observable events, $N_\mathrm{nobs}$, from $0$ to $\infty$:
\begin{align} \label{eq:hier_likelihood_v4}
    p(&\{h_i\},\{\theta_i\}|\beta) \propto \nonumber\\
    &e^{-N_\beta} \left[\prod_{i=1}^{N_\mathrm{obs}} p(h_i|\theta_i)\,r(\theta_i|\beta)\right] \sum_{N_\mathrm{nobs}=0}^\infty \frac{(N_\beta^\mathrm{ndet})^{N_\mathrm{nobs}}}{N_\mathrm{nobs}!} \nonumber\\
    \propto &\,\, e^{(N_\beta^\mathrm{ndet}-N_\beta)} \prod_{i=1}^{N_\mathrm{obs}} p(h_i|\theta_i)\,r(\theta_i|\beta) \nonumber\\
    \propto &\,\, N_\beta^{N_\mathrm{obs}}\,e^{-N_\beta^\mathrm{det}} \prod_{i=1}^{N_\mathrm{obs}} p(h_i|\theta_i)\,p(\theta_i|\beta),
\end{align}
where,
\begin{align}
    N_\beta^\mathrm{det} &= \int\int_{\{h\in[\mathrm{detection]}\}} dhd\theta\, p(h|\theta)\, r(\theta|\beta) \nonumber\\
    &= \int d\theta\, p_\mathrm{det}(\theta)\, r(\theta|\beta)\nonumber\\
    &= N_\beta \times \int d\theta\, p_\mathrm{det}(\theta)\, p(\theta|\beta)\nonumber\\
    &= N_\beta\,\times\,\epsilon_\beta
\end{align}
is the expected number of detected events in a population model with hyper-parameters $\beta$, such that $N_\beta = N_\beta^\mathrm{det} + N_\beta^\mathrm{ndet}$. The probability of detection as a function of binary parameters is given by $p_\mathrm{det}(\theta)$ from \autoref{sec:gw_param_est}. The efficiency $\epsilon_\beta = \int d\theta\, p_\mathrm{det}(\theta)\, p(\theta|\beta)$ denotes the fraction of merging systems that are detectable for a given hyper-parameter coordinate.

Equation \ref{eq:hier_likelihood_v4} is appropriate if we fully model all factors influencing the number and distribution of detectable GW events, such as the local merger-rate density, the duty cycle of the detectors, etc. In our analysis we construct $r(\theta|\beta)$ from population synthesis simulations, from which we record the fraction of initialized stars that were evolved to become merging BH-BH systems. We do not want to make our analysis sensitive to duty-cycle choices or poorly-constrained scaling parameters that could affect rates, so we marginalize over such factors \citep{2011PhRvD..83d4036S,2012PhRvD..86b3502T,2012PhRvD..85b3535T,2017PhRvD..95l4046G}. This is done by marginalizing over $N_\beta$ with the prior $p(N_\beta)\propto 1/N_\beta$, such that \citep{2018ApJ...863L..41F}
\begin{equation} \label{eq:hier_likelihood_v5}
    p(\{h_i\},\{\theta_i\}|\beta) \propto (N_\mathrm{obs}-1)! \prod_{i=1}^{N_\mathrm{obs}} \frac{p(h_i|\theta_i)\,p(\theta_i|\beta)}{\epsilon_\beta}.
\end{equation}
The first term in the numerator is the single-event likelihood used for GW parameter-estimation. We do not want to repeat all of the effort that went into reducing the raw detector output to a set of likelihood evaluations. Rather, we assume a GW catalog will eventually be provided in the form of a set of posterior samples for each event: 
\begin{equation} \label{eq:event_samples}
p(\theta_i|h_i,\overline{\beta}) = \frac{p(h_i|\theta_i)p(\theta_i|\overline{\beta})}{p(h_i|\overline{\beta})},
\end{equation}
where $\overline{\beta}$ denotes the prior for the BH/NS parameters chosen by the issuers of the catalog (e.g. uniform in component masses, comoving volume, etc.). Plugging \autoref{eq:event_samples} into \autoref{eq:hier_likelihood_v4}, and Monte Carlo integrating over the posterior distribution of event parameters with \autoref{eq:mcmc_marg} gives
\begin{equation} \label{eq:hier_likelihood_v6}
    p(\{h_i\}|\beta) \propto Z_{\overline{\beta}}\,\times N_\beta^{N_\mathrm{obs}}\,e^{-N_\beta \epsilon_\beta} \prod_{i=1}^{N_\mathrm{obs}} \left\langle \frac{p(\theta_i|\beta)}{p(\theta_i|\overline{\beta})} \right\rangle_{\mathrm{post},i},
\end{equation}
where $Z_{\overline{\beta}}$ is the evidence for the interim prior model using the data from all observed events. This is a constant and can thus be ignored. The expectation value in \autoref{eq:hier_likelihood_v6} is taken over samples drawn from the joint posterior distribution of each event in the GW catalog, while the argument is the ratio of the rate of detected-event parameters under our new parametrized model (constructed from simulations) versus the interim prior (used in the catalog construction). Dividing out the influence of the interim prior used in the original event analysis is crucial (e.g. Ref.\ \cite{2017PhRvL.119y1103V}), since our goal is to re-analyze the entire catalog under the new parametrized prior that has been constructed from simulations. For the examples reported in this paper, we approximate the interim prior as being uniform over the region of parameter space with likelihood support, so that we can safely ignore this subtlety.

\begin{figure}
\begin{center}
\includegraphics[width=0.5\textwidth]{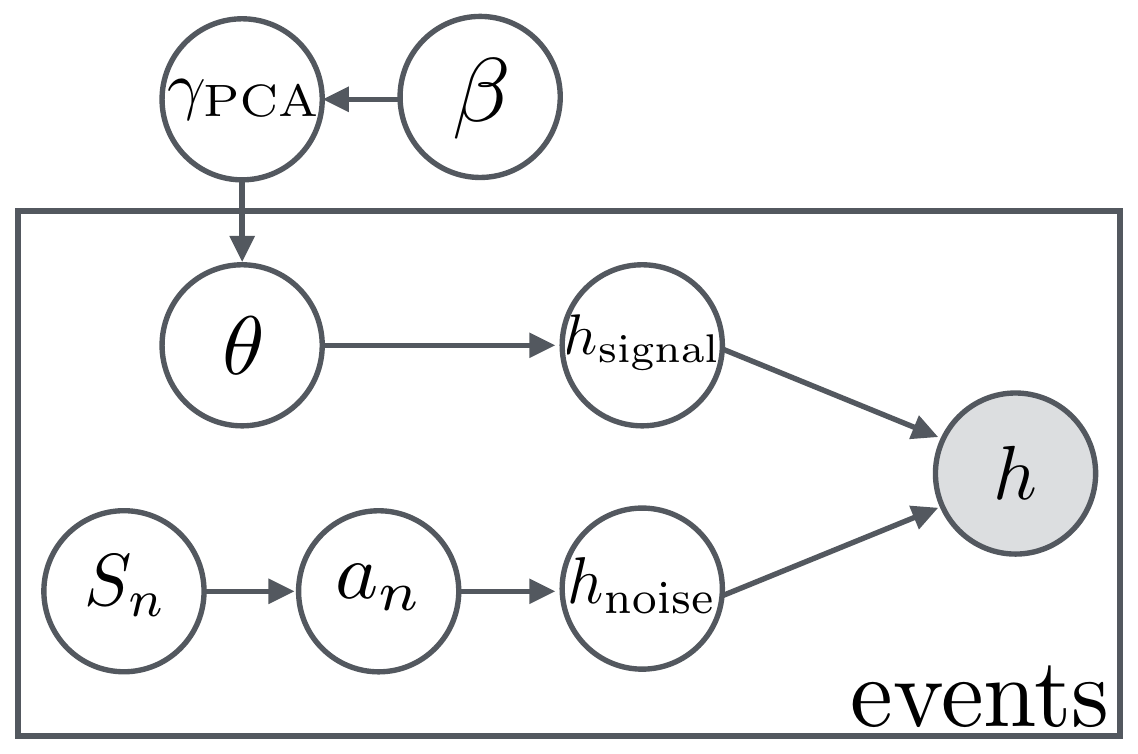}
\end{center}
\caption{A probabilistic graphical model illustrating \autoref{eq:hier_likelihood}. The detector output, $h$, depends on noise and signal processes. The noise may be decomposed onto a Fourier basis with coefficients, $a_n$, whose variance in turn may be constrained by a model for the power-spectral density, $S_n$. The strain induced by each signal depends on the intrinsic and extrinsic parameters of each binary $\theta$. We place a parametrized prior on a subset of these parameters, given by orthogonal basis distributions determined from PCA of population synthesis simulations, $\gamma_{\rm PCA}$. The amplitude of each basis distribution has a Gaussian prior from GP training on these simulations, informed by some hyper-parameters, $\beta$.}
\label{fig:pgm}
\end{figure}

Monte Carlo integrating over the posterior distribution of event parameters in \autoref{eq:hier_likelihood_v5} gives
\begin{equation} \label{eq:hier_likelihood_v7}
    p(\{h_i\}|\beta) \propto Z_{\overline{\beta}}\,\times  (N_\mathrm{obs}-1)! \prod_{i=1}^{N_\mathrm{obs}} \frac{1}{\epsilon_\beta}\left\langle \frac{p(\theta_i|\beta)}{p(\theta_i|\overline{\beta})} \right\rangle_{\mathrm{post},i}.
\end{equation}
The rate, $N_\beta$, and distribution, $p(\theta|\beta)$, are constructed using the simulation and emulation scheme described in \autoref{sec:stats}, where the former is found by training on the fraction of ZAMS stars that form merging BH-BH systems. \autoref{fig:pgm} shows the probabilistic graphical model for our inference framework, and illustrates the chain of conditional dependencies for constraining the parameters of each event with a prior that is a function of progenitor and evolutionary properties. We use both \autoref{eq:hier_likelihood_v6} and \autoref{eq:hier_likelihood_v7} in the following test cases.

\section{Results}
\label{sec:results}

We now implement our new framework on three case studies. These case studies begin with a toy model (\autoref{sec:toymodel}), then increase in complexity and astrophysical realism using both public data (\autoref{sec:compasmodel}) and tailored simulations (\autoref{sec:popcasestudy}) to showcase how one might use our findings in practice.

\subsection{Toy Model}
\label{sec:toymodel}

\begin{figure*}
\begin{center}
	\includegraphics[width=0.49\textwidth]{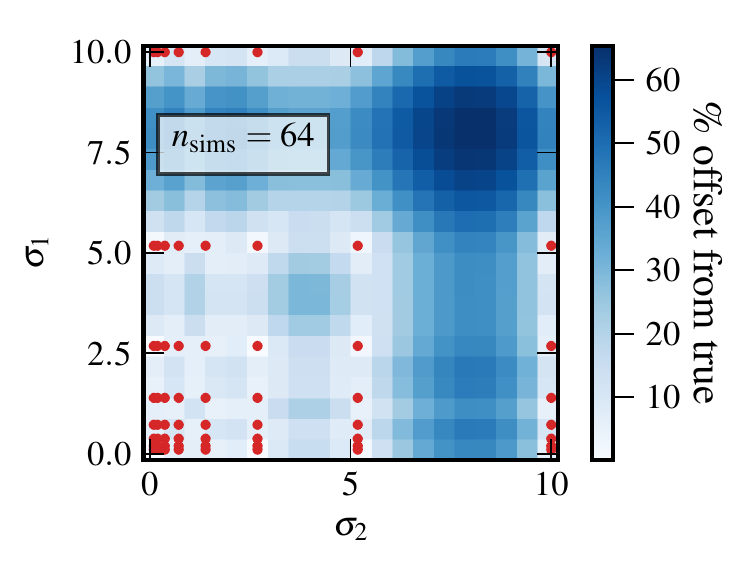}
	\includegraphics[width=0.49\textwidth]{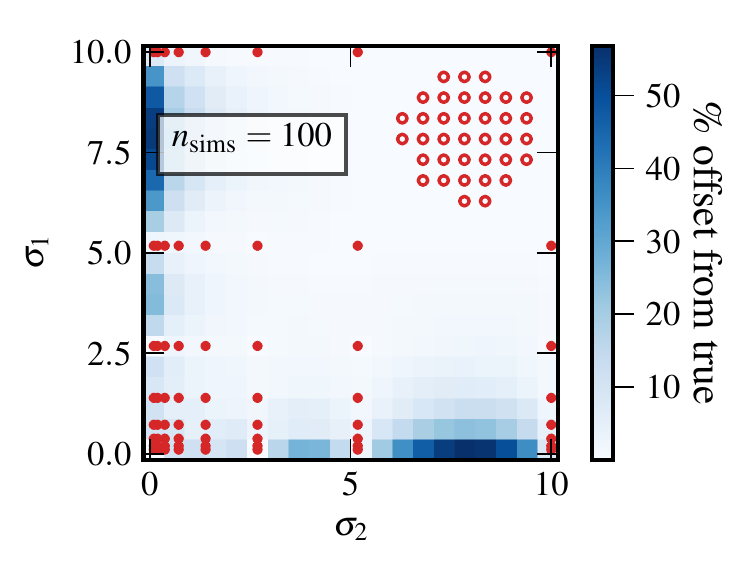}
\end{center}
\caption{Testing the accuracy of our GP emulator for the model of  \autoref{eq:gauss_prod}. In the left panel we create training data on an evenly-spaced $8\times8$ grid in $\log_{10}\sigma_{1,2}$ space (red points). We achieve a data compression factor of $\sim 500$, then train a GP in each of the reduced basis features. The GP prediction is compared to the analytic result across $\sigma_{1,2}$ space by taking the GP-mean (offset by $1$ $\sigma$), rotating back to the full $z_{1,2}$ basis, then finding the maximum difference from the analytic value in any $z_{1,2}$ bin. Low accuracy locations are used to inform the positions at which new simulations are performed. These additional points are shown in the right panel as empty circles, where we see that their addition improves accuracy across the entire hyper-parameter space.}
\label{fig:toymodel_accuracy}
\end{figure*}

\begin{figure}
\begin{center}
	\includegraphics[width=0.5\textwidth]{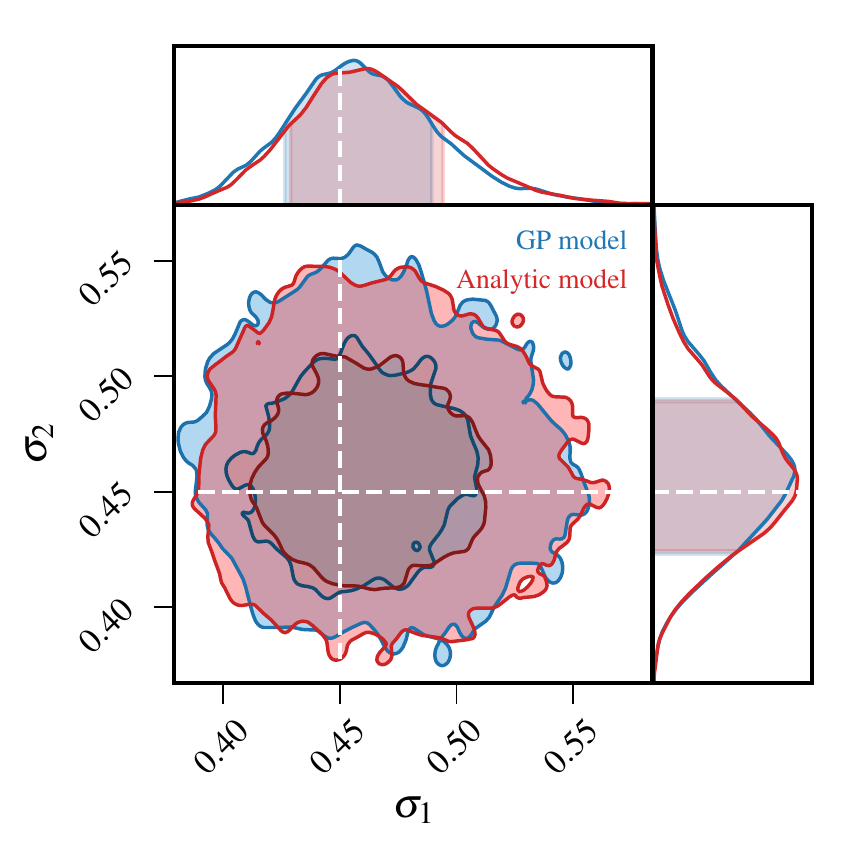}
	\caption{Comparison of posterior recoveries of population hyper-parameters from a catalog of $100$ sources with spin-alignment distribution given by \autoref{eq:gauss_prod} \citep{2017PhRvD..96b3012T}. The true hyper-parameter coordinate, $\{\sigma_1=0.45, \sigma_2=0.45\}$ is indicated via intersecting white dashed lines.}
	\label{fig:testpop_modelcompare}
\end{center}
\end{figure}

Our first demonstration corresponds to the inference of binary spin-alignment distributions. Spin alignments are indeed recognized as one of the cleanest indicators for constraining BH formation and evolutionary processes \cite{2013PhRvD..87j4028G,2014PhRvD..89l4025G,2016PhRvD..93d4071T,2016ApJ...832L...2R,2017MNRAS.471.2801S,2017PhRvD..96b3012T,2018ApJ...854L...9F,2017Natur.548..426F,2018arXiv180601285A}.
Here we implement the approach developed by Talbot and Thrane \cite{2017PhRvD..96b3012T}. The observed quantities in this model are  the projection of each binary component's spin onto the orbital angular momentum vector: 
\begin{equation}
z_1 = \hat{L}\cdot\hat{S}_1, \quad z_2 = \hat{L}\cdot\hat{S}_2,
\end{equation}
where $z_{\{1,2\}} \in [-1,1]$. Dynamical capture mechanisms in, e.g., a globular cluster are expected to produce an isotropic distribution of spin alignments
\begin{equation}
p_0(z_1,z_2)=\frac{1}{4}.
\end{equation}
For field binaries, the evolutionary path of each progenitor star (in particular natal kicks during supernova) is assumed to produce a truncated Gaussian distribution of alignments. Two hyper-parameters $\sigma_1$ and $\sigma_2$ control the degree with which (anti-)alignment is favored:
\begin{equation} \label{eq:gauss_prod}
p_1(z_1,z_2) = \frac{2}{\pi}\frac{1}{\sigma_1}\frac{e^{-(z_1-1)^2 / 2\sigma_1^2}}{\mathrm{erf}(\sqrt{2}\sigma_1)} \frac{1}{\sigma_2}\frac{e^{-(z_2-1)^2 / 2\sigma_2^2}}{\mathrm{erf}(\sqrt{2}\sigma_2)}. 
\end{equation}
In this model, $\sigma=0$ produces perfect alignment, while $\sigma=\infty$ tends to the dynamic-capture distribution.

We use $p_1(z_1,z_2)$ as the test destribution to be inferred. This probability function has hard-edges at $[z_1=\pm1,z_2=\pm 1]$, making it challenging to learn and thus an excellent  testbed to test our framework. The observed parameters from each GW binary event are $\theta\in\{z_1,z_2\}$, and the hyper-parameters of the population are $\beta\in\{\sigma_1,\sigma_2\}$. The parameter probabilities are represented on a $40\times40$ binning in $\{z_1,z_2\}$ space.

We generate training data using \autoref{eq:gauss_prod} for a range of $\sigma_{1,2}\in[0.1,10]$ values, sampled uniformly in log-space. To examine how many training datasets are needed, we create grids of training data with different densities in hyper-parameter (i.e. $\beta$) space. We find a compressed basis representation of the training-data distributions, then train a GP at each bin in the compressed parameter space. In all cases we find that the initial parameter binning can be compressed by a factor of $\sim 500$ with high-fidelity\footnote{We compute the normalized inner product of the training data (flattened to be the vector of all samples in the dataset) with the compressed data (which has been rotated back into the full parameter basis). With only $3$ reduced basis distributions, corresponding to a compression of $(40\times40) / 3 \approx 533$, we achieve discrepancies from true that are of $\mathcal{O}(10^{-16})$.}. In this case the compression and training is performed on the logarithm of the training data, since this reduces the dynamic range of values across parameter space and ensures that the predicted proability values will always be positive. We can now predict the distribution values in compressed parameter space, and rotate this back into the full parameter space to construct the final predictions.

Figure~\ref{fig:toymodel_accuracy} shows validation studies for different numbers of initial training data. For an evenly-spaced grid of $8\times8=64$ training datasets in hyper-parameter space, we achieve an accuracy of better than $\sim 50\%$ across the majority of the space. The worst performance occurs in parts of hyper-parameter space that are voids of simulations. We find the $36$ worst accuracy locations, and add these as additional simulations to improve accuracy to better than $10\%$. Similar accuracy is given by an Latin-hypercube design of $100$ training datasets. 

We now test our framework on a simulated population, consisting of $100$ sources drawn from $p(z_1,z_2)$ with $\beta=\left\{\sigma_1=0.45,\sigma_2=0.45\right\}$. A comparison of the joint posterior probability distribution of $\{\sigma_1,\sigma_2\}$ as recovered by the analytic model [\autoref{eq:gauss_prod}] and the GP framework is shown in \autoref{fig:testpop_modelcompare}. The GP framework is trained on $100$ simulations from a Latin-hypercube design; we use this design because it is our standard approach for efficiently sampling the high-dimensional hyper-parameter space of binary stellar evolution, and it gives similar emulation accuracy to the adaptive design in the right panel of Fig.~\ref{fig:toymodel_accuracy}. In this analysis, we have propagated all uncertainties from the GP prediction and the hyper-parameters of the trained GP covariance function into the final model. The agreement is excellent, with the true hyper-parameter coordinate lying well within the $68\%$ credible region of both techniques. We have not incorporated the effect of individual event measurement uncertainties, which will be explored in the next examples. 
\subsection{COMPAS Populations}
\label{sec:compasmodel}

We now test our framework on an example with greater astrophysical realism. We take publicly-available populations\footnote{Populations available at \href{http://www.sr.bham.ac.uk/compas/data}{http://www.sr.bham.ac.uk/compas/data}.} of synthesized binary BHs from \citet{2017NatCo...814906S} as training data. In the aforementioned paper, the authors introduce \texttt{COMPAS}: a code (broadly similar to \texttt{BSE}) for evolving zero-age-main-sequence (ZAMS) binary star systems through classical isolated evolution (i.e. including common-envelope stages). By simulating low-metallicity populations and following the binary evolution, the authors find that all three initial Advanced LIGO events (GW$150914$, GW$151226$, and LVT$151012$) could have been formed with a single model in an environment with $Z\sim0.05Z_\odot$. In Ref.~\cite{2017NatCo...814906S}, the statistic for checking whether a given simulated binary was consistent with forming each individual detected GW event was whether the simulated binary's total mass (chirp mass) fell within the quoted $90\%$ credible region for GW$150914$ (GW$151226$, LVT$151012$), and whether the mass-ratio exceeded the quoted $90\%$ credible lower bound. While this is a reasonable measure of consistency, it does not provide a corresponding measure of statistical credibility for the inferred progenitor metallicities. By contrast, our framework allows the posterior probability distribution of progenitor metallicities to be recovered.

We use populations produced with fiducial assumptions under different metallicities, corresponding to $Z = \{0.05, 0.1, 0.25\} Z_\odot$. In this example, $Z$ is the only hyper-parameter that we aim to infer. All binaries reported merge within a Hubble time, and we incorporate detector selection effects using the detection probability mentioned in \autoref{sec:gw_param_est}. In principle we would use the binary component masses, spin information, and redshift to discriminate progenitor properties and evolutionary paths. But since there is only a limited amount of information that can be inferred based on these three training populations, we opt for simplicity and only use the chirp mass information from each binary. We do not consider rate information either, such that our likelihood is given by \autoref{eq:hier_likelihood_v7}. By using these publicly-available populations as training data, we implicitly approximate all BH systems as forming from progenitors with a common metallicity.

We compress histograms of each population's chirp masses from $80$ initial bins down to a PCA basis of size $2$ (which is set by the small number of training populations). The compressed training data is then interpolated over metallicity using a GP with a squared-exponential kernel. This procedure gives a model for the distribution of {detectable} chirp masses as a function of metallicity. 

\begin{table} 
	\begin{center}
	\caption{\label{tab:real_bbh} The existing catalog of binary BH detections from Advanced-LIGO--Advanced-Virgo, with measured source-frame chirp masses and merger redshifts reported as median values and associated $90\%$ credible bounds.}
	\clearpage{}

\centering
\begin{tabular}{@{\hskip 0.07in}c@{\hskip 0.07in}@{\hskip 0.07in}c@{\hskip 0.07in}@{\hskip 0.07in}c@{\hskip 0.07in}@{\hskip 0.07in}c@{\hskip 0.07in}}

    \hline\hline
		Event & Chirp mass $\mathcal{M}$ & Merger redshift $z$ & Refs.\\
        \hline\\
    GW150914&  $28.1\substack{+1.8 \\ -1.5}\,M_\odot$ & $0.09\substack{+0.029 \\ -0.036}$ & \cite{2016PhRvL.116x1102A,2016PhRvX...6d1015A}\\[1ex]
    LVT151012& $15.1\substack{+1.4 \\ -1.1}\,M_\odot$  & $0.201\substack{+0.086 \\ -0.091}$ & \cite{2016PhRvX...6d1015A} \\[1ex]
    GW151226&  $8.88\substack{+0.33 \\ -0.28}\,M_\odot$ & $0.094\substack{+0.035 \\ -0.039}$ & \cite{2016PhRvL.116x1103A,2016PhRvX...6d1015A}  \\[1ex]
    GW170104&  $21.1\substack{+2.4 \\ -2.7}\,M_\odot$ & $0.18\substack{+0.08 \\ -0.07}$ & \cite{2017PhRvL.118v1101A}\\[1ex]
    GW170608&  $7.9\substack{+0.2 \\ -0.2}\,M_\odot$ & $0.07\substack{+0.03 \\ -0.03}$ &\cite{2017ApJ...851L..35A}\\[1ex]
    GW170814&  $24.1\substack{+1.4 \\ -1.1}\,M_\odot$ & $0.11\substack{+0.03 \\ -0.04}$ &\cite{2017PhRvL.119n1101A}\\[1ex]
    \hline
\end{tabular}
\clearpage{}
    \end{center}
\end{table}
We perform a simple test using chirp-mass and redshift information from the catalog of existing BH detections, see \autoref{tab:real_bbh}. We make the very simple approximation that the source-frame chirp mass and merger redshift posterior distributions are Gaussian and uncorrelated, from which we can easily draw posterior samples.  We draw $100$ independent posterior samples for each event and use these samples to propagate parameter-estimation uncertainty into our population hyper-parameter inference. This is obviously a highly simplified representation of the real event posteriors, but it outlines the scheme one would use when provided with the samples from the true GW catalog. 

Another subtlety that we do not consider here (but that must be accounted for in a real analysis) is the influence of the original priors from the parameter-estimation analysis of each individual event (c.f.\ \autoref{sec:hierarchical}). In the current Advanced-LIGO--Advanced-Virgo searches, the component mass priors are uniform, while the luminosity distance prior assumes the mergers occur uniformly in comoving volume. These choices do not translate to uniform priors in chirp mass or redshift, so that we should re-weight the posterior samples from each event to reflect the likelihood, then apply our newly-formulated parameter priors (as a function of population hyper-parameters) to the entire detected event catalog. In this analysis, we simply assume that the chirp mass and redshift priors were uniform in the analysis of each GW event.

\begin{figure}
\begin{center}
	\includegraphics[width=\columnwidth]{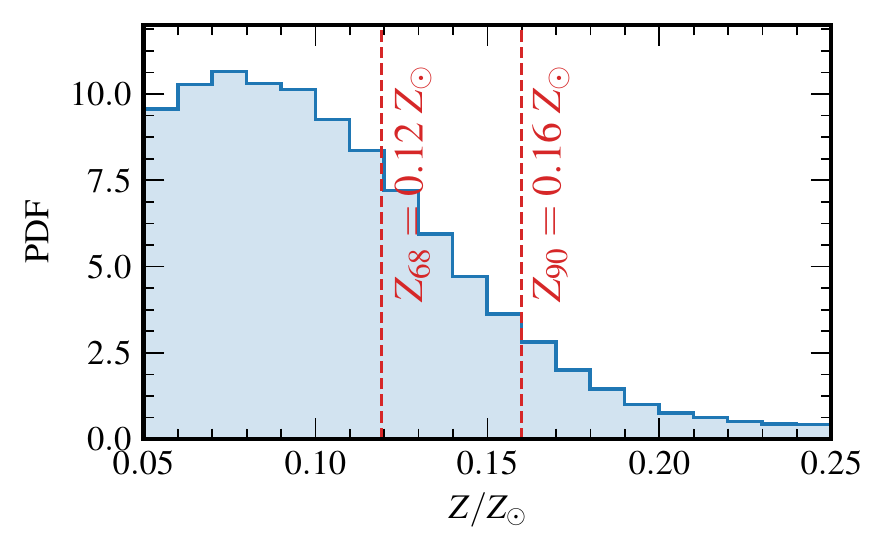}
\end{center}
\caption{Posterior probability distribution of progenitor metallicity $Z$, as inferred by an analysis of the current BH catalog in \autoref{tab:real_bbh} using a model for the chirp mass distribution that is conditioned on simulations from \cite{2017NatCo...814906S}. Dashed vertical lines marks the 68\% and 90\% confidence intervals.}
\label{fig:compas_pop1}
\end{figure}

\begin{figure*}
\begin{center}
	\includegraphics[width=0.85\textwidth]{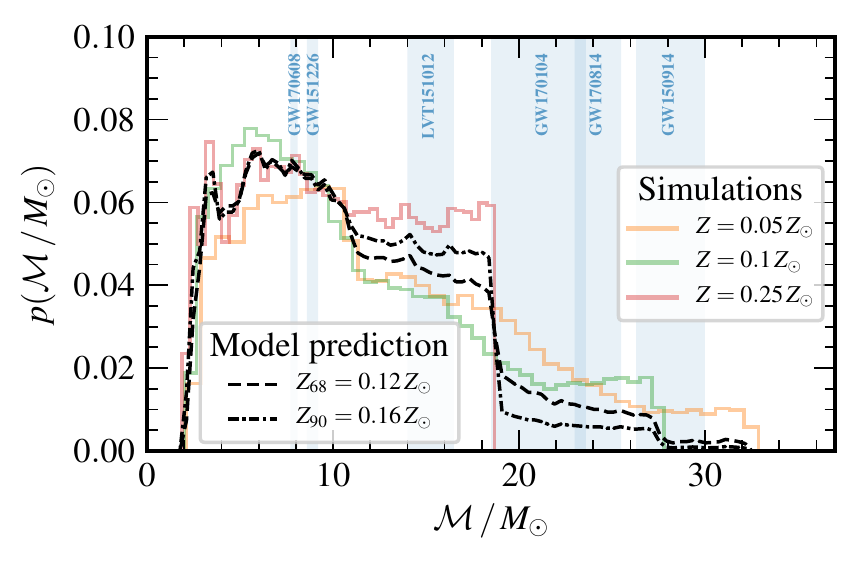}
\end{center}
\caption{Intrinsic distribution of BH binary chirp masses for progenitor metallicity values corresponding  the simulations by \cite{2017NatCo...814906S} (colored lines) and the $68\%$ and $90\%$ upper limits from an analysis of the current GW catalog (black dashed lines). The chirp masses of the GW events in the catalog are shown with vertical blue bands.}
\label{fig:compas_pop2}
\end{figure*}

The resulting posterior distribution for progenitor metallicity is shown  \autoref{fig:compas_pop1}, where the $68\%$ and $90\%$ upper limits are found to be $Z<0.12\,Z_\odot$ and $Z<0.16\,Z_\odot$, respectively. This is in broad agreement with \citet{2017NatCo...814906S}, who found that the three events required $Z\simeq0.05\,Z_\odot$. Our constraints reflect uncertainties in the GP model prediction and the parameter estimation of each event. In \autoref{fig:compas_pop2} we also show the reconstructed {intrinsic} chirp-mass distribution of binary BHs at metallicities corresponding to our credible limits, as well as the original training distributions. We see that our model correctly interpolates the physical behavior on which it was trained (including some sharp features), namely that the distribution of chirp masses shifts to smaller values as the progenitor metallicity is increased. Physically, this is because stellar winds are weaker in stars with lower metallicity, that thus tend to form heavier BHs like the ones detected by Advanced LIGO \cite{2016Natur.534..512B,2016ApJ...818L..22A,2016MNRAS.463L..31L,2017NatCo...814906S}.  The events of the current binary BH catalog are shown as vertical bands corresponding to the $90\%$ credible region of chirp mass.
 
\subsection{BSE Population Synthesis}
\label{sec:popcasestudy}

To further showcase the effectiveness of our statistical framework, we now consider a more elaborate set of input data. We perform a dedicated program of population-synthesis simulations to predict properties of BH binaries from isolated binary stars. 

We use a modified version of the public population synthesis code \texttt{BSE} \cite{2002MNRAS.329..897H,2000MNRAS.315..543H}. The modifications implemented here are the same described in Refs.\ \cite{2016MNRAS.463L..31L,2018arXiv180103099L}: wind mass loss prescriptions according to Ref.\ \cite{2010ApJ...714.1217B} and core-collapse remnant mass relationship following Ref.\ \cite{2008ApJS..174..223B}. These minimal updates are necessary to generate any BHs of masses $\gtrsim 10 M_\odot$ like the ones that are now detected, and thus to attempt a comparison with the Advanced-LIGO--Advanced-Virgo data. We stress, however, that this study is not meant to rival with the full complexity of state-of-the-art binary evolution codes, but rather highlight the potential of our inference pipeline. 

\texttt{BSE} requires us to specify distributions of binary stars on their zero-age main sequence (ZAMS), and a variety of flags encoding assumptions of the underlying stellar physics. We distribute primary masses $m_1$ from an initial mass function $p(m_1)\propto m_1^{-2.3}$ in $[5,100] M_\odot$; mass ratios $q=m_2/m_1$ uniformly in $[0,1]$; initial separations $R$ uniformly in $\log_{10}$ in $[10,10^5] R_\odot$; eccentricities $e$ from a thermal distribution $p(e)\propto e$; and redshifts $z$ uniformly in comoving volume  using the Plank cosmology  \cite{2016A&A...594A..13P} (c.f. Ref.\ \cite{2017ApJ...846...82Z} for similar choices).

The evolutionary flags are the quantities that should be treated as hyper-parameters, and that could potentially be constrained with current and future catalogs of GW events. 
For simplicity, we present results considering a 3-dimensional hyper-parameter space, but our method is fully generalizable and scalable to higher dimensions. We fix all flags to their default value in BSE, except for the following three:
\begin{enumerate}
\item \emph{Metallicity of the ZAMS star: $Z$}. As already highlighted above, the progenitor metallicity has a large impact on the properties of the resulting BHs. Metallicity strongly affects massive star winds and thus the mass that remains available to form the final compact object \cite{2010ApJ...716..615O,2010ApJ...714.1217B,2010ApJ...708..117B,2018MNRAS.474.2959G,2018arXiv180105433K,2015MNRAS.452.1068C,2012ApJ...749...91F}. Here we consider a metallicity range $0.0001\leq Z \leq 0.03$ where $Z_\odot=0.02$ \cite{2002MNRAS.329..897H}.
\item \emph{Kicks imparted to BHs at formation: $\sigma_{\rm k}$}. As stars collapse (perhaps exploding into supernovae), asymmetries in the emitted material and neutrinos may impart a recoil to the newly formed compact object (e.g.\ Ref.\ \cite{2013MNRAS.434.1355J}). Observations of galactic pulsar proper motions suggest that NS recoils are well modeled by a single Maxwellian distribution with 1D root-mean-square $\sigma_{\rm k}\sim265$ km/s \cite{2005MNRAS.360..974H,2017MNRAS.467..298R}. Whether BHs receive any kick at formation is still a matter of debate. On the one hand, X-ray binary measurements hint at large kick velocities \cite{2012MNRAS.425.2799R} (c.f.\ also Ref.\ \cite{2017PhRvL.119a1101O} for a GW constraint). Conversely, theoretical arguments and simulations suggest that kicks for BHs might be suppressed because of material falling back after the explosion \cite{2013MNRAS.434.1355J,1999ApJ...522..413F,2001ApJ...554..548F}. This is a clear case where a method like ours, allowing for a direct estimate of $\sigma_{\rm k}$, might show its potential. We consider BH recoils in the range $0\, {\rm km/s}\leq\sigma_{\rm k}\leq 265 \, {\rm km/s}$ independently of BH mass or other parameters (see Ref.\ \cite{2018PhRvD..97d3014W} for a discussion of this point).
\item \emph{Efficiency of the common envelope: $\alpha_{\rm ce}$}. After the first star collapses, the binary system consists of a BH and an extended star. As this second star expands into a supergiant, it may overflow its Roche Lobe and undergo unstable mass transfer to the BH \cite{1976IAUS...73...75P,1993PASP..105.1373I,2000ARA&A..38..113T,2013A&ARv..21...59I}. The envelope of the giant engulfs the companion BH. In this process, known as the common-envelope stage, a fraction $\alpha_{\rm ce}$ of the binary's orbital energy is transferred to the envelope, thus hardening the binary. In the standard evolutionary channel considered here, common envelope evolution is the key stage to produce BHs able to merge within a Hubble time. The details of the common envelope phase are still very uncertain \cite{2010ApJ...716..114X,2011ApJ...743...49L,2014MNRAS.442.1980Z,2012ApJ...759...52D}, and are arguably one of the most important stellar \mbox{(hyper-)parameters} that can potentially be measured with GW data. Here we vary $\alpha_{\rm ce}$ in $[0.001,10.0]$.
\end{enumerate}

We use $\{ Z, \sigma_{\rm k}, \alpha_{\rm ce}\}$ as hyper-parameters, thus  implicitly assuming that all stars in the same simulated universe share common values of those quantities. While this might be a good working assumption for, e.g., $\alpha_{\rm ce}$, it is surely not true for other parameters like the metallicity. That said, our methods can be straightforwardly generalized to a distribution of metallicities with parameters that can be treated as hyper-parameters in our inference instead of $Z$ itself (much like $\sigma_{\rm k}$, which is a parameter in the Maxwellian kick distribution, not the kick velocity itself).

We perform $125$ BSE simulations distributing $\log_{10} Z$, $\sigma_{\rm k}$, and $\log_{10} \alpha_{\rm ce}$ on a Latin hyper-cube as described in \autoref{sec:simdesign} and drawing $N=10^7$ ZAMS binaries at each point in hyper-parameter space. Each of these $125\times N$ simulated stars is filtered according to two criteria: $(i)$ a BH binary is formed, and $(ii)$ it merges before $z=0$. Binaries passing these cuts are assigned an Advanced LIGO detection probability, $p_{\rm det}$ (c.f. \autoref{sec:gw_param_est}). 

\begin{figure} 
\begin{center}
	\includegraphics[width=0.48\textwidth]{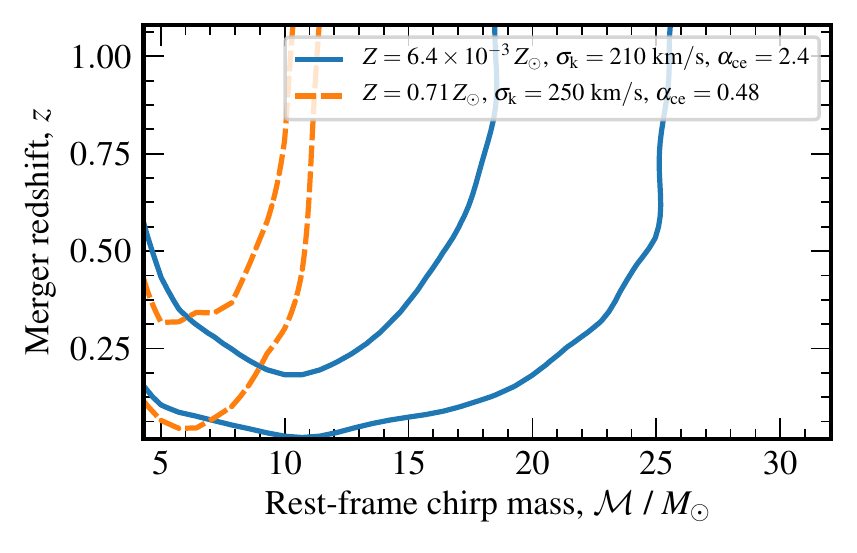}
\end{center}
\caption{An example of two \texttt{BSE} training simulations, showing the intrinsic $\{\mathcal{M},z\}$ distribution of merging BH binaries. Contours enclose $68\%$ and $90\%$ of simulated binaries, where the blue solid lines are for a very low metallicity progenitor scenario, while the orange dashed lines are for a simulation close to solar metallicity.}
\label{fig:detratefig1}
\end{figure}

\begin{figure} 
\begin{center}
	\includegraphics[width=0.48\textwidth]{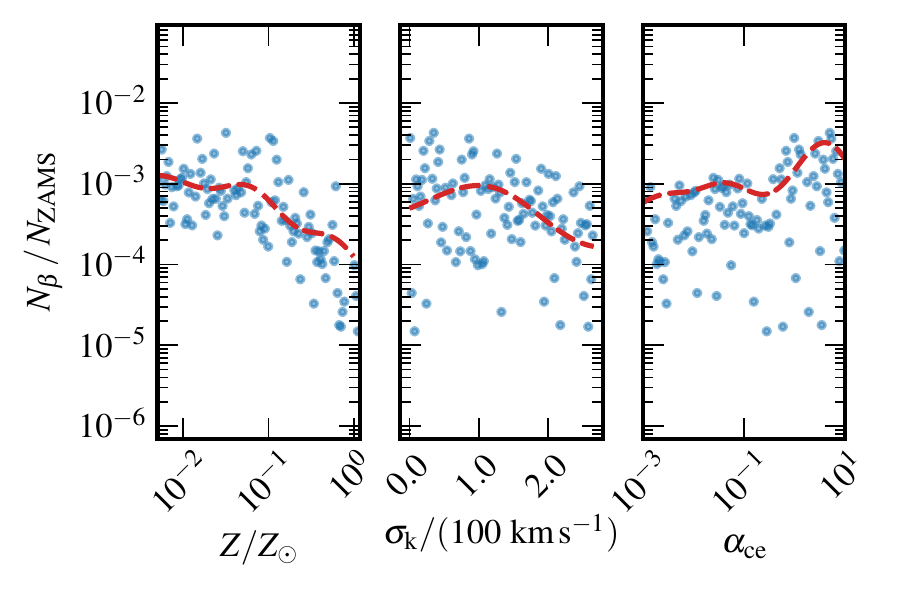}
\end{center}
\caption{Fraction of ZAMS stars that form merging binary BH systems. The three panels show fractions in each of our three hyper-parameters: metallicity $Z$, natal kicks $\sigma_{\rm k}$ and common-envelope efficiency $\alpha_{\rm ce}$. The dashed lines in each panel show predictions from a GP that has been trained on these rates, with only the hyper-parameter relevant to the panel varied in the prediction.}
\label{fig:detratefig2}
\end{figure}

Each BSE simulation returns a population of BH binaries  characterized by their masses and merger redshifts, which we use as measured event parameters in our statistical inference. Examples of the intrinsic $\{\mathcal{M},z\}$ distribution for two of these simulations are shown in \autoref{fig:detratefig1}, where low $Z$ values ensure stars are able to form massive BHs. The relative merger rate (i.e. the fraction of ZAMS stars that form merging BH binaries) is shown in \autoref{fig:detratefig2}. The rate decreases with Z because (i) fewer BHs are formed in favor of NSs (which are not considered here for simplicity) and (ii) stars become puffier at large Z and are more likely to merge earlier in the evolution (e.g. Ref.~\citep{2016MNRAS.463L..31L}). The rate also decreases with $\sigma_{\rm k}$ because strong kicks more easily unbind binaries (e.g.\ Ref.\ \cite{1999A&A...346...91B,2018PhRvD..97d3014W}).

We do not know a-priori how many ZAMS stars survive as merging BH binaries. Some points in hyper-parameter space lead to only a handful of events, giving a jagged distribution in parameter space that suffered from finiteness. To counter this, we require a simulation to provide at least $500$ systems in order to be included in our training and validation procedures. This leaves $115$ out of the original $125$ hyper-parameter coordinates. Even though this renders our simulation coordinates no longer a perfect LH design, it creates a training set with smoother and more robust parameter distributions. Out of the surviving $115$ simulations, we train our GP emulator on a randomly chosen $100$, with another $14$ selected for independent validation of the GP, and the final simulation left as a test population for the full hierarchical Bayesian pipeline.

For each training simulation, we create a normalized KDE-smoothed\footnote{We use the \texttt{scipy.stats} implementation of Gaussian kernel density estimation  with a bandwidth selected by Scott's Rule \citep{2015mdet.book.....S}.} $2$D distribution in intrinsic chirp mass, $\mathcal{M}$, and merger redshift, $z$, with a common $20\times20$ binning scheme. The distributions are PCA-compressed by a factor of $8$, with a compression fidelity of better than $0.01\%$. The remaining $50$ features (or ``bins'') in the compressed distributions are each interpolated over the three-dimensional hyper-parameter space of $\beta=\{\log_{10}Z, \sigma_{\rm k}, \log_{10}\alpha_{\rm ce}\}$ using GPs with squared-exponential kernels. We denote a match statistic that is the normalized inner product of the bin heights in the validation distribution with the GP-predicted distribution. With maximum a-posteriori GP kernel parameters, the $14$ validation distributions (KDE-smoothed and normalized) all match their GP-predicted distributions to better than $7\%$. We also train a separate GP on the fraction of ZAMS stars that survive as merging binary BH systems, which was used to make the smooth rate curves in \autoref{fig:detratefig2}. This is convolved with detector selection effects to compute the fraction of merging systems that are detectable in Advanced LIGO. 

We still have one population that was held out of the GP emulator training and validation, which we now use as data for a test of the entire hierarchical Bayesian pipeline. The hyper-parameters of this population are $\beta=\{Z = 7.3\times10^{-4}, \sigma_{\rm k}=100\,\mathrm{km}/\mathrm{s}, \alpha_{\rm ce}=0.021\}$. We weight each system in the population by its detection probably, then randomly select $100$ to be our catalog, corresponding to (depending on duty-cycle and sensitivity assumptions) a few years of Advanced Advanced-LIGO--Advanced-Virgo observations. The evaluated match statistic between this distribution and our GP prediction is $\sim 0.5\%$. We take two approaches to analyze this catalog: 
\begin{itemize}
\item [(i)] using only the information given by the $\{\mathcal{M},z\}$ distribution of sources, see \autoref{eq:hier_likelihood_v7};  
\item [(ii)] artificially scaling the rate-GP to predict $100$ detected events for the test hyper-parameters so that we can use a Poisson likelihood, see \autoref{eq:hier_likelihood_v6}.
\end{itemize}

\begin{figure} 
\begin{center}
	\includegraphics[width=0.5\textwidth]{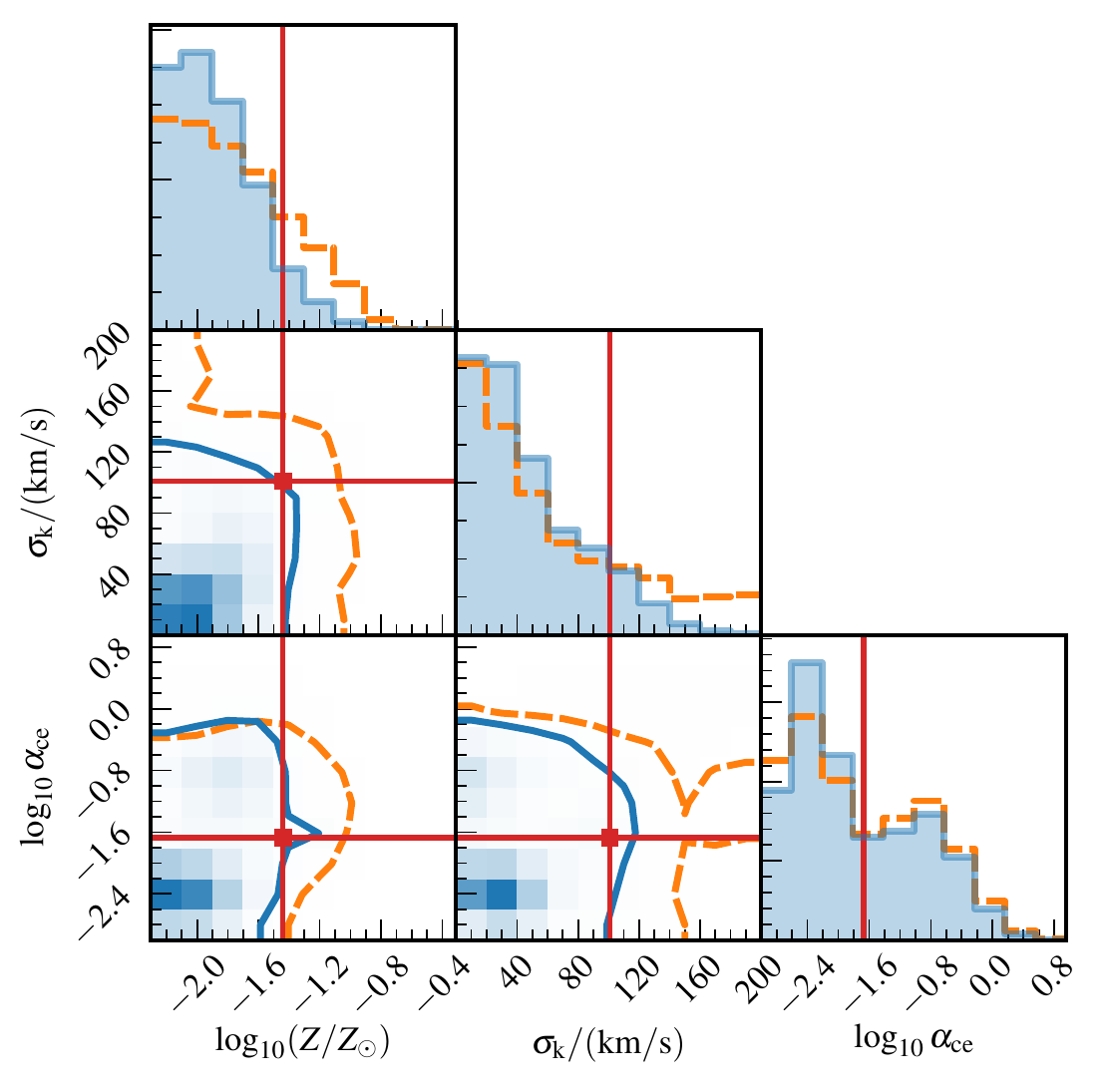} \\
\end{center}
\caption{Corner plot showing $1$D-marginalized posterior distributions of binary BH population hyper-parameters along the diagonal, and pairwise $2$D-marginalized posterior distributions in the lower axes (lines denote $90\%$ credible regions). The true hyper-parameters are indicated with red lines. The data were $100$ binary BHs from a population simulated with \texttt{BSE} that was held out of our GP emulator training. Results are for a distribution-only likelihood [\textit{orange dashed}, \autoref{eq:hier_likelihood_v7}], and a re-scaled Poisson-rate likelihood [\textit{blue solid}, \autoref{eq:hier_likelihood_v6}].}
\label{fig:abse_results_summary_mapgpkern1}
\end{figure}

The recovered posterior probability distributions of population hyper-parameters are shown in \autoref{fig:abse_results_summary_mapgpkern1}, where all are consistent with the true values. 
We have not marginalized over GP kernel posteriors or the GP prediction uncertainties so that we may see the effect (or in this case lack thereof) of systematic offsets from interpolation errors. 
We have also not modeled parameter uncertainties in the cataloged events, but these can be straightforwardly incorporated.

\begin{figure}
\begin{center}
    \includegraphics[width=0.5\textwidth]{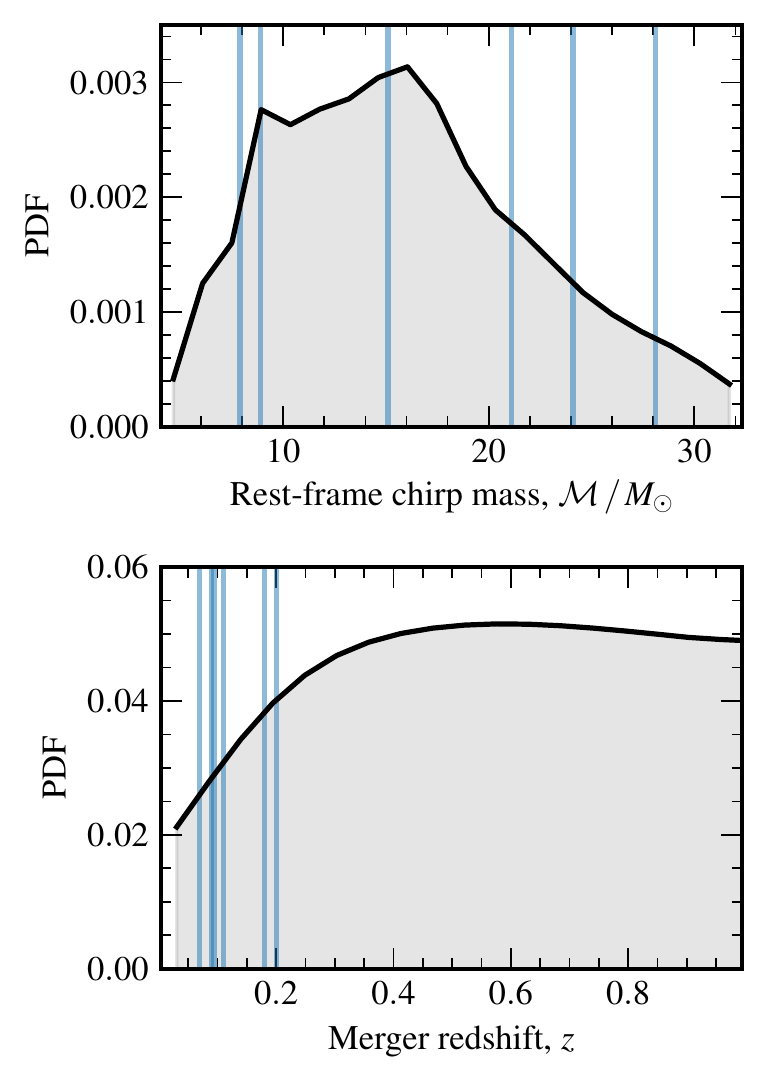}
\end{center}
\caption{Marginalized binary BH population distributions of rest-frame chirp mass and redshift for the maximum a-posteriori hyper-parameters from an analysis of the current Advanced-LIGO--Advanced-Virgo catalog. These are the intrinsic merger distributions, rather than convolved with detector selection effects. The blue vertical lines indicate the parameters of cataloged events.}
\label{fig:abse_ligocatalog}
\end{figure}

As a final test, we analyze the current Advanced-LIGO--Advanced-Virgo catalog from \autoref{tab:real_bbh}, following the same assumptions as in \autoref{sec:compasmodel}. We use \autoref{eq:hier_likelihood_v7}, and marginalize over all cataloged event parameter uncertainties. As expected, with only $6$ events the posterior distributions for $\sigma_\mathrm{k}$ and $\alpha_\mathrm{ce}$ are broad and do not significantly update their priors. However, we place a constraint on progenitor metallicity corresponding to $Z < 0.09\,Z_\odot$ at $90\%$ credibility. The marginalized mass and redshift distribution of binary BH mergers for the maximum a-posterioi hyper-parameters from this analysis are shown in \autoref{fig:abse_ligocatalog}. We stress again that all constraints are subject to our assumptions and minimal updates of \texttt{BSE}, which are only intended to show the capabilities of our approach. 

\section{Conclusions}
\label{sec:conclusions}

We have developed a new hierarchical Bayesian framework that is capable of recovering posterior probability distributions of compact-binary population hyper-parameters. These hyper-parameters encode details of stellar evolution, progenitor conditions, and the evolutionary paths taken to form systems that are detected by ground-based GW instruments such as Advanced LIGO and Advancdd Virgo. 

Our methods fuse non-parametric (i.e.\ agnostic) modeling of GW parameter distributions with population synthesis simulations. Given a collection of population synthesis simulations of potential GW events, we first formed smoothed histograms of the binary parameters, stacked the vectors of histogram bin heights, then performed PCA to compress the bins into ``features''. This allowed significant dimensionality reduction while preserving the original distributions to high fidelity. We then trained GPs to interpolate the weights of these features across hyper-parameter space, so that we could emulate parameter distributions at any choice of population hyper-parameters between the simulated values. Using a GP allowed uncertainties in the interpolation training to be propagated through to subsequent statistical analyses. Other interpolant choices are possible; in future work we will explore the ability for a deep neural network to learn compact-binary distributions, and for such a network to be embedded in a population inference pipeline.

Having constructed a model for GW parameter distributions, we incorporated it into a hierarchical inference pipeline that used information from the distribution and rate of binary BH mergers in parameter space to discriminate compact-binary progenitor and evolutionary scenarios. We tested our pipeline on three case studies that successively increased in complexity and astrophysical realism. These ranged from a toy analytic model of binary component spin alignments, to publicly available population simulations, and finally to our own custom population synthesis simulations using a modified version of the publicly-available \texttt{BSE} code. In our final study, we trained Gaussian processes on the $2$-D distribution of binary BH chirp masses and redshifts across the hyper-parameter space of progenitor stellar metallicity, BH natal kicks, and common-envelope hardening efficiency. The recovered hyper-parameter posteriors were fully consistent with the injected values. We also performed a simple analysis on the existing Advanced-LIGO--Advanced-Virgo binary BH catalog, where we incorporated parameter measurement uncertainties to constrain progenitor metallicity to be $Z < 0.09\,Z_\odot$ at $90\%$ credibility. (However, there are many caveats to this, and we quote it only to demonstrate the capabilities of our framework.)

The framework introduced here can be expanded and refined in many different ways. Further study is needed to understand how hyper-parameter measurement uncertainties will scale with the number of detected binaries, and how these compare with Fisher matrix approaches \citep{2018MNRAS.477.4685B}. Furthermore, while we have carried out studies in controlled circumstances, full production-level analysis of real GW catalogs will require that several conditions be met: e.g.\ $(i)$ the number of required training simulations should be determined through an iterative process, where GP uncertainties are investigated across hyper-parameter space to motivate new simulation locations; $(ii)$ the number of binaries in each simulation should be large enough (ideally $\gtrsim 10^3$) to construct smoothed distributions that are representative of a large population. These refinements are important since we found that sampling the hyper-parameter space was challenging in large-event catalogs.

In this paper we mainly focused on binary BH systems, but our approach can be easily generalized to incorporate the relative observed fraction of BH-BH, NS-BH, and NS-NS systems as another means of discriminating evolutionary and progenitor conditions. Likewise, we only considered classical isolated binary evolution as the mechanism of compact-binary formation, but our framework could be applied to dynamical formation scenarios, allowing the details of many-body scattering in dense stellar clusters to be revealed. A mixture model would allow us to tease apart the sub-populations within a GW catalog that have evolved through each mechanism. With this method, the mixing fractions are just other hyper-parameters that can be estimated together with those describing the various channels. Unfortunately, the public version of \texttt{BSE} that we used does not provide information on component BH spins. We stress that inclusion of spins (and other parameters in general, like eccentricity) can be easily accommodated within our framework by carrying out informative training simulations.

We are entering a new source-rich era of GW astronomy, where catalogs of compact binary coalescences will reveal much about stellar astrophysics, including the processes underlying stellar evolution and the dynamics of dense stellar clusters. As third-generation ground-based detectors become a reality, so too will the opportunity to probe star formation rates across cosmic time, constrain cosmological parameters, understand the equation-of-state of nuclear matter, and use the huge event rates to limit modifications to GR. Furthermore, a space-based detector such as LISA will catalog hundreds of massive BH mergers, permitting reconstruction of massive BH seed formation scenarios and accretion efficiencies over cosmic time. Incorporating the detailed physics of population simulations into GW catalog analysis will allow for powerful statistical inference of the aforementioned processes. We hope that our framework lays the foundation for this exciting endeavor.

The code used to perform all analyses in this paper is publicly-available at \href{https://github.com/stevertaylor/gw_catalog_mining}{github.com/stevertaylor/gw\_catalog\_mining}, along with an example \texttt{jupyter} notebook for the toy model analysis in \autoref{sec:toymodel}. 
\acknowledgments
The authors thank Michele Vallisneri and Will Farr for useful discussions regarding Bayesian hierarchical modeling. We are grateful to Astrid Lamberts and Drew Clausen for providing us with a modified version of the BSE population synthesis code. S.R.T. acknowledges support from the NANOGrav project which receives support from NSF Physics Frontier Center award number 1430284. S.R.T thanks Erika Salomon for fruitful discussions. D.G. is supported by NASA through Einstein Postdoctoral Fellowship Grant No. PF6-170152 awarded by the Chandra X-ray Center, which is operated by the Smithsonian Astrophysical Observatory for NASA under Contract NAS8-03060.  
A majority of the computational work was performed on Caltech computer cluster ``Wheeler'' supported by
the Sherman Fairchild Foundation and Caltech. Some of the computational work was performed on the Nemo cluster at UWM supported by NSF grant No.~0923409. 

\bibliography{references}

\begin{thebibliography}{115}%
\makeatletter
\providecommand \@ifxundefined [1]{%
 \@ifx{#1\undefined}
}%
\providecommand \@ifnum [1]{%
 \ifnum #1\expandafter \@firstoftwo
 \else \expandafter \@secondoftwo
 \fi
}%
\providecommand \@ifx [1]{%
 \ifx #1\expandafter \@firstoftwo
 \else \expandafter \@secondoftwo
 \fi
}%
\providecommand \natexlab [1]{#1}%
\providecommand \enquote  [1]{``#1''}%
\providecommand \bibnamefont  [1]{#1}%
\providecommand \bibfnamefont [1]{#1}%
\providecommand \citenamefont [1]{#1}%
\providecommand \href@noop [0]{\@secondoftwo}%
\providecommand \href [0]{\begingroup \@sanitize@url \@href}%
\providecommand \@href[1]{\@@startlink{#1}\@@href}%
\providecommand \@@href[1]{\endgroup#1\@@endlink}%
\providecommand \@sanitize@url [0]{\catcode `\\12\catcode `\$12\catcode
  `\&12\catcode `\#12\catcode `\^12\catcode `\_12\catcode `\%12\relax}%
\providecommand \@@startlink[1]{}%
\providecommand \@@endlink[0]{}%
\providecommand \url  [0]{\begingroup\@sanitize@url \@url }%
\providecommand \@url [1]{\endgroup\@href {#1}{\urlprefix }}%
\providecommand \urlprefix  [0]{URL }%
\providecommand \Eprint [0]{\href }%
\providecommand \doibase [0]{http://dx.doi.org/}%
\providecommand \selectlanguage [0]{\@gobble}%
\providecommand \bibinfo  [0]{\@secondoftwo}%
\providecommand \bibfield  [0]{\@secondoftwo}%
\providecommand \translation [1]{[#1]}%
\providecommand \BibitemOpen [0]{}%
\providecommand \bibitemStop [0]{}%
\providecommand \bibitemNoStop [0]{.\EOS\space}%
\providecommand \EOS [0]{\spacefactor3000\relax}%
\providecommand \BibitemShut  [1]{\csname bibitem#1\endcsname}%
\let\auto@bib@innerbib\@empty
\bibitem [{\citenamefont {{B.~P.~Abbott {\it et al.} (LIGO and Virgo
  Collaboration)}}(2016{\natexlab{a}})}]{2016PhRvL.116f1102A}%
  \BibitemOpen
  \bibfield  {author} {\bibinfo {author} {\bibnamefont {{B.~P.~Abbott {\it et
  al.} (LIGO and Virgo Collaboration)}}},\ }\href {\doibase
  10.1103/PhysRevLett.116.061102} {\bibfield  {journal} {\bibinfo  {journal}
  {\prl}\ }\textbf {\bibinfo {volume} {116}},\ \bibinfo {eid} {061102}
  (\bibinfo {year} {2016}{\natexlab{a}})},\ \Eprint
  {http://arxiv.org/abs/1602.03837} {arXiv:1602.03837 [gr-qc]} \BibitemShut
  {NoStop}%
\bibitem [{\citenamefont {{B.~P.~Abbott {\it et al.} (LIGO and Virgo
  Collaboration)}}(2016{\natexlab{b}})}]{2016PhRvX...6d1015A}%
  \BibitemOpen
  \bibfield  {author} {\bibinfo {author} {\bibnamefont {{B.~P.~Abbott {\it et
  al.} (LIGO and Virgo Collaboration)}}},\ }\href {\doibase
  10.1103/PhysRevX.6.041015} {\bibfield  {journal} {\bibinfo  {journal} {\prx}\
  }\textbf {\bibinfo {volume} {6}},\ \bibinfo {eid} {041015} (\bibinfo {year}
  {2016}{\natexlab{b}})},\ \Eprint {http://arxiv.org/abs/1606.04856}
  {arXiv:1606.04856 [gr-qc]} \BibitemShut {NoStop}%
\bibitem [{\citenamefont {{B.~P.~Abbott {\it et al.} (LIGO and Virgo
  Collaboration)}}(2016{\natexlab{c}})}]{2016PhRvL.116x1103A}%
  \BibitemOpen
  \bibfield  {author} {\bibinfo {author} {\bibnamefont {{B.~P.~Abbott {\it et
  al.} (LIGO and Virgo Collaboration)}}},\ }\href {\doibase
  10.1103/PhysRevLett.116.241103} {\bibfield  {journal} {\bibinfo  {journal}
  {\prl}\ }\textbf {\bibinfo {volume} {116}},\ \bibinfo {eid} {241103}
  (\bibinfo {year} {2016}{\natexlab{c}})},\ \Eprint
  {http://arxiv.org/abs/1606.04855} {arXiv:1606.04855 [gr-qc]} \BibitemShut
  {NoStop}%
\bibitem [{\citenamefont {{B.~P.~Abbott {\it et al.} (LIGO and Virgo
  Collaboration)}}(2017{\natexlab{a}})}]{2017PhRvL.118v1101A}%
  \BibitemOpen
  \bibfield  {author} {\bibinfo {author} {\bibnamefont {{B.~P.~Abbott {\it et
  al.} (LIGO and Virgo Collaboration)}}},\ }\href {\doibase
  10.1103/PhysRevLett.118.221101} {\bibfield  {journal} {\bibinfo  {journal}
  {\prl}\ }\textbf {\bibinfo {volume} {118}},\ \bibinfo {eid} {221101}
  (\bibinfo {year} {2017}{\natexlab{a}})},\ \Eprint
  {http://arxiv.org/abs/1706.01812} {arXiv:1706.01812 [gr-qc]} \BibitemShut
  {NoStop}%
\bibitem [{\citenamefont {{B.~P.~Abbott {\it et al.} (LIGO and Virgo
  Collaboration)}}(2017{\natexlab{b}})}]{2017ApJ...851L..35A}%
  \BibitemOpen
  \bibfield  {author} {\bibinfo {author} {\bibnamefont {{B.~P.~Abbott {\it et
  al.} (LIGO and Virgo Collaboration)}}},\ }\href {\doibase
  10.3847/2041-8213/aa9f0c} {\bibfield  {journal} {\bibinfo  {journal} {\apjl}\
  }\textbf {\bibinfo {volume} {851}},\ \bibinfo {eid} {L35} (\bibinfo {year}
  {2017}{\natexlab{b}})},\ \Eprint {http://arxiv.org/abs/1711.05578}
  {arXiv:1711.05578 [astro-ph.HE]} \BibitemShut {NoStop}%
\bibitem [{\citenamefont {{B.~P.~Abbott {\it et al.} (LIGO and Virgo
  Collaboration)}}(2017{\natexlab{c}})}]{2017PhRvL.119n1101A}%
  \BibitemOpen
  \bibfield  {author} {\bibinfo {author} {\bibnamefont {{B.~P.~Abbott {\it et
  al.} (LIGO and Virgo Collaboration)}}},\ }\href {\doibase
  10.1103/PhysRevLett.119.141101} {\bibfield  {journal} {\bibinfo  {journal}
  {\prl}\ }\textbf {\bibinfo {volume} {119}},\ \bibinfo {eid} {141101}
  (\bibinfo {year} {2017}{\natexlab{c}})},\ \Eprint
  {http://arxiv.org/abs/1709.09660} {arXiv:1709.09660 [gr-qc]} \BibitemShut
  {NoStop}%
\bibitem [{\citenamefont {{B.~P.~Abbott {\it et al.} (LIGO and Virgo
  Collaboration)}}(2017{\natexlab{d}})}]{2017PhRvL.119p1101A}%
  \BibitemOpen
  \bibfield  {author} {\bibinfo {author} {\bibnamefont {{B.~P.~Abbott {\it et
  al.} (LIGO and Virgo Collaboration)}}},\ }\href {\doibase
  10.1103/PhysRevLett.119.161101} {\bibfield  {journal} {\bibinfo  {journal}
  {\prl}\ }\textbf {\bibinfo {volume} {119}},\ \bibinfo {eid} {161101}
  (\bibinfo {year} {2017}{\natexlab{d}})},\ \Eprint
  {http://arxiv.org/abs/1710.05832} {arXiv:1710.05832 [gr-qc]} \BibitemShut
  {NoStop}%
\bibitem [{\citenamefont {{B.~P.~Abbott {\it et al.} (LIGO and Virgo
  Collaboration)}}(2017{\natexlab{e}})}]{2017ApJ...848L..12A}%
  \BibitemOpen
  \bibfield  {author} {\bibinfo {author} {\bibnamefont {{B.~P.~Abbott {\it et
  al.} (LIGO and Virgo Collaboration)}}},\ }\href {\doibase
  10.3847/2041-8213/aa91c9} {\bibfield  {journal} {\bibinfo  {journal} {\apjl}\
  }\textbf {\bibinfo {volume} {848}},\ \bibinfo {eid} {L12} (\bibinfo {year}
  {2017}{\natexlab{e}})},\ \Eprint {http://arxiv.org/abs/1710.05833}
  {arXiv:1710.05833 [astro-ph.HE]} \BibitemShut {NoStop}%
\bibitem [{\citenamefont {{Sathyaprakash}}\ \emph {et~al.}(2012)\citenamefont
  {{Sathyaprakash}} \emph {et~al.}}]{2012CQGra..29l4013S}%
  \BibitemOpen
  \bibfield  {author} {\bibinfo {author} {\bibfnamefont {B.}~\bibnamefont
  {{Sathyaprakash}}} \emph {et~al.},\ }\href {\doibase
  10.1088/0264-9381/29/12/124013} {\bibfield  {journal} {\bibinfo  {journal}
  {\cqg}\ }\textbf {\bibinfo {volume} {29}},\ \bibinfo {eid} {124013} (\bibinfo
  {year} {2012})},\ \Eprint {http://arxiv.org/abs/1206.0331} {arXiv:1206.0331
  [gr-qc]} \BibitemShut {NoStop}%
\bibitem [{\citenamefont {{Yunes}}\ \emph {et~al.}(2016)\citenamefont
  {{Yunes}}, \citenamefont {{Yagi}},\ and\ \citenamefont
  {{Pretorius}}}]{2016PhRvD..94h4002Y}%
  \BibitemOpen
  \bibfield  {author} {\bibinfo {author} {\bibfnamefont {N.}~\bibnamefont
  {{Yunes}}}, \bibinfo {author} {\bibfnamefont {K.}~\bibnamefont {{Yagi}}}, \
  and\ \bibinfo {author} {\bibfnamefont {F.}~\bibnamefont {{Pretorius}}},\
  }\href {\doibase 10.1103/PhysRevD.94.084002} {\bibfield  {journal} {\bibinfo
  {journal} {\prd}\ }\textbf {\bibinfo {volume} {94}},\ \bibinfo {eid} {084002}
  (\bibinfo {year} {2016})},\ \Eprint {http://arxiv.org/abs/1603.08955}
  {arXiv:1603.08955 [gr-qc]} \BibitemShut {NoStop}%
\bibitem [{\citenamefont {{B.~P.~Abbott {\it et al.} (LIGO and Virgo
  Collaboration)}}(2016{\natexlab{d}})}]{2016PhRvL.116v1101A}%
  \BibitemOpen
  \bibfield  {author} {\bibinfo {author} {\bibnamefont {{B.~P.~Abbott {\it et
  al.} (LIGO and Virgo Collaboration)}}},\ }\href {\doibase
  10.1103/PhysRevLett.116.221101} {\bibfield  {journal} {\bibinfo  {journal}
  {\prl}\ }\textbf {\bibinfo {volume} {116}},\ \bibinfo {eid} {221101}
  (\bibinfo {year} {2016}{\natexlab{d}})},\ \Eprint
  {http://arxiv.org/abs/1602.03841} {arXiv:1602.03841 [gr-qc]} \BibitemShut
  {NoStop}%
\bibitem [{\citenamefont {{De}}\ \emph {et~al.}(2018)\citenamefont {{De}},
  \citenamefont {{Finstad}}, \citenamefont {{Lattimer}}, \citenamefont
  {{Brown}}, \citenamefont {{Berger}},\ and\ \citenamefont
  {{Biwer}}}]{2018arXiv180408583D}%
  \BibitemOpen
  \bibfield  {author} {\bibinfo {author} {\bibfnamefont {S.}~\bibnamefont
  {{De}}}, \bibinfo {author} {\bibfnamefont {D.}~\bibnamefont {{Finstad}}},
  \bibinfo {author} {\bibfnamefont {J.~M.}\ \bibnamefont {{Lattimer}}},
  \bibinfo {author} {\bibfnamefont {D.~A.}\ \bibnamefont {{Brown}}}, \bibinfo
  {author} {\bibfnamefont {E.}~\bibnamefont {{Berger}}}, \ and\ \bibinfo
  {author} {\bibfnamefont {C.~M.}\ \bibnamefont {{Biwer}}},\ }\href@noop {} {\
  (\bibinfo {year} {2018})},\ \Eprint {http://arxiv.org/abs/1804.08583}
  {arXiv:1804.08583 [astro-ph.HE]} \BibitemShut {NoStop}%
\bibitem [{\citenamefont {{B.~P.~Abbott {\it et al.} (LIGO and Virgo
  Collaboration)}}(2018{\natexlab{a}})}]{2018arXiv180511581T}%
  \BibitemOpen
  \bibfield  {author} {\bibinfo {author} {\bibnamefont {{B.~P.~Abbott {\it et
  al.} (LIGO and Virgo Collaboration)}}},\ }\href@noop {} {\  (\bibinfo {year}
  {2018}{\natexlab{a}})},\ \Eprint {http://arxiv.org/abs/1805.11581}
  {arXiv:1805.11581 [gr-qc]} \BibitemShut {NoStop}%
\bibitem [{\citenamefont {{B.~P.~Abbott {\it et al.} (LIGO and Virgo
  Collaboration)}}(2017{\natexlab{f}})}]{2017ApJ...848L..13A}%
  \BibitemOpen
  \bibfield  {author} {\bibinfo {author} {\bibnamefont {{B.~P.~Abbott {\it et
  al.} (LIGO and Virgo Collaboration)}}},\ }\href {\doibase
  10.3847/2041-8213/aa920c} {\bibfield  {journal} {\bibinfo  {journal} {\apjl}\
  }\textbf {\bibinfo {volume} {848}},\ \bibinfo {eid} {L13} (\bibinfo {year}
  {2017}{\natexlab{f}})},\ \Eprint {http://arxiv.org/abs/1710.05834}
  {arXiv:1710.05834 [astro-ph.HE]} \BibitemShut {NoStop}%
\bibitem [{\citenamefont {{B.~P.~Abbott {\it et al.} (LIGO and Virgo
  Collaboration)}}(2017{\natexlab{g}})}]{2017Natur.551...85A}%
  \BibitemOpen
  \bibfield  {author} {\bibinfo {author} {\bibnamefont {{B.~P.~Abbott {\it et
  al.} (LIGO and Virgo Collaboration)}}},\ }\href {\doibase
  10.1038/nature24471} {\bibfield  {journal} {\bibinfo  {journal} {\nat}\
  }\textbf {\bibinfo {volume} {551}},\ \bibinfo {pages} {85} (\bibinfo {year}
  {2017}{\natexlab{g}})},\ \Eprint {http://arxiv.org/abs/1710.05835}
  {arXiv:1710.05835} \BibitemShut {NoStop}%
\bibitem [{\citenamefont {{Apostolatos}}\ \emph {et~al.}(1994)\citenamefont
  {{Apostolatos}}, \citenamefont {{Cutler}}, \citenamefont {{Sussman}},\ and\
  \citenamefont {{Thorne}}}]{1994PhRvD..49.6274A}%
  \BibitemOpen
  \bibfield  {author} {\bibinfo {author} {\bibfnamefont {T.~A.}\ \bibnamefont
  {{Apostolatos}}}, \bibinfo {author} {\bibfnamefont {C.}~\bibnamefont
  {{Cutler}}}, \bibinfo {author} {\bibfnamefont {G.~J.}\ \bibnamefont
  {{Sussman}}}, \ and\ \bibinfo {author} {\bibfnamefont {K.~S.}\ \bibnamefont
  {{Thorne}}},\ }\href {\doibase 10.1103/PhysRevD.49.6274} {\bibfield
  {journal} {\bibinfo  {journal} {\prd}\ }\textbf {\bibinfo {volume} {49}},\
  \bibinfo {pages} {6274} (\bibinfo {year} {1994})}\BibitemShut {NoStop}%
\bibitem [{\citenamefont {{Postnov}}\ and\ \citenamefont
  {{Yungelson}}(2014)}]{2014LRR....17....3P}%
  \BibitemOpen
  \bibfield  {author} {\bibinfo {author} {\bibfnamefont {K.~A.}\ \bibnamefont
  {{Postnov}}}\ and\ \bibinfo {author} {\bibfnamefont {L.~R.}\ \bibnamefont
  {{Yungelson}}},\ }\href {\doibase 10.12942/lrr-2014-3} {\bibfield  {journal}
  {\bibinfo  {journal} {\llr}\ }\textbf {\bibinfo {volume} {17}},\ \bibinfo
  {eid} {3} (\bibinfo {year} {2014})},\ \Eprint
  {http://arxiv.org/abs/1403.4754} {arXiv:1403.4754 [astro-ph.HE]} \BibitemShut
  {NoStop}%
\bibitem [{\citenamefont {{Hurley}}\ \emph {et~al.}(2002)\citenamefont
  {{Hurley}}, \citenamefont {{Tout}},\ and\ \citenamefont
  {{Pols}}}]{2002MNRAS.329..897H}%
  \BibitemOpen
  \bibfield  {author} {\bibinfo {author} {\bibfnamefont {J.~R.}\ \bibnamefont
  {{Hurley}}}, \bibinfo {author} {\bibfnamefont {C.~A.}\ \bibnamefont
  {{Tout}}}, \ and\ \bibinfo {author} {\bibfnamefont {O.~R.}\ \bibnamefont
  {{Pols}}},\ }\href {\doibase 10.1046/j.1365-8711.2002.05038.x} {\bibfield
  {journal} {\bibinfo  {journal} {\mnras}\ }\textbf {\bibinfo {volume} {329}},\
  \bibinfo {pages} {897} (\bibinfo {year} {2002})},\ \Eprint
  {http://arxiv.org/abs/astro-ph/0201220} {astro-ph/0201220} \BibitemShut
  {NoStop}%
\bibitem [{\citenamefont {{Izzard}}\ \emph {et~al.}(2004)\citenamefont
  {{Izzard}}, \citenamefont {{Tout}}, \citenamefont {{Karakas}},\ and\
  \citenamefont {{Pols}}}]{2004MNRAS.350..407I}%
  \BibitemOpen
  \bibfield  {author} {\bibinfo {author} {\bibfnamefont {R.~G.}\ \bibnamefont
  {{Izzard}}}, \bibinfo {author} {\bibfnamefont {C.~A.}\ \bibnamefont
  {{Tout}}}, \bibinfo {author} {\bibfnamefont {A.~I.}\ \bibnamefont
  {{Karakas}}}, \ and\ \bibinfo {author} {\bibfnamefont {O.~R.}\ \bibnamefont
  {{Pols}}},\ }\href {\doibase 10.1111/j.1365-2966.2004.07446.x} {\bibfield
  {journal} {\bibinfo  {journal} {\mnras}\ }\textbf {\bibinfo {volume} {350}},\
  \bibinfo {pages} {407} (\bibinfo {year} {2004})},\ \Eprint
  {http://arxiv.org/abs/astro-ph/0402403} {astro-ph/0402403} \BibitemShut
  {NoStop}%
\bibitem [{\citenamefont {{Belczynski}}\ \emph {et~al.}(2008)\citenamefont
  {{Belczynski}}, \citenamefont {{Kalogera}}, \citenamefont {{Rasio}},
  \citenamefont {{Taam}}, \citenamefont {{Zezas}}, \citenamefont {{Bulik}},
  \citenamefont {{Maccarone}},\ and\ \citenamefont
  {{Ivanova}}}]{2008ApJS..174..223B}%
  \BibitemOpen
  \bibfield  {author} {\bibinfo {author} {\bibfnamefont {K.}~\bibnamefont
  {{Belczynski}}}, \bibinfo {author} {\bibfnamefont {V.}~\bibnamefont
  {{Kalogera}}}, \bibinfo {author} {\bibfnamefont {F.~A.}\ \bibnamefont
  {{Rasio}}}, \bibinfo {author} {\bibfnamefont {R.~E.}\ \bibnamefont {{Taam}}},
  \bibinfo {author} {\bibfnamefont {A.}~\bibnamefont {{Zezas}}}, \bibinfo
  {author} {\bibfnamefont {T.}~\bibnamefont {{Bulik}}}, \bibinfo {author}
  {\bibfnamefont {T.~J.}\ \bibnamefont {{Maccarone}}}, \ and\ \bibinfo {author}
  {\bibfnamefont {N.}~\bibnamefont {{Ivanova}}},\ }\href {\doibase
  10.1086/521026} {\bibfield  {journal} {\bibinfo  {journal} {\apjs}\ }\textbf
  {\bibinfo {volume} {174}},\ \bibinfo {pages} {223} (\bibinfo {year}
  {2008})},\ \Eprint {http://arxiv.org/abs/astro-ph/0511811} {astro-ph/0511811}
  \BibitemShut {NoStop}%
\bibitem [{\citenamefont {{Spera}}\ \emph {et~al.}(2015)\citenamefont
  {{Spera}}, \citenamefont {{Mapelli}},\ and\ \citenamefont
  {{Bressan}}}]{2015MNRAS.451.4086S}%
  \BibitemOpen
  \bibfield  {author} {\bibinfo {author} {\bibfnamefont {M.}~\bibnamefont
  {{Spera}}}, \bibinfo {author} {\bibfnamefont {M.}~\bibnamefont {{Mapelli}}},
  \ and\ \bibinfo {author} {\bibfnamefont {A.}~\bibnamefont {{Bressan}}},\
  }\href {\doibase 10.1093/mnras/stv1161} {\bibfield  {journal} {\bibinfo
  {journal} {\mnras}\ }\textbf {\bibinfo {volume} {451}},\ \bibinfo {pages}
  {4086} (\bibinfo {year} {2015})},\ \Eprint {http://arxiv.org/abs/1505.05201}
  {arXiv:1505.05201 [astro-ph.SR]} \BibitemShut {NoStop}%
\bibitem [{\citenamefont {{Giacobbo}}\ \emph {et~al.}(2018)\citenamefont
  {{Giacobbo}}, \citenamefont {{Mapelli}},\ and\ \citenamefont
  {{Spera}}}]{2018MNRAS.474.2959G}%
  \BibitemOpen
  \bibfield  {author} {\bibinfo {author} {\bibfnamefont {N.}~\bibnamefont
  {{Giacobbo}}}, \bibinfo {author} {\bibfnamefont {M.}~\bibnamefont
  {{Mapelli}}}, \ and\ \bibinfo {author} {\bibfnamefont {M.}~\bibnamefont
  {{Spera}}},\ }\href {\doibase 10.1093/mnras/stx2933} {\bibfield  {journal}
  {\bibinfo  {journal} {\mnras}\ }\textbf {\bibinfo {volume} {474}},\ \bibinfo
  {pages} {2959} (\bibinfo {year} {2018})},\ \Eprint
  {http://arxiv.org/abs/1711.03556} {arXiv:1711.03556 [astro-ph.SR]}
  \BibitemShut {NoStop}%
\bibitem [{\citenamefont {{Breivik}}\ \emph {et~al.}(2018)\citenamefont
  {{Breivik}}, \citenamefont {{Kremer}}, \citenamefont {{Bueno}}, \citenamefont
  {{Larson}}, \citenamefont {{Coughlin}},\ and\ \citenamefont
  {{Kalogera}}}]{2018ApJ...854L...1B}%
  \BibitemOpen
  \bibfield  {author} {\bibinfo {author} {\bibfnamefont {K.}~\bibnamefont
  {{Breivik}}}, \bibinfo {author} {\bibfnamefont {K.}~\bibnamefont {{Kremer}}},
  \bibinfo {author} {\bibfnamefont {M.}~\bibnamefont {{Bueno}}}, \bibinfo
  {author} {\bibfnamefont {S.~L.}\ \bibnamefont {{Larson}}}, \bibinfo {author}
  {\bibfnamefont {S.}~\bibnamefont {{Coughlin}}}, \ and\ \bibinfo {author}
  {\bibfnamefont {V.}~\bibnamefont {{Kalogera}}},\ }\href {\doibase
  10.3847/2041-8213/aaaa23} {\bibfield  {journal} {\bibinfo  {journal} {\apjl}\
  }\textbf {\bibinfo {volume} {854}},\ \bibinfo {eid} {L1} (\bibinfo {year}
  {2018})},\ \Eprint {http://arxiv.org/abs/1710.08370} {arXiv:1710.08370
  [astro-ph.SR]} \BibitemShut {NoStop}%
\bibitem [{\citenamefont {{Kruckow}}\ \emph {et~al.}(2018)\citenamefont
  {{Kruckow}}, \citenamefont {{Tauris}}, \citenamefont {{Langer}},
  \citenamefont {{Kramer}},\ and\ \citenamefont
  {{Izzard}}}]{2018arXiv180105433K}%
  \BibitemOpen
  \bibfield  {author} {\bibinfo {author} {\bibfnamefont {M.~U.}\ \bibnamefont
  {{Kruckow}}}, \bibinfo {author} {\bibfnamefont {T.~M.}\ \bibnamefont
  {{Tauris}}}, \bibinfo {author} {\bibfnamefont {N.}~\bibnamefont {{Langer}}},
  \bibinfo {author} {\bibfnamefont {M.}~\bibnamefont {{Kramer}}}, \ and\
  \bibinfo {author} {\bibfnamefont {R.~G.}\ \bibnamefont {{Izzard}}},\
  }\href@noop {} {\  (\bibinfo {year} {2018})},\ \Eprint
  {http://arxiv.org/abs/1801.05433} {arXiv:1801.05433 [astro-ph.SR]}
  \BibitemShut {NoStop}%
\bibitem [{\citenamefont {{Stevenson}}\ \emph
  {et~al.}(2017{\natexlab{a}})\citenamefont {{Stevenson}}, \citenamefont
  {{Vigna-G{\'o}mez}}, \citenamefont {{Mandel}}, \citenamefont {{Barrett}},
  \citenamefont {{Neijssel}}, \citenamefont {{Perkins}},\ and\ \citenamefont
  {{de Mink}}}]{2017NatCo...814906S}%
  \BibitemOpen
  \bibfield  {author} {\bibinfo {author} {\bibfnamefont {S.}~\bibnamefont
  {{Stevenson}}}, \bibinfo {author} {\bibfnamefont {A.}~\bibnamefont
  {{Vigna-G{\'o}mez}}}, \bibinfo {author} {\bibfnamefont {I.}~\bibnamefont
  {{Mandel}}}, \bibinfo {author} {\bibfnamefont {J.~W.}\ \bibnamefont
  {{Barrett}}}, \bibinfo {author} {\bibfnamefont {C.~J.}\ \bibnamefont
  {{Neijssel}}}, \bibinfo {author} {\bibfnamefont {D.}~\bibnamefont
  {{Perkins}}}, \ and\ \bibinfo {author} {\bibfnamefont {S.~E.}\ \bibnamefont
  {{de Mink}}},\ }\href {\doibase 10.1038/ncomms14906} {\bibfield  {journal}
  {\bibinfo  {journal} {Nature Comm.}\ }\textbf {\bibinfo {volume} {8}},\
  \bibinfo {eid} {14906} (\bibinfo {year} {2017}{\natexlab{a}})},\ \Eprint
  {http://arxiv.org/abs/1704.01352} {arXiv:1704.01352 [astro-ph.HE]}
  \BibitemShut {NoStop}%
\bibitem [{\citenamefont {{Benacquista}}\ and\ \citenamefont
  {{Downing}}(2013)}]{2013LRR....16....4B}%
  \BibitemOpen
  \bibfield  {author} {\bibinfo {author} {\bibfnamefont {M.~J.}\ \bibnamefont
  {{Benacquista}}}\ and\ \bibinfo {author} {\bibfnamefont {J.~M.~B.}\
  \bibnamefont {{Downing}}},\ }\href {\doibase 10.12942/lrr-2013-4} {\bibfield
  {journal} {\bibinfo  {journal} {\llr}\ }\textbf {\bibinfo {volume} {16}},\
  \bibinfo {eid} {4} (\bibinfo {year} {2013})},\ \Eprint
  {http://arxiv.org/abs/1110.4423} {arXiv:1110.4423 [astro-ph.SR]} \BibitemShut
  {NoStop}%
\bibitem [{\citenamefont {{Taylor}}\ \emph {et~al.}(2012)\citenamefont
  {{Taylor}}, \citenamefont {{Gair}},\ and\ \citenamefont
  {{Mandel}}}]{2012PhRvD..85b3535T}%
  \BibitemOpen
  \bibfield  {author} {\bibinfo {author} {\bibfnamefont {S.~R.}\ \bibnamefont
  {{Taylor}}}, \bibinfo {author} {\bibfnamefont {J.~R.}\ \bibnamefont
  {{Gair}}}, \ and\ \bibinfo {author} {\bibfnamefont {I.}~\bibnamefont
  {{Mandel}}},\ }\href {\doibase 10.1103/PhysRevD.85.023535} {\bibfield
  {journal} {\bibinfo  {journal} {\prd}\ }\textbf {\bibinfo {volume} {85}},\
  \bibinfo {eid} {023535} (\bibinfo {year} {2012})},\ \Eprint
  {http://arxiv.org/abs/1108.5161} {arXiv:1108.5161 [gr-qc]} \BibitemShut
  {NoStop}%
\bibitem [{\citenamefont {{Zhu}}\ \emph {et~al.}(2017)\citenamefont {{Zhu}},
  \citenamefont {{Thrane}}, \citenamefont {{Os{\l}owski}}, \citenamefont
  {{Levin}},\ and\ \citenamefont {{Lasky}}}]{2017arXiv171109226Z}%
  \BibitemOpen
  \bibfield  {author} {\bibinfo {author} {\bibfnamefont {X.-J.}\ \bibnamefont
  {{Zhu}}}, \bibinfo {author} {\bibfnamefont {E.}~\bibnamefont {{Thrane}}},
  \bibinfo {author} {\bibfnamefont {S.}~\bibnamefont {{Os{\l}owski}}}, \bibinfo
  {author} {\bibfnamefont {Y.}~\bibnamefont {{Levin}}}, \ and\ \bibinfo
  {author} {\bibfnamefont {P.~D.}\ \bibnamefont {{Lasky}}},\ }\href@noop {} {\
  (\bibinfo {year} {2017})},\ \Eprint {http://arxiv.org/abs/1711.09226}
  {arXiv:1711.09226 [astro-ph.HE]} \BibitemShut {NoStop}%
\bibitem [{\citenamefont {{Zevin}}\ \emph {et~al.}(2017)\citenamefont
  {{Zevin}}, \citenamefont {{Pankow}}, \citenamefont {{Rodriguez}},
  \citenamefont {{Sampson}}, \citenamefont {{Chase}}, \citenamefont
  {{Kalogera}},\ and\ \citenamefont {{Rasio}}}]{2017ApJ...846...82Z}%
  \BibitemOpen
  \bibfield  {author} {\bibinfo {author} {\bibfnamefont {M.}~\bibnamefont
  {{Zevin}}}, \bibinfo {author} {\bibfnamefont {C.}~\bibnamefont {{Pankow}}},
  \bibinfo {author} {\bibfnamefont {C.~L.}\ \bibnamefont {{Rodriguez}}},
  \bibinfo {author} {\bibfnamefont {L.}~\bibnamefont {{Sampson}}}, \bibinfo
  {author} {\bibfnamefont {E.}~\bibnamefont {{Chase}}}, \bibinfo {author}
  {\bibfnamefont {V.}~\bibnamefont {{Kalogera}}}, \ and\ \bibinfo {author}
  {\bibfnamefont {F.~A.}\ \bibnamefont {{Rasio}}},\ }\href {\doibase
  10.3847/1538-4357/aa8408} {\bibfield  {journal} {\bibinfo  {journal} {\apj}\
  }\textbf {\bibinfo {volume} {846}},\ \bibinfo {eid} {82} (\bibinfo {year}
  {2017})},\ \Eprint {http://arxiv.org/abs/1704.07379} {arXiv:1704.07379
  [astro-ph.HE]} \BibitemShut {NoStop}%
\bibitem [{\citenamefont {{Farr}}\ \emph {et~al.}(2018)\citenamefont {{Farr}},
  \citenamefont {{Holz}},\ and\ \citenamefont {{Farr}}}]{2018ApJ...854L...9F}%
  \BibitemOpen
  \bibfield  {author} {\bibinfo {author} {\bibfnamefont {B.}~\bibnamefont
  {{Farr}}}, \bibinfo {author} {\bibfnamefont {D.~E.}\ \bibnamefont {{Holz}}},
  \ and\ \bibinfo {author} {\bibfnamefont {W.~M.}\ \bibnamefont {{Farr}}},\
  }\href {\doibase 10.3847/2041-8213/aaaa64} {\bibfield  {journal} {\bibinfo
  {journal} {\apjl}\ }\textbf {\bibinfo {volume} {854}},\ \bibinfo {eid} {L9}
  (\bibinfo {year} {2018})},\ \Eprint {http://arxiv.org/abs/1709.07896}
  {arXiv:1709.07896 [astro-ph.HE]} \BibitemShut {NoStop}%
\bibitem [{\citenamefont {{Wysocki}}\ \emph
  {et~al.}(2018{\natexlab{a}})\citenamefont {{Wysocki}}, \citenamefont
  {{Lange}},\ and\ \citenamefont {{O'Shaughnessy}}}]{2018arXiv180506442W}%
  \BibitemOpen
  \bibfield  {author} {\bibinfo {author} {\bibfnamefont {D.}~\bibnamefont
  {{Wysocki}}}, \bibinfo {author} {\bibfnamefont {J.}~\bibnamefont {{Lange}}},
  \ and\ \bibinfo {author} {\bibfnamefont {R.}~\bibnamefont
  {{O'Shaughnessy}}},\ }\href@noop {} {\  (\bibinfo {year}
  {2018}{\natexlab{a}})},\ \Eprint {http://arxiv.org/abs/1805.06442}
  {arXiv:1805.06442 [gr-qc]} \BibitemShut {NoStop}%
\bibitem [{\citenamefont {{Talbot}}\ and\ \citenamefont
  {{Thrane}}(2018)}]{2018ApJ...856..173T}%
  \BibitemOpen
  \bibfield  {author} {\bibinfo {author} {\bibfnamefont {C.}~\bibnamefont
  {{Talbot}}}\ and\ \bibinfo {author} {\bibfnamefont {E.}~\bibnamefont
  {{Thrane}}},\ }\href {\doibase 10.3847/1538-4357/aab34c} {\bibfield
  {journal} {\bibinfo  {journal} {\apj}\ }\textbf {\bibinfo {volume} {856}},\
  \bibinfo {eid} {173} (\bibinfo {year} {2018})},\ \Eprint
  {http://arxiv.org/abs/1801.02699} {arXiv:1801.02699 [astro-ph.HE]}
  \BibitemShut {NoStop}%
\bibitem [{\citenamefont {{Roulet}}\ and\ \citenamefont
  {{Zaldarriaga}}(2018)}]{2018arXiv180610610R}%
  \BibitemOpen
  \bibfield  {author} {\bibinfo {author} {\bibfnamefont {J.}~\bibnamefont
  {{Roulet}}}\ and\ \bibinfo {author} {\bibfnamefont {M.}~\bibnamefont
  {{Zaldarriaga}}},\ }\href@noop {} {\bibfield  {journal} {\bibinfo  {journal}
  {ArXiv e-prints}\ } (\bibinfo {year} {2018})},\ \Eprint
  {http://arxiv.org/abs/1806.10610} {arXiv:1806.10610 [astro-ph.HE]}
  \BibitemShut {NoStop}%
\bibitem [{\citenamefont {{Belczynski}}\ \emph {et~al.}(2016)\citenamefont
  {{Belczynski}}, \citenamefont {{Holz}}, \citenamefont {{Bulik}},\ and\
  \citenamefont {{O'Shaughnessy}}}]{2016Natur.534..512B}%
  \BibitemOpen
  \bibfield  {author} {\bibinfo {author} {\bibfnamefont {K.}~\bibnamefont
  {{Belczynski}}}, \bibinfo {author} {\bibfnamefont {D.~E.}\ \bibnamefont
  {{Holz}}}, \bibinfo {author} {\bibfnamefont {T.}~\bibnamefont {{Bulik}}}, \
  and\ \bibinfo {author} {\bibfnamefont {R.}~\bibnamefont {{O'Shaughnessy}}},\
  }\href {\doibase 10.1038/nature18322} {\bibfield  {journal} {\bibinfo
  {journal} {\nat}\ }\textbf {\bibinfo {volume} {534}},\ \bibinfo {pages} {512}
  (\bibinfo {year} {2016})},\ \Eprint {http://arxiv.org/abs/1602.04531}
  {arXiv:1602.04531 [astro-ph.HE]} \BibitemShut {NoStop}%
\bibitem [{\citenamefont {{B.~P.~Abbott {\it et al.} (LIGO and Virgo
  Collaboration)}}(2016{\natexlab{e}})}]{2016ApJ...818L..22A}%
  \BibitemOpen
  \bibfield  {author} {\bibinfo {author} {\bibnamefont {{B.~P.~Abbott {\it et
  al.} (LIGO and Virgo Collaboration)}}},\ }\href {\doibase
  10.3847/2041-8205/818/2/L22} {\bibfield  {journal} {\bibinfo  {journal}
  {\apjl}\ }\textbf {\bibinfo {volume} {818}},\ \bibinfo {eid} {L22} (\bibinfo
  {year} {2016}{\natexlab{e}})},\ \Eprint {http://arxiv.org/abs/1602.03846}
  {arXiv:1602.03846 [astro-ph.HE]} \BibitemShut {NoStop}%
\bibitem [{\citenamefont {{Lamberts}}\ \emph {et~al.}(2016)\citenamefont
  {{Lamberts}}, \citenamefont {{Garrison-Kimmel}}, \citenamefont {{Clausen}},\
  and\ \citenamefont {{Hopkins}}}]{2016MNRAS.463L..31L}%
  \BibitemOpen
  \bibfield  {author} {\bibinfo {author} {\bibfnamefont {A.}~\bibnamefont
  {{Lamberts}}}, \bibinfo {author} {\bibfnamefont {S.}~\bibnamefont
  {{Garrison-Kimmel}}}, \bibinfo {author} {\bibfnamefont {D.~R.}\ \bibnamefont
  {{Clausen}}}, \ and\ \bibinfo {author} {\bibfnamefont {P.~F.}\ \bibnamefont
  {{Hopkins}}},\ }\href {\doibase 10.1093/mnrasl/slw152} {\bibfield  {journal}
  {\bibinfo  {journal} {\mnras}\ }\textbf {\bibinfo {volume} {463}},\ \bibinfo
  {pages} {L31} (\bibinfo {year} {2016})},\ \Eprint
  {http://arxiv.org/abs/1605.08783} {arXiv:1605.08783 [astro-ph.HE]}
  \BibitemShut {NoStop}%
\bibitem [{\citenamefont {{Stevenson}}\ \emph {et~al.}(2015)\citenamefont
  {{Stevenson}}, \citenamefont {{Ohme}},\ and\ \citenamefont
  {{Fairhurst}}}]{2015ApJ...810...58S}%
  \BibitemOpen
  \bibfield  {author} {\bibinfo {author} {\bibfnamefont {S.}~\bibnamefont
  {{Stevenson}}}, \bibinfo {author} {\bibfnamefont {F.}~\bibnamefont {{Ohme}}},
  \ and\ \bibinfo {author} {\bibfnamefont {S.}~\bibnamefont {{Fairhurst}}},\
  }\href {\doibase 10.1088/0004-637X/810/1/58} {\bibfield  {journal} {\bibinfo
  {journal} {\apj}\ }\textbf {\bibinfo {volume} {810}},\ \bibinfo {eid} {58}
  (\bibinfo {year} {2015})},\ \Eprint {http://arxiv.org/abs/1504.07802}
  {arXiv:1504.07802 [astro-ph.HE]} \BibitemShut {NoStop}%
\bibitem [{\citenamefont {{Gerosa}}\ and\ \citenamefont
  {{Berti}}(2017)}]{2017PhRvD..95l4046G}%
  \BibitemOpen
  \bibfield  {author} {\bibinfo {author} {\bibfnamefont {D.}~\bibnamefont
  {{Gerosa}}}\ and\ \bibinfo {author} {\bibfnamefont {E.}~\bibnamefont
  {{Berti}}},\ }\href {\doibase 10.1103/PhysRevD.95.124046} {\bibfield
  {journal} {\bibinfo  {journal} {\prd}\ }\textbf {\bibinfo {volume} {95}},\
  \bibinfo {eid} {124046} (\bibinfo {year} {2017})},\ \Eprint
  {http://arxiv.org/abs/1703.06223} {arXiv:1703.06223 [gr-qc]} \BibitemShut
  {NoStop}%
\bibitem [{\citenamefont {{Stevenson}}\ \emph
  {et~al.}(2017{\natexlab{b}})\citenamefont {{Stevenson}}, \citenamefont
  {{Berry}},\ and\ \citenamefont {{Mandel}}}]{2017MNRAS.471.2801S}%
  \BibitemOpen
  \bibfield  {author} {\bibinfo {author} {\bibfnamefont {S.}~\bibnamefont
  {{Stevenson}}}, \bibinfo {author} {\bibfnamefont {C.~P.~L.}\ \bibnamefont
  {{Berry}}}, \ and\ \bibinfo {author} {\bibfnamefont {I.}~\bibnamefont
  {{Mandel}}},\ }\href {\doibase 10.1093/mnras/stx1764} {\bibfield  {journal}
  {\bibinfo  {journal} {\mnras}\ }\textbf {\bibinfo {volume} {471}},\ \bibinfo
  {pages} {2801} (\bibinfo {year} {2017}{\natexlab{b}})},\ \Eprint
  {http://arxiv.org/abs/1703.06873} {arXiv:1703.06873 [astro-ph.HE]}
  \BibitemShut {NoStop}%
\bibitem [{\citenamefont {{Wysocki}}\ \emph
  {et~al.}(2018{\natexlab{b}})\citenamefont {{Wysocki}}, \citenamefont
  {{Gerosa}}, \citenamefont {{O'Shaughnessy}}, \citenamefont {{Belczynski}},
  \citenamefont {{Gladysz}}, \citenamefont {{Berti}}, \citenamefont
  {{Kesden}},\ and\ \citenamefont {{Holz}}}]{2018PhRvD..97d3014W}%
  \BibitemOpen
  \bibfield  {author} {\bibinfo {author} {\bibfnamefont {D.}~\bibnamefont
  {{Wysocki}}}, \bibinfo {author} {\bibfnamefont {D.}~\bibnamefont {{Gerosa}}},
  \bibinfo {author} {\bibfnamefont {R.}~\bibnamefont {{O'Shaughnessy}}},
  \bibinfo {author} {\bibfnamefont {K.}~\bibnamefont {{Belczynski}}}, \bibinfo
  {author} {\bibfnamefont {W.}~\bibnamefont {{Gladysz}}}, \bibinfo {author}
  {\bibfnamefont {E.}~\bibnamefont {{Berti}}}, \bibinfo {author} {\bibfnamefont
  {M.}~\bibnamefont {{Kesden}}}, \ and\ \bibinfo {author} {\bibfnamefont
  {D.~E.}\ \bibnamefont {{Holz}}},\ }\href {\doibase
  10.1103/PhysRevD.97.043014} {\bibfield  {journal} {\bibinfo  {journal}
  {\prd}\ }\textbf {\bibinfo {volume} {97}},\ \bibinfo {eid} {043014} (\bibinfo
  {year} {2018}{\natexlab{b}})},\ \Eprint {http://arxiv.org/abs/1709.01943}
  {arXiv:1709.01943 [astro-ph.HE]} \BibitemShut {NoStop}%
\bibitem [{\citenamefont {{Mandel}}\ \emph {et~al.}(2017)\citenamefont
  {{Mandel}}, \citenamefont {{Farr}}, \citenamefont {{Colonna}}, \citenamefont
  {{Stevenson}}, \citenamefont {{Ti{\v n}o}},\ and\ \citenamefont
  {{Veitch}}}]{2017MNRAS.465.3254M}%
  \BibitemOpen
  \bibfield  {author} {\bibinfo {author} {\bibfnamefont {I.}~\bibnamefont
  {{Mandel}}}, \bibinfo {author} {\bibfnamefont {W.~M.}\ \bibnamefont
  {{Farr}}}, \bibinfo {author} {\bibfnamefont {A.}~\bibnamefont {{Colonna}}},
  \bibinfo {author} {\bibfnamefont {S.}~\bibnamefont {{Stevenson}}}, \bibinfo
  {author} {\bibfnamefont {P.}~\bibnamefont {{Ti{\v n}o}}}, \ and\ \bibinfo
  {author} {\bibfnamefont {J.}~\bibnamefont {{Veitch}}},\ }\href {\doibase
  10.1093/mnras/stw2883} {\bibfield  {journal} {\bibinfo  {journal} {\mnras}\
  }\textbf {\bibinfo {volume} {465}},\ \bibinfo {pages} {3254} (\bibinfo {year}
  {2017})},\ \Eprint {http://arxiv.org/abs/1608.08223} {arXiv:1608.08223
  [astro-ph.HE]} \BibitemShut {NoStop}%
\bibitem [{\citenamefont {{Heitmann}}\ \emph {et~al.}(2006)\citenamefont
  {{Heitmann}}, \citenamefont {{Higdon}}, \citenamefont {{Nakhleh}},\ and\
  \citenamefont {{Habib}}}]{2006ApJ...646L...1H}%
  \BibitemOpen
  \bibfield  {author} {\bibinfo {author} {\bibfnamefont {K.}~\bibnamefont
  {{Heitmann}}}, \bibinfo {author} {\bibfnamefont {D.}~\bibnamefont
  {{Higdon}}}, \bibinfo {author} {\bibfnamefont {C.}~\bibnamefont {{Nakhleh}}},
  \ and\ \bibinfo {author} {\bibfnamefont {S.}~\bibnamefont {{Habib}}},\ }\href
  {\doibase 10.1086/506448} {\bibfield  {journal} {\bibinfo  {journal} {\apjl}\
  }\textbf {\bibinfo {volume} {646}},\ \bibinfo {pages} {L1} (\bibinfo {year}
  {2006})},\ \Eprint {http://arxiv.org/abs/astro-ph/0606154} {astro-ph/0606154}
  \BibitemShut {NoStop}%
\bibitem [{\citenamefont {{Habib}}\ \emph {et~al.}(2007)\citenamefont
  {{Habib}}, \citenamefont {{Heitmann}}, \citenamefont {{Higdon}},
  \citenamefont {{Nakhleh}},\ and\ \citenamefont
  {{Williams}}}]{2007PhRvD..76h3503H}%
  \BibitemOpen
  \bibfield  {author} {\bibinfo {author} {\bibfnamefont {S.}~\bibnamefont
  {{Habib}}}, \bibinfo {author} {\bibfnamefont {K.}~\bibnamefont {{Heitmann}}},
  \bibinfo {author} {\bibfnamefont {D.}~\bibnamefont {{Higdon}}}, \bibinfo
  {author} {\bibfnamefont {C.}~\bibnamefont {{Nakhleh}}}, \ and\ \bibinfo
  {author} {\bibfnamefont {B.}~\bibnamefont {{Williams}}},\ }\href {\doibase
  10.1103/PhysRevD.76.083503} {\bibfield  {journal} {\bibinfo  {journal}
  {\prd}\ }\textbf {\bibinfo {volume} {76}},\ \bibinfo {eid} {083503} (\bibinfo
  {year} {2007})},\ \Eprint {http://arxiv.org/abs/astro-ph/0702348}
  {astro-ph/0702348} \BibitemShut {NoStop}%
\bibitem [{\citenamefont {{Taylor}}\ \emph {et~al.}(2017)\citenamefont
  {{Taylor}}, \citenamefont {{Simon}},\ and\ \citenamefont
  {{Sampson}}}]{2017PhRvL.118r1102T}%
  \BibitemOpen
  \bibfield  {author} {\bibinfo {author} {\bibfnamefont {S.~R.}\ \bibnamefont
  {{Taylor}}}, \bibinfo {author} {\bibfnamefont {J.}~\bibnamefont {{Simon}}}, \
  and\ \bibinfo {author} {\bibfnamefont {L.}~\bibnamefont {{Sampson}}},\ }\href
  {\doibase 10.1103/PhysRevLett.118.181102} {\bibfield  {journal} {\bibinfo
  {journal} {\prl}\ }\textbf {\bibinfo {volume} {118}},\ \bibinfo {eid}
  {181102} (\bibinfo {year} {2017})},\ \Eprint
  {http://arxiv.org/abs/1612.02817} {arXiv:1612.02817} \BibitemShut {NoStop}%
\bibitem [{\citenamefont {{Z~ Arzoumanian {\it et al.} (NANOGrav
  Collaboration)}}(2018)}]{2018ApJ...859...47A}%
  \BibitemOpen
  \bibfield  {author} {\bibinfo {author} {\bibnamefont {{Z~ Arzoumanian {\it et
  al.} (NANOGrav Collaboration)}}},\ }\href {\doibase 10.3847/1538-4357/aabd3b}
  {\bibfield  {journal} {\bibinfo  {journal} {\apj}\ }\textbf {\bibinfo
  {volume} {859}},\ \bibinfo {eid} {47} (\bibinfo {year} {2018})},\ \Eprint
  {http://arxiv.org/abs/1801.02617} {arXiv:1801.02617 [astro-ph.HE]}
  \BibitemShut {NoStop}%
\bibitem [{\citenamefont {{Barrett}}\ \emph {et~al.}(2017)\citenamefont
  {{Barrett}}, \citenamefont {{Mandel}}, \citenamefont {{Neijssel}},
  \citenamefont {{Stevenson}},\ and\ \citenamefont
  {{Vigna-G{\'o}mez}}}]{2017IAUS..325...46B}%
  \BibitemOpen
  \bibfield  {author} {\bibinfo {author} {\bibfnamefont {J.~W.}\ \bibnamefont
  {{Barrett}}}, \bibinfo {author} {\bibfnamefont {I.}~\bibnamefont {{Mandel}}},
  \bibinfo {author} {\bibfnamefont {C.~J.}\ \bibnamefont {{Neijssel}}},
  \bibinfo {author} {\bibfnamefont {S.}~\bibnamefont {{Stevenson}}}, \ and\
  \bibinfo {author} {\bibfnamefont {A.}~\bibnamefont {{Vigna-G{\'o}mez}}},\
  }in\ \href {\doibase 10.1017/S1743921317000059} {\emph {\bibinfo {booktitle}
  {Astroinformatics}}},\ \bibinfo {series} {IAU Symposium}, Vol.\ \bibinfo
  {volume} {325}\ (\bibinfo {year} {2017})\ pp.\ \bibinfo {pages} {46--50},\
  \Eprint {http://arxiv.org/abs/1704.03781} {arXiv:1704.03781 [astro-ph.HE]}
  \BibitemShut {NoStop}%
\bibitem [{\citenamefont {McKay}\ \emph {et~al.}(1979)\citenamefont {McKay},
  \citenamefont {Beckman},\ and\ \citenamefont {Conover}}]{mbc79}%
  \BibitemOpen
  \bibfield  {author} {\bibinfo {author} {\bibfnamefont {M.~D.}\ \bibnamefont
  {McKay}}, \bibinfo {author} {\bibfnamefont {R.~J.}\ \bibnamefont {Beckman}},
  \ and\ \bibinfo {author} {\bibfnamefont {W.~J.}\ \bibnamefont {Conover}},\
  }\href {http://www.jstor.org/stable/1268522} {\bibfield  {journal} {\bibinfo
  {journal} {Technometrics}\ }\textbf {\bibinfo {volume} {21}},\ \bibinfo
  {pages} {239} (\bibinfo {year} {1979})}\BibitemShut {NoStop}%
\bibitem [{\citenamefont {Singh}()}]{pyDOE}%
  \BibitemOpen
  \bibfield  {author} {\bibinfo {author} {\bibfnamefont {A.}~\bibnamefont
  {Singh}},\ }\href@noop {} {\bibinfo  {journal} {{\it pyDOE: The experimental
  design package for python}
  \href{https://github.com/tisimst/pyDOE}{github.com/tisimst/pyDOE}}\
  }\BibitemShut {NoStop}%
\bibitem [{\citenamefont {{Rasmussen}}\ and\ \citenamefont
  {{Williams}}(2006)}]{2006gpml.book.....R}%
  \BibitemOpen
\bibfield  {journal} {  }\bibfield  {author} {\bibinfo {author} {\bibfnamefont
  {C.~E.}\ \bibnamefont {{Rasmussen}}}\ and\ \bibinfo {author} {\bibfnamefont
  {C.~K.~I.}\ \bibnamefont {{Williams}}},\ }\href@noop {} {\emph {\bibinfo
  {title} {{Gaussian Processes for Machine Learning}}}}\ (\bibinfo  {publisher}
  {MIT Press},\ \bibinfo {year} {2006})\BibitemShut {NoStop}%
\bibitem [{\citenamefont {MacKay}(1998)}]{m98}%
  \BibitemOpen
  \bibfield  {author} {\bibinfo {author} {\bibfnamefont {D.~J.}\ \bibnamefont
  {MacKay}},\ }\href@noop {} {\bibfield  {journal} {\bibinfo  {journal} {NATO
  ASI F}\ }\textbf {\bibinfo {volume} {168}},\ \bibinfo {pages} {133} (\bibinfo
  {year} {1998})}\BibitemShut {NoStop}%
\bibitem [{\citenamefont {{Ebden}}(2015)}]{2015arXiv150502965E}%
  \BibitemOpen
  \bibfield  {author} {\bibinfo {author} {\bibfnamefont {M.}~\bibnamefont
  {{Ebden}}},\ }\href@noop {} {\  (\bibinfo {year} {2015})},\ \Eprint
  {http://arxiv.org/abs/1505.02965} {arXiv:1505.02965 [math.ST]} \BibitemShut
  {NoStop}%
\bibitem [{\citenamefont {{Gair}}\ and\ \citenamefont
  {{Moore}}(2015)}]{2015PhRvD..91l4062G}%
  \BibitemOpen
  \bibfield  {author} {\bibinfo {author} {\bibfnamefont {J.~R.}\ \bibnamefont
  {{Gair}}}\ and\ \bibinfo {author} {\bibfnamefont {C.~J.}\ \bibnamefont
  {{Moore}}},\ }\href {\doibase 10.1103/PhysRevD.91.124062} {\bibfield
  {journal} {\bibinfo  {journal} {\prd}\ }\textbf {\bibinfo {volume} {91}},\
  \bibinfo {eid} {124062} (\bibinfo {year} {2015})},\ \Eprint
  {http://arxiv.org/abs/1504.02767} {arXiv:1504.02767 [gr-qc]} \BibitemShut
  {NoStop}%
\bibitem [{\citenamefont {{Moore}}\ \emph {et~al.}(2016)\citenamefont
  {{Moore}}, \citenamefont {{Berry}}, \citenamefont {{Chua}},\ and\
  \citenamefont {{Gair}}}]{2016PhRvD..93f4001M}%
  \BibitemOpen
  \bibfield  {author} {\bibinfo {author} {\bibfnamefont {C.~J.}\ \bibnamefont
  {{Moore}}}, \bibinfo {author} {\bibfnamefont {C.~P.~L.}\ \bibnamefont
  {{Berry}}}, \bibinfo {author} {\bibfnamefont {A.~J.~K.}\ \bibnamefont
  {{Chua}}}, \ and\ \bibinfo {author} {\bibfnamefont {J.~R.}\ \bibnamefont
  {{Gair}}},\ }\href {\doibase 10.1103/PhysRevD.93.064001} {\bibfield
  {journal} {\bibinfo  {journal} {\prd}\ }\textbf {\bibinfo {volume} {93}},\
  \bibinfo {eid} {064001} (\bibinfo {year} {2016})},\ \Eprint
  {http://arxiv.org/abs/1509.04066} {arXiv:1509.04066 [gr-qc]} \BibitemShut
  {NoStop}%
\bibitem [{\citenamefont {{O'Shaughnessy}}(2013)}]{2013PhRvD..88h4061O}%
  \BibitemOpen
  \bibfield  {author} {\bibinfo {author} {\bibfnamefont {R.}~\bibnamefont
  {{O'Shaughnessy}}},\ }\href {\doibase 10.1103/PhysRevD.88.084061} {\bibfield
  {journal} {\bibinfo  {journal} {\prd}\ }\textbf {\bibinfo {volume} {88}},\
  \bibinfo {eid} {084061} (\bibinfo {year} {2013})},\ \Eprint
  {http://arxiv.org/abs/1204.3117} {arXiv:1204.3117 [astro-ph.CO]} \BibitemShut
  {NoStop}%
\bibitem [{\citenamefont {{van Haasteren}}\ and\ \citenamefont
  {{Vallisneri}}(2014)}]{2014PhRvD..90j4012V}%
  \BibitemOpen
  \bibfield  {author} {\bibinfo {author} {\bibfnamefont {R.}~\bibnamefont {{van
  Haasteren}}}\ and\ \bibinfo {author} {\bibfnamefont {M.}~\bibnamefont
  {{Vallisneri}}},\ }\href {\doibase 10.1103/PhysRevD.90.104012} {\bibfield
  {journal} {\bibinfo  {journal} {\prd}\ }\textbf {\bibinfo {volume} {90}},\
  \bibinfo {eid} {104012} (\bibinfo {year} {2014})},\ \Eprint
  {http://arxiv.org/abs/1407.1838} {arXiv:1407.1838 [gr-qc]} \BibitemShut
  {NoStop}%
\bibitem [{\citenamefont {{Ambikasaran}}\ \emph {et~al.}(2015)\citenamefont
  {{Ambikasaran}}, \citenamefont {{Foreman-Mackey}}, \citenamefont
  {{Greengard}}, \citenamefont {{Hogg}},\ and\ \citenamefont
  {{O'Neil}}}]{2015ITPAM..38..252A}%
  \BibitemOpen
  \bibfield  {author} {\bibinfo {author} {\bibfnamefont {S.}~\bibnamefont
  {{Ambikasaran}}}, \bibinfo {author} {\bibfnamefont {D.}~\bibnamefont
  {{Foreman-Mackey}}}, \bibinfo {author} {\bibfnamefont {L.}~\bibnamefont
  {{Greengard}}}, \bibinfo {author} {\bibfnamefont {D.~W.}\ \bibnamefont
  {{Hogg}}}, \ and\ \bibinfo {author} {\bibfnamefont {M.}~\bibnamefont
  {{O'Neil}}},\ }\href@noop {} {\bibfield  {journal} {\bibinfo  {journal} {IEEE
  Transactions on Pattern Analysis and Machine Intelligence}\ }\textbf
  {\bibinfo {volume} {38}} (\bibinfo {year} {2015})},\ \Eprint
  {http://arxiv.org/abs/1403.6015} {arXiv:1403.6015 [math.NA]} \BibitemShut
  {NoStop}%
\bibitem [{\citenamefont {{Bayes}}(1763)}]{1763RSPT...53..370B}%
  \BibitemOpen
  \bibfield  {author} {\bibinfo {author} {\bibfnamefont {T.}~\bibnamefont
  {{Bayes}}},\ }\href {\doibase 10.1098/rstl.1763.0053} {\bibfield  {journal}
  {\bibinfo  {journal} {Philosophical Transactions of the Royal Society of
  London Series I}\ }\textbf {\bibinfo {volume} {53}},\ \bibinfo {pages} {370}
  (\bibinfo {year} {1763})}\BibitemShut {NoStop}%
\bibitem [{\citenamefont {{Foreman-Mackey}}\ \emph {et~al.}(2013)\citenamefont
  {{Foreman-Mackey}}, \citenamefont {{Hogg}}, \citenamefont {{Lang}},\ and\
  \citenamefont {{Goodman}}}]{2013PASP..125..306F}%
  \BibitemOpen
  \bibfield  {author} {\bibinfo {author} {\bibfnamefont {D.}~\bibnamefont
  {{Foreman-Mackey}}}, \bibinfo {author} {\bibfnamefont {D.~W.}\ \bibnamefont
  {{Hogg}}}, \bibinfo {author} {\bibfnamefont {D.}~\bibnamefont {{Lang}}}, \
  and\ \bibinfo {author} {\bibfnamefont {J.}~\bibnamefont {{Goodman}}},\ }\href
  {\doibase 10.1086/670067} {\bibfield  {journal} {\bibinfo  {journal} {\pasp}\
  }\textbf {\bibinfo {volume} {125}},\ \bibinfo {pages} {306} (\bibinfo {year}
  {2013})},\ \Eprint {http://arxiv.org/abs/1202.3665} {arXiv:1202.3665
  [astro-ph.IM]} \BibitemShut {NoStop}%
\bibitem [{\citenamefont {{Shoemaker}}\ \emph {et~al.}()\citenamefont
  {{Shoemaker}} \emph {et~al.}}]{LIGOcurve}%
  \BibitemOpen
  \bibfield  {author} {\bibinfo {author} {\bibfnamefont {D.}~\bibnamefont
  {{Shoemaker}}} \emph {et~al.},\ }\href@noop {} {\bibinfo  {journal}
  {\href{https://dcc.ligo.org/LIGO-T0900288/public}{dcc.ligo.org/LIGO-T0900288}}\
  }\BibitemShut {NoStop}%
\bibitem [{\citenamefont {{B.~P.~Abbott {\it et al.} (LIGO and Virgo
  Collaboration)}}(2018{\natexlab{b}})}]{2018arXiv180511579T}%
  \BibitemOpen
\bibfield  {journal} {  }\bibfield  {author} {\bibinfo {author} {\bibnamefont
  {{B.~P.~Abbott {\it et al.} (LIGO and Virgo Collaboration)}}},\ }\href@noop
  {} {\  (\bibinfo {year} {2018}{\natexlab{b}})},\ \Eprint
  {http://arxiv.org/abs/1805.11579} {arXiv:1805.11579 [gr-qc]} \BibitemShut
  {NoStop}%
\bibitem [{\citenamefont {{Vitale}}\ \emph {et~al.}(2017)\citenamefont
  {{Vitale}}, \citenamefont {{Gerosa}}, \citenamefont {{Haster}}, \citenamefont
  {{Chatziioannou}},\ and\ \citenamefont {{Zimmerman}}}]{2017PhRvL.119y1103V}%
  \BibitemOpen
  \bibfield  {author} {\bibinfo {author} {\bibfnamefont {S.}~\bibnamefont
  {{Vitale}}}, \bibinfo {author} {\bibfnamefont {D.}~\bibnamefont {{Gerosa}}},
  \bibinfo {author} {\bibfnamefont {C.-J.}\ \bibnamefont {{Haster}}}, \bibinfo
  {author} {\bibfnamefont {K.}~\bibnamefont {{Chatziioannou}}}, \ and\ \bibinfo
  {author} {\bibfnamefont {A.}~\bibnamefont {{Zimmerman}}},\ }\href {\doibase
  10.1103/PhysRevLett.119.251103} {\bibfield  {journal} {\bibinfo  {journal}
  {\prl}\ }\textbf {\bibinfo {volume} {119}},\ \bibinfo {eid} {251103}
  (\bibinfo {year} {2017})},\ \Eprint {http://arxiv.org/abs/1707.04637}
  {arXiv:1707.04637 [gr-qc]} \BibitemShut {NoStop}%
\bibitem [{\citenamefont {{Khan}}\ \emph {et~al.}(2016)\citenamefont {{Khan}},
  \citenamefont {{Husa}}, \citenamefont {{Hannam}}, \citenamefont {{Ohme}},
  \citenamefont {{P{\"u}rrer}}, \citenamefont {{Forteza}},\ and\ \citenamefont
  {{Boh{\'e}}}}]{2016PhRvD..93d4007K}%
  \BibitemOpen
  \bibfield  {author} {\bibinfo {author} {\bibfnamefont {S.}~\bibnamefont
  {{Khan}}}, \bibinfo {author} {\bibfnamefont {S.}~\bibnamefont {{Husa}}},
  \bibinfo {author} {\bibfnamefont {M.}~\bibnamefont {{Hannam}}}, \bibinfo
  {author} {\bibfnamefont {F.}~\bibnamefont {{Ohme}}}, \bibinfo {author}
  {\bibfnamefont {M.}~\bibnamefont {{P{\"u}rrer}}}, \bibinfo {author}
  {\bibfnamefont {X.~J.}\ \bibnamefont {{Forteza}}}, \ and\ \bibinfo {author}
  {\bibfnamefont {A.}~\bibnamefont {{Boh{\'e}}}},\ }\href {\doibase
  10.1103/PhysRevD.93.044007} {\bibfield  {journal} {\bibinfo  {journal}
  {\prd}\ }\textbf {\bibinfo {volume} {93}},\ \bibinfo {eid} {044007} (\bibinfo
  {year} {2016})},\ \Eprint {http://arxiv.org/abs/1508.07253} {arXiv:1508.07253
  [gr-qc]} \BibitemShut {NoStop}%
\bibitem [{\citenamefont {{Ng}}\ \emph {et~al.}(2018)\citenamefont {{Ng}},
  \citenamefont {{Vitale}}, \citenamefont {{Zimmerman}}, \citenamefont
  {{Chatziioannou}}, \citenamefont {{Gerosa}},\ and\ \citenamefont
  {{Haster}}}]{2018arXiv180503046N}%
  \BibitemOpen
  \bibfield  {author} {\bibinfo {author} {\bibfnamefont {K.~K.~Y.}\
  \bibnamefont {{Ng}}}, \bibinfo {author} {\bibfnamefont {S.}~\bibnamefont
  {{Vitale}}}, \bibinfo {author} {\bibfnamefont {A.}~\bibnamefont
  {{Zimmerman}}}, \bibinfo {author} {\bibfnamefont {K.}~\bibnamefont
  {{Chatziioannou}}}, \bibinfo {author} {\bibfnamefont {D.}~\bibnamefont
  {{Gerosa}}}, \ and\ \bibinfo {author} {\bibfnamefont {C.-J.}\ \bibnamefont
  {{Haster}}},\ }\href@noop {} {\  (\bibinfo {year} {2018})},\ \Eprint
  {http://arxiv.org/abs/1805.03046} {arXiv:1805.03046 [gr-qc]} \BibitemShut
  {NoStop}%
\bibitem [{\citenamefont {{Dal Canton}}\ \emph {et~al.}(2014)\citenamefont
  {{Dal Canton}}, \citenamefont {{Nitz}}, \citenamefont {{Lundgren}},
  \citenamefont {{Nielsen}}, \citenamefont {{Brown}}, \citenamefont {{Dent}},
  \citenamefont {{Harry}}, \citenamefont {{Krishnan}}, \citenamefont
  {{Miller}}, \citenamefont {{Wette}}, \citenamefont {{Wiesner}},\ and\
  \citenamefont {{Willis}}}]{2014PhRvD..90h2004D}%
  \BibitemOpen
  \bibfield  {author} {\bibinfo {author} {\bibfnamefont {T.}~\bibnamefont {{Dal
  Canton}}}, \bibinfo {author} {\bibfnamefont {A.~H.}\ \bibnamefont {{Nitz}}},
  \bibinfo {author} {\bibfnamefont {A.~P.}\ \bibnamefont {{Lundgren}}},
  \bibinfo {author} {\bibfnamefont {A.~B.}\ \bibnamefont {{Nielsen}}}, \bibinfo
  {author} {\bibfnamefont {D.~A.}\ \bibnamefont {{Brown}}}, \bibinfo {author}
  {\bibfnamefont {T.}~\bibnamefont {{Dent}}}, \bibinfo {author} {\bibfnamefont
  {I.~W.}\ \bibnamefont {{Harry}}}, \bibinfo {author} {\bibfnamefont
  {B.}~\bibnamefont {{Krishnan}}}, \bibinfo {author} {\bibfnamefont {A.~J.}\
  \bibnamefont {{Miller}}}, \bibinfo {author} {\bibfnamefont {K.}~\bibnamefont
  {{Wette}}}, \bibinfo {author} {\bibfnamefont {K.}~\bibnamefont {{Wiesner}}},
  \ and\ \bibinfo {author} {\bibfnamefont {J.~L.}\ \bibnamefont {{Willis}}},\
  }\href {\doibase 10.1103/PhysRevD.90.082004} {\bibfield  {journal} {\bibinfo
  {journal} {\prd}\ }\textbf {\bibinfo {volume} {90}},\ \bibinfo {eid} {082004}
  (\bibinfo {year} {2014})},\ \Eprint {http://arxiv.org/abs/1405.6731}
  {arXiv:1405.6731 [gr-qc]} \BibitemShut {NoStop}%
\bibitem [{\citenamefont {{Usman}}\ \emph {et~al.}(2016)\citenamefont
  {{Usman}}, \citenamefont {{Nitz}}, \citenamefont {{Harry}}, \citenamefont
  {{Biwer}}, \citenamefont {{Brown}}, \citenamefont {{Cabero}}, \citenamefont
  {{Capano}}, \citenamefont {{Dal Canton}}, \citenamefont {{Dent}},
  \citenamefont {{Fairhurst}}, \citenamefont {{Kehl}}, \citenamefont
  {{Keppel}}, \citenamefont {{Krishnan}}, \citenamefont {{Lenon}},
  \citenamefont {{Lundgren}}, \citenamefont {{Nielsen}}, \citenamefont
  {{Pekowsky}}, \citenamefont {{Pfeiffer}}, \citenamefont {{Saulson}},
  \citenamefont {{West}},\ and\ \citenamefont
  {{Willis}}}]{2016CQGra..33u5004U}%
  \BibitemOpen
  \bibfield  {author} {\bibinfo {author} {\bibfnamefont {S.~A.}\ \bibnamefont
  {{Usman}}}, \bibinfo {author} {\bibfnamefont {A.~H.}\ \bibnamefont {{Nitz}}},
  \bibinfo {author} {\bibfnamefont {I.~W.}\ \bibnamefont {{Harry}}}, \bibinfo
  {author} {\bibfnamefont {C.~M.}\ \bibnamefont {{Biwer}}}, \bibinfo {author}
  {\bibfnamefont {D.~A.}\ \bibnamefont {{Brown}}}, \bibinfo {author}
  {\bibfnamefont {M.}~\bibnamefont {{Cabero}}}, \bibinfo {author}
  {\bibfnamefont {C.~D.}\ \bibnamefont {{Capano}}}, \bibinfo {author}
  {\bibfnamefont {T.}~\bibnamefont {{Dal Canton}}}, \bibinfo {author}
  {\bibfnamefont {T.}~\bibnamefont {{Dent}}}, \bibinfo {author} {\bibfnamefont
  {S.}~\bibnamefont {{Fairhurst}}}, \bibinfo {author} {\bibfnamefont {M.~S.}\
  \bibnamefont {{Kehl}}}, \bibinfo {author} {\bibfnamefont {D.}~\bibnamefont
  {{Keppel}}}, \bibinfo {author} {\bibfnamefont {B.}~\bibnamefont
  {{Krishnan}}}, \bibinfo {author} {\bibfnamefont {A.}~\bibnamefont {{Lenon}}},
  \bibinfo {author} {\bibfnamefont {A.}~\bibnamefont {{Lundgren}}}, \bibinfo
  {author} {\bibfnamefont {A.~B.}\ \bibnamefont {{Nielsen}}}, \bibinfo {author}
  {\bibfnamefont {L.~P.}\ \bibnamefont {{Pekowsky}}}, \bibinfo {author}
  {\bibfnamefont {H.~P.}\ \bibnamefont {{Pfeiffer}}}, \bibinfo {author}
  {\bibfnamefont {P.~R.}\ \bibnamefont {{Saulson}}}, \bibinfo {author}
  {\bibfnamefont {M.}~\bibnamefont {{West}}}, \ and\ \bibinfo {author}
  {\bibfnamefont {J.~L.}\ \bibnamefont {{Willis}}},\ }\href {\doibase
  10.1088/0264-9381/33/21/215004} {\bibfield  {journal} {\bibinfo  {journal}
  {\cqg}\ }\textbf {\bibinfo {volume} {33}},\ \bibinfo {eid} {215004} (\bibinfo
  {year} {2016})},\ \Eprint {http://arxiv.org/abs/1508.02357} {arXiv:1508.02357
  [gr-qc]} \BibitemShut {NoStop}%
\bibitem [{\citenamefont {{Sathyaprakash}}\ and\ \citenamefont
  {{Schutz}}(2009)}]{2009LRR....12....2S}%
  \BibitemOpen
  \bibfield  {author} {\bibinfo {author} {\bibfnamefont {B.~S.}\ \bibnamefont
  {{Sathyaprakash}}}\ and\ \bibinfo {author} {\bibfnamefont {B.~F.}\
  \bibnamefont {{Schutz}}},\ }\href {\doibase 10.12942/lrr-2009-2} {\bibfield
  {journal} {\bibinfo  {journal} {\llr}\ }\textbf {\bibinfo {volume} {12}},\
  \bibinfo {eid} {2} (\bibinfo {year} {2009})},\ \Eprint
  {http://arxiv.org/abs/0903.0338} {arXiv:0903.0338 [gr-qc]} \BibitemShut
  {NoStop}%
\bibitem [{\citenamefont {{Finn}}\ and\ \citenamefont
  {{Chernoff}}(1993)}]{1993PhRvD..47.2198F}%
  \BibitemOpen
  \bibfield  {author} {\bibinfo {author} {\bibfnamefont {L.~S.}\ \bibnamefont
  {{Finn}}}\ and\ \bibinfo {author} {\bibfnamefont {D.~F.}\ \bibnamefont
  {{Chernoff}}},\ }\href {\doibase 10.1103/PhysRevD.47.2198} {\bibfield
  {journal} {\bibinfo  {journal} {\prd}\ }\textbf {\bibinfo {volume} {47}},\
  \bibinfo {pages} {2198} (\bibinfo {year} {1993})},\ \Eprint
  {http://arxiv.org/abs/gr-qc/9301003} {gr-qc/9301003} \BibitemShut {NoStop}%
\bibitem [{\citenamefont {{Finn}}(1996)}]{1996PhRvD..53.2878F}%
  \BibitemOpen
  \bibfield  {author} {\bibinfo {author} {\bibfnamefont {L.~S.}\ \bibnamefont
  {{Finn}}},\ }\href {\doibase 10.1103/PhysRevD.53.2878} {\bibfield  {journal}
  {\bibinfo  {journal} {\prd}\ }\textbf {\bibinfo {volume} {53}},\ \bibinfo
  {pages} {2878} (\bibinfo {year} {1996})},\ \Eprint
  {http://arxiv.org/abs/gr-qc/9601048} {gr-qc/9601048} \BibitemShut {NoStop}%
\bibitem [{\citenamefont {{Dominik}}\ \emph {et~al.}(2015)\citenamefont
  {{Dominik}}, \citenamefont {{Berti}}, \citenamefont {{O'Shaughnessy}},
  \citenamefont {{Mandel}}, \citenamefont {{Belczynski}}, \citenamefont
  {{Fryer}}, \citenamefont {{Holz}}, \citenamefont {{Bulik}},\ and\
  \citenamefont {{Pannarale}}}]{2015ApJ...806..263D}%
  \BibitemOpen
  \bibfield  {author} {\bibinfo {author} {\bibfnamefont {M.}~\bibnamefont
  {{Dominik}}}, \bibinfo {author} {\bibfnamefont {E.}~\bibnamefont {{Berti}}},
  \bibinfo {author} {\bibfnamefont {R.}~\bibnamefont {{O'Shaughnessy}}},
  \bibinfo {author} {\bibfnamefont {I.}~\bibnamefont {{Mandel}}}, \bibinfo
  {author} {\bibfnamefont {K.}~\bibnamefont {{Belczynski}}}, \bibinfo {author}
  {\bibfnamefont {C.}~\bibnamefont {{Fryer}}}, \bibinfo {author} {\bibfnamefont
  {D.~E.}\ \bibnamefont {{Holz}}}, \bibinfo {author} {\bibfnamefont
  {T.}~\bibnamefont {{Bulik}}}, \ and\ \bibinfo {author} {\bibfnamefont
  {F.}~\bibnamefont {{Pannarale}}},\ }\href {\doibase
  10.1088/0004-637X/806/2/263} {\bibfield  {journal} {\bibinfo  {journal}
  {\apj}\ }\textbf {\bibinfo {volume} {806}},\ \bibinfo {eid} {263} (\bibinfo
  {year} {2015})},\ \Eprint {http://arxiv.org/abs/1405.7016} {arXiv:1405.7016
  [astro-ph.HE]} \BibitemShut {NoStop}%
\bibitem [{\citenamefont {{J. Abadie {\it et al.} (LIGO and Virgo
  Collaboration)}}(2010)}]{2010CQGra..27q3001A}%
  \BibitemOpen
  \bibfield  {author} {\bibinfo {author} {\bibnamefont {{J. Abadie {\it et al.}
  (LIGO and Virgo Collaboration)}}},\ }\href {\doibase
  10.1088/0264-9381/27/17/173001} {\bibfield  {journal} {\bibinfo  {journal}
  {\cqg}\ }\textbf {\bibinfo {volume} {27}},\ \bibinfo {eid} {173001} (\bibinfo
  {year} {2010})},\ \Eprint {http://arxiv.org/abs/1003.2480} {arXiv:1003.2480
  [astro-ph.HE]} \BibitemShut {NoStop}%
\bibitem [{\citenamefont {{Gerosa}}()}]{davide_gerosa_2017_889966}%
  \BibitemOpen
  \bibfield  {author} {\bibinfo {author} {\bibfnamefont {D.}~\bibnamefont
  {{Gerosa}}},\ }\href@noop {} {\bibinfo  {journal} {{\it gwdet: Detectability
  of gravitational-wave signals from compact binary coalescences}
  \href{https://doi.org/10.5281/zenodo.889966}{doi.org/10.5281/zenodo.889966}}\
  }\BibitemShut {NoStop}%
\bibitem [{\citenamefont {{Mandel}}(2010)}]{2010PhRvD..81h4029M}%
  \BibitemOpen
\bibfield  {journal} {  }\bibfield  {author} {\bibinfo {author} {\bibfnamefont
  {I.}~\bibnamefont {{Mandel}}},\ }\href {\doibase 10.1103/PhysRevD.81.084029}
  {\bibfield  {journal} {\bibinfo  {journal} {\prd}\ }\textbf {\bibinfo
  {volume} {81}},\ \bibinfo {eid} {084029} (\bibinfo {year} {2010})},\ \Eprint
  {http://arxiv.org/abs/0912.5531} {arXiv:0912.5531 [astro-ph.HE]} \BibitemShut
  {NoStop}%
\bibitem [{\citenamefont {{Hogg}}\ \emph {et~al.}(2010)\citenamefont {{Hogg}},
  \citenamefont {{Myers}},\ and\ \citenamefont {{Bovy}}}]{2010ApJ...725.2166H}%
  \BibitemOpen
  \bibfield  {author} {\bibinfo {author} {\bibfnamefont {D.~W.}\ \bibnamefont
  {{Hogg}}}, \bibinfo {author} {\bibfnamefont {A.~D.}\ \bibnamefont {{Myers}}},
  \ and\ \bibinfo {author} {\bibfnamefont {J.}~\bibnamefont {{Bovy}}},\ }\href
  {\doibase 10.1088/0004-637X/725/2/2166} {\bibfield  {journal} {\bibinfo
  {journal} {\apj}\ }\textbf {\bibinfo {volume} {725}},\ \bibinfo {pages}
  {2166} (\bibinfo {year} {2010})},\ \Eprint {http://arxiv.org/abs/1008.4146}
  {arXiv:1008.4146 [astro-ph.SR]} \BibitemShut {NoStop}%
\bibitem [{\citenamefont {{Adams}}\ \emph {et~al.}(2012)\citenamefont
  {{Adams}}, \citenamefont {{Cornish}},\ and\ \citenamefont
  {{Littenberg}}}]{2012PhRvD..86l4032A}%
  \BibitemOpen
  \bibfield  {author} {\bibinfo {author} {\bibfnamefont {M.~R.}\ \bibnamefont
  {{Adams}}}, \bibinfo {author} {\bibfnamefont {N.~J.}\ \bibnamefont
  {{Cornish}}}, \ and\ \bibinfo {author} {\bibfnamefont {T.~B.}\ \bibnamefont
  {{Littenberg}}},\ }\href {\doibase 10.1103/PhysRevD.86.124032} {\bibfield
  {journal} {\bibinfo  {journal} {\prd}\ }\textbf {\bibinfo {volume} {86}},\
  \bibinfo {eid} {124032} (\bibinfo {year} {2012})},\ \Eprint
  {http://arxiv.org/abs/1209.6286} {arXiv:1209.6286 [gr-qc]} \BibitemShut
  {NoStop}%
\bibitem [{\citenamefont {{Foreman-Mackey}}\ \emph {et~al.}(2014)\citenamefont
  {{Foreman-Mackey}}, \citenamefont {{Hogg}},\ and\ \citenamefont
  {{Morton}}}]{2014ApJ...795...64F}%
  \BibitemOpen
  \bibfield  {author} {\bibinfo {author} {\bibfnamefont {D.}~\bibnamefont
  {{Foreman-Mackey}}}, \bibinfo {author} {\bibfnamefont {D.~W.}\ \bibnamefont
  {{Hogg}}}, \ and\ \bibinfo {author} {\bibfnamefont {T.~D.}\ \bibnamefont
  {{Morton}}},\ }\href {\doibase 10.1088/0004-637X/795/1/64} {\bibfield
  {journal} {\bibinfo  {journal} {\apj}\ }\textbf {\bibinfo {volume} {795}},\
  \bibinfo {eid} {64} (\bibinfo {year} {2014})},\ \Eprint
  {http://arxiv.org/abs/1406.3020} {arXiv:1406.3020 [astro-ph.EP]} \BibitemShut
  {NoStop}%
\bibitem [{\citenamefont {{Taylor}}\ and\ \citenamefont
  {{Gair}}(2012)}]{2012PhRvD..86b3502T}%
  \BibitemOpen
  \bibfield  {author} {\bibinfo {author} {\bibfnamefont {S.~R.}\ \bibnamefont
  {{Taylor}}}\ and\ \bibinfo {author} {\bibfnamefont {J.~R.}\ \bibnamefont
  {{Gair}}},\ }\href {\doibase 10.1103/PhysRevD.86.023502} {\bibfield
  {journal} {\bibinfo  {journal} {\prd}\ }\textbf {\bibinfo {volume} {86}},\
  \bibinfo {eid} {023502} (\bibinfo {year} {2012})},\ \Eprint
  {http://arxiv.org/abs/1204.6739} {arXiv:1204.6739 [astro-ph.CO]} \BibitemShut
  {NoStop}%
\bibitem [{\citenamefont {{Pitkin}}\ \emph {et~al.}(2018)\citenamefont
  {{Pitkin}}, \citenamefont {{Messenger}},\ and\ \citenamefont
  {{Fan}}}]{2018arXiv180706726P}%
  \BibitemOpen
  \bibfield  {author} {\bibinfo {author} {\bibfnamefont {M.}~\bibnamefont
  {{Pitkin}}}, \bibinfo {author} {\bibfnamefont {C.}~\bibnamefont
  {{Messenger}}}, \ and\ \bibinfo {author} {\bibfnamefont {X.}~\bibnamefont
  {{Fan}}},\ }\href@noop {} {\bibfield  {journal} {\bibinfo  {journal} {ArXiv
  e-prints}\ } (\bibinfo {year} {2018})},\ \Eprint
  {http://arxiv.org/abs/1807.06726} {arXiv:1807.06726 [astro-ph.IM]}
  \BibitemShut {NoStop}%
\bibitem [{\citenamefont {{Mandel}}\ \emph {et~al.}(2018)\citenamefont
  {{Mandel}}, \citenamefont {{Farr}},\ and\ \citenamefont {{Gair}}}]{mfg18}%
  \BibitemOpen
  \bibfield  {author} {\bibinfo {author} {\bibfnamefont {I.}~\bibnamefont
  {{Mandel}}}, \bibinfo {author} {\bibfnamefont {W.}~\bibnamefont {{Farr}}}, \
  and\ \bibinfo {author} {\bibfnamefont {J.~R.}\ \bibnamefont {{Gair}}},\
  }\href@noop {} {\bibfield  {journal} {\bibinfo  {journal} {ArXiv e-prints}\ }
  (\bibinfo {year} {2018})},\ \Eprint {http://arxiv.org/abs/1809.02063}
  {arXiv:1809.02063 [astro-ph.physics.data-an]} \BibitemShut {NoStop}%
\bibitem [{\citenamefont {{Farr}}\ \emph {et~al.}()\citenamefont {{Farr}},
  \citenamefont {{Mandel}},\ and\ \citenamefont {{Gair}}}]{selectioneffects}%
  \BibitemOpen
  \bibfield  {author} {\bibinfo {author} {\bibfnamefont {W.}~\bibnamefont
  {{Farr}}}, \bibinfo {author} {\bibfnamefont {I.}~\bibnamefont {{Mandel}}}, \
  and\ \bibinfo {author} {\bibfnamefont {J.~R.}\ \bibnamefont {{Gair}}},\
  }\href@noop {} {\bibinfo  {journal} {{\it Selection effects}
  \href{https://github.com/farr/SelectionExample}{github.com/farr/SelectionExample}}\
  }\BibitemShut {NoStop}%
\bibitem [{\citenamefont {{Sesana}}\ \emph {et~al.}(2011)\citenamefont
  {{Sesana}}, \citenamefont {{Gair}}, \citenamefont {{Berti}},\ and\
  \citenamefont {{Volonteri}}}]{2011PhRvD..83d4036S}%
  \BibitemOpen
\bibfield  {journal} {  }\bibfield  {author} {\bibinfo {author} {\bibfnamefont
  {A.}~\bibnamefont {{Sesana}}}, \bibinfo {author} {\bibfnamefont
  {J.}~\bibnamefont {{Gair}}}, \bibinfo {author} {\bibfnamefont
  {E.}~\bibnamefont {{Berti}}}, \ and\ \bibinfo {author} {\bibfnamefont
  {M.}~\bibnamefont {{Volonteri}}},\ }\href {\doibase
  10.1103/PhysRevD.83.044036} {\bibfield  {journal} {\bibinfo  {journal}
  {\prd}\ }\textbf {\bibinfo {volume} {83}},\ \bibinfo {eid} {044036} (\bibinfo
  {year} {2011})},\ \Eprint {http://arxiv.org/abs/1011.5893} {arXiv:1011.5893
  [astro-ph.CO]} \BibitemShut {NoStop}%
\bibitem [{\citenamefont {{Fishbach}}\ \emph {et~al.}(2018)\citenamefont
  {{Fishbach}}, \citenamefont {{Holz}},\ and\ \citenamefont
  {{Farr}}}]{2018ApJ...863L..41F}%
  \BibitemOpen
  \bibfield  {author} {\bibinfo {author} {\bibfnamefont {M.}~\bibnamefont
  {{Fishbach}}}, \bibinfo {author} {\bibfnamefont {D.~E.}\ \bibnamefont
  {{Holz}}}, \ and\ \bibinfo {author} {\bibfnamefont {W.~M.}\ \bibnamefont
  {{Farr}}},\ }\href {\doibase 10.3847/2041-8213/aad800} {\bibfield  {journal}
  {\bibinfo  {journal} {\apj}\ }\textbf {\bibinfo {volume} {863}},\ \bibinfo
  {eid} {L41} (\bibinfo {year} {2018})}\BibitemShut {NoStop}%
\bibitem [{\citenamefont {{Talbot}}\ and\ \citenamefont
  {{Thrane}}(2017)}]{2017PhRvD..96b3012T}%
  \BibitemOpen
  \bibfield  {author} {\bibinfo {author} {\bibfnamefont {C.}~\bibnamefont
  {{Talbot}}}\ and\ \bibinfo {author} {\bibfnamefont {E.}~\bibnamefont
  {{Thrane}}},\ }\href {\doibase 10.1103/PhysRevD.96.023012} {\bibfield
  {journal} {\bibinfo  {journal} {\prd}\ }\textbf {\bibinfo {volume} {96}},\
  \bibinfo {eid} {023012} (\bibinfo {year} {2017})},\ \Eprint
  {http://arxiv.org/abs/1704.08370} {arXiv:1704.08370 [astro-ph.HE]}
  \BibitemShut {NoStop}%
\bibitem [{\citenamefont {{Gerosa}}\ \emph {et~al.}(2013)\citenamefont
  {{Gerosa}}, \citenamefont {{Kesden}}, \citenamefont {{Berti}}, \citenamefont
  {{O'Shaughnessy}},\ and\ \citenamefont {{Sperhake}}}]{2013PhRvD..87j4028G}%
  \BibitemOpen
  \bibfield  {author} {\bibinfo {author} {\bibfnamefont {D.}~\bibnamefont
  {{Gerosa}}}, \bibinfo {author} {\bibfnamefont {M.}~\bibnamefont {{Kesden}}},
  \bibinfo {author} {\bibfnamefont {E.}~\bibnamefont {{Berti}}}, \bibinfo
  {author} {\bibfnamefont {R.}~\bibnamefont {{O'Shaughnessy}}}, \ and\ \bibinfo
  {author} {\bibfnamefont {U.}~\bibnamefont {{Sperhake}}},\ }\href {\doibase
  10.1103/PhysRevD.87.104028} {\bibfield  {journal} {\bibinfo  {journal}
  {\prd}\ }\textbf {\bibinfo {volume} {87}},\ \bibinfo {eid} {104028} (\bibinfo
  {year} {2013})},\ \Eprint {http://arxiv.org/abs/1302.4442} {arXiv:1302.4442
  [gr-qc]} \BibitemShut {NoStop}%
\bibitem [{\citenamefont {{Gerosa}}\ \emph {et~al.}(2014)\citenamefont
  {{Gerosa}}, \citenamefont {{O'Shaughnessy}}, \citenamefont {{Kesden}},
  \citenamefont {{Berti}},\ and\ \citenamefont
  {{Sperhake}}}]{2014PhRvD..89l4025G}%
  \BibitemOpen
  \bibfield  {author} {\bibinfo {author} {\bibfnamefont {D.}~\bibnamefont
  {{Gerosa}}}, \bibinfo {author} {\bibfnamefont {R.}~\bibnamefont
  {{O'Shaughnessy}}}, \bibinfo {author} {\bibfnamefont {M.}~\bibnamefont
  {{Kesden}}}, \bibinfo {author} {\bibfnamefont {E.}~\bibnamefont {{Berti}}}, \
  and\ \bibinfo {author} {\bibfnamefont {U.}~\bibnamefont {{Sperhake}}},\
  }\href {\doibase 10.1103/PhysRevD.89.124025} {\bibfield  {journal} {\bibinfo
  {journal} {\prd}\ }\textbf {\bibinfo {volume} {89}},\ \bibinfo {eid} {124025}
  (\bibinfo {year} {2014})},\ \Eprint {http://arxiv.org/abs/1403.7147}
  {arXiv:1403.7147 [gr-qc]} \BibitemShut {NoStop}%
\bibitem [{\citenamefont {{Trifir{\`o}}}\ \emph {et~al.}(2016)\citenamefont
  {{Trifir{\`o}}}, \citenamefont {{O'Shaughnessy}}, \citenamefont {{Gerosa}},
  \citenamefont {{Berti}}, \citenamefont {{Kesden}}, \citenamefont
  {{Littenberg}},\ and\ \citenamefont {{Sperhake}}}]{2016PhRvD..93d4071T}%
  \BibitemOpen
  \bibfield  {author} {\bibinfo {author} {\bibfnamefont {D.}~\bibnamefont
  {{Trifir{\`o}}}}, \bibinfo {author} {\bibfnamefont {R.}~\bibnamefont
  {{O'Shaughnessy}}}, \bibinfo {author} {\bibfnamefont {D.}~\bibnamefont
  {{Gerosa}}}, \bibinfo {author} {\bibfnamefont {E.}~\bibnamefont {{Berti}}},
  \bibinfo {author} {\bibfnamefont {M.}~\bibnamefont {{Kesden}}}, \bibinfo
  {author} {\bibfnamefont {T.}~\bibnamefont {{Littenberg}}}, \ and\ \bibinfo
  {author} {\bibfnamefont {U.}~\bibnamefont {{Sperhake}}},\ }\href {\doibase
  10.1103/PhysRevD.93.044071} {\bibfield  {journal} {\bibinfo  {journal}
  {\prd}\ }\textbf {\bibinfo {volume} {93}},\ \bibinfo {eid} {044071} (\bibinfo
  {year} {2016})},\ \Eprint {http://arxiv.org/abs/1507.05587} {arXiv:1507.05587
  [gr-qc]} \BibitemShut {NoStop}%
\bibitem [{\citenamefont {{Rodriguez}}\ \emph {et~al.}(2016)\citenamefont
  {{Rodriguez}}, \citenamefont {{Zevin}}, \citenamefont {{Pankow}},
  \citenamefont {{Kalogera}},\ and\ \citenamefont
  {{Rasio}}}]{2016ApJ...832L...2R}%
  \BibitemOpen
  \bibfield  {author} {\bibinfo {author} {\bibfnamefont {C.~L.}\ \bibnamefont
  {{Rodriguez}}}, \bibinfo {author} {\bibfnamefont {M.}~\bibnamefont
  {{Zevin}}}, \bibinfo {author} {\bibfnamefont {C.}~\bibnamefont {{Pankow}}},
  \bibinfo {author} {\bibfnamefont {V.}~\bibnamefont {{Kalogera}}}, \ and\
  \bibinfo {author} {\bibfnamefont {F.~A.}\ \bibnamefont {{Rasio}}},\ }\href
  {\doibase 10.3847/2041-8205/832/1/L2} {\bibfield  {journal} {\bibinfo
  {journal} {\apjl}\ }\textbf {\bibinfo {volume} {832}},\ \bibinfo {eid} {L2}
  (\bibinfo {year} {2016})},\ \Eprint {http://arxiv.org/abs/1609.05916}
  {arXiv:1609.05916 [astro-ph.HE]} \BibitemShut {NoStop}%
\bibitem [{\citenamefont {{Farr}}\ \emph {et~al.}(2017)\citenamefont {{Farr}},
  \citenamefont {{Stevenson}}, \citenamefont {{Miller}}, \citenamefont
  {{Mandel}}, \citenamefont {{Farr}},\ and\ \citenamefont
  {{Vecchio}}}]{2017Natur.548..426F}%
  \BibitemOpen
  \bibfield  {author} {\bibinfo {author} {\bibfnamefont {W.~M.}\ \bibnamefont
  {{Farr}}}, \bibinfo {author} {\bibfnamefont {S.}~\bibnamefont {{Stevenson}}},
  \bibinfo {author} {\bibfnamefont {M.~C.}\ \bibnamefont {{Miller}}}, \bibinfo
  {author} {\bibfnamefont {I.}~\bibnamefont {{Mandel}}}, \bibinfo {author}
  {\bibfnamefont {B.}~\bibnamefont {{Farr}}}, \ and\ \bibinfo {author}
  {\bibfnamefont {A.}~\bibnamefont {{Vecchio}}},\ }\href {\doibase
  10.1038/nature23453} {\bibfield  {journal} {\bibinfo  {journal} {\nat}\
  }\textbf {\bibinfo {volume} {548}},\ \bibinfo {pages} {426} (\bibinfo {year}
  {2017})},\ \Eprint {http://arxiv.org/abs/1706.01385} {arXiv:1706.01385
  [astro-ph.HE]} \BibitemShut {NoStop}%
\bibitem [{\citenamefont {{Arca Sedda}}\ and\ \citenamefont
  {{Benacquista}}(2018)}]{2018arXiv180601285A}%
  \BibitemOpen
  \bibfield  {author} {\bibinfo {author} {\bibfnamefont {M.}~\bibnamefont
  {{Arca Sedda}}}\ and\ \bibinfo {author} {\bibfnamefont {M.}~\bibnamefont
  {{Benacquista}}},\ }\href@noop {} {\bibfield  {journal} {\bibinfo  {journal}
  {ArXiv e-prints}\ } (\bibinfo {year} {2018})},\ \Eprint
  {http://arxiv.org/abs/1806.01285} {arXiv:1806.01285} \BibitemShut {NoStop}%
\bibitem [{\citenamefont {{Abbott}}\ \emph {et~al.}(2016)\citenamefont
  {{Abbott}}, \citenamefont {{Abbott}}, \citenamefont {{Abbott}}, \citenamefont
  {{Abernathy}}, \citenamefont {{Acernese}}, \citenamefont {{Ackley}},
  \citenamefont {{Adams}}, \citenamefont {{Adams}}, \citenamefont {{Addesso}},
  \citenamefont {{Adhikari}},\ and\ \citenamefont
  {et~al.}}]{2016PhRvL.116x1102A}%
  \BibitemOpen
  \bibfield  {author} {\bibinfo {author} {\bibfnamefont {B.~P.}\ \bibnamefont
  {{Abbott}}}, \bibinfo {author} {\bibfnamefont {R.}~\bibnamefont {{Abbott}}},
  \bibinfo {author} {\bibfnamefont {T.~D.}\ \bibnamefont {{Abbott}}}, \bibinfo
  {author} {\bibfnamefont {M.~R.}\ \bibnamefont {{Abernathy}}}, \bibinfo
  {author} {\bibfnamefont {F.}~\bibnamefont {{Acernese}}}, \bibinfo {author}
  {\bibfnamefont {K.}~\bibnamefont {{Ackley}}}, \bibinfo {author}
  {\bibfnamefont {C.}~\bibnamefont {{Adams}}}, \bibinfo {author} {\bibfnamefont
  {T.}~\bibnamefont {{Adams}}}, \bibinfo {author} {\bibfnamefont
  {P.}~\bibnamefont {{Addesso}}}, \bibinfo {author} {\bibfnamefont {R.~X.}\
  \bibnamefont {{Adhikari}}}, \ and\ \bibinfo {author} {\bibnamefont
  {et~al.}},\ }\href {\doibase 10.1103/PhysRevLett.116.241102} {\bibfield
  {journal} {\bibinfo  {journal} {Physical Review Letters}\ }\textbf {\bibinfo
  {volume} {116}},\ \bibinfo {eid} {241102} (\bibinfo {year} {2016})},\ \Eprint
  {http://arxiv.org/abs/1602.03840} {arXiv:1602.03840 [gr-qc]} \BibitemShut
  {NoStop}%
\bibitem [{\citenamefont {{Hurley}}\ \emph {et~al.}(2000)\citenamefont
  {{Hurley}}, \citenamefont {{Pols}},\ and\ \citenamefont
  {{Tout}}}]{2000MNRAS.315..543H}%
  \BibitemOpen
  \bibfield  {author} {\bibinfo {author} {\bibfnamefont {J.~R.}\ \bibnamefont
  {{Hurley}}}, \bibinfo {author} {\bibfnamefont {O.~R.}\ \bibnamefont
  {{Pols}}}, \ and\ \bibinfo {author} {\bibfnamefont {C.~A.}\ \bibnamefont
  {{Tout}}},\ }\href {\doibase 10.1046/j.1365-8711.2000.03426.x} {\bibfield
  {journal} {\bibinfo  {journal} {\mnras}\ }\textbf {\bibinfo {volume} {315}},\
  \bibinfo {pages} {543} (\bibinfo {year} {2000})},\ \Eprint
  {http://arxiv.org/abs/astro-ph/0001295} {astro-ph/0001295} \BibitemShut
  {NoStop}%
\bibitem [{\citenamefont {{Lamberts}}\ \emph {et~al.}(2018)\citenamefont
  {{Lamberts}}, \citenamefont {{Garrison-Kimmel}}, \citenamefont {{Hopkins}},
  \citenamefont {{Quataert}}, \citenamefont {{Bullock}}, \citenamefont
  {{Faucher-Gigu{\`e}re}}, \citenamefont {{Wetzel}}, \citenamefont {{Keres}},
  \citenamefont {{Drango}},\ and\ \citenamefont
  {{Sanderson}}}]{2018arXiv180103099L}%
  \BibitemOpen
  \bibfield  {author} {\bibinfo {author} {\bibfnamefont {A.}~\bibnamefont
  {{Lamberts}}}, \bibinfo {author} {\bibfnamefont {S.}~\bibnamefont
  {{Garrison-Kimmel}}}, \bibinfo {author} {\bibfnamefont {P.}~\bibnamefont
  {{Hopkins}}}, \bibinfo {author} {\bibfnamefont {E.}~\bibnamefont
  {{Quataert}}}, \bibinfo {author} {\bibfnamefont {J.}~\bibnamefont
  {{Bullock}}}, \bibinfo {author} {\bibfnamefont {C.-A.}\ \bibnamefont
  {{Faucher-Gigu{\`e}re}}}, \bibinfo {author} {\bibfnamefont {A.}~\bibnamefont
  {{Wetzel}}}, \bibinfo {author} {\bibfnamefont {D.}~\bibnamefont {{Keres}}},
  \bibinfo {author} {\bibfnamefont {K.}~\bibnamefont {{Drango}}}, \ and\
  \bibinfo {author} {\bibfnamefont {R.}~\bibnamefont {{Sanderson}}},\
  }\href@noop {} {\  (\bibinfo {year} {2018})},\ \Eprint
  {http://arxiv.org/abs/1801.03099} {arXiv:1801.03099} \BibitemShut {NoStop}%
\bibitem [{\citenamefont {{Belczynski}}\ \emph
  {et~al.}(2010{\natexlab{a}})\citenamefont {{Belczynski}}, \citenamefont
  {{Bulik}}, \citenamefont {{Fryer}}, \citenamefont {{Ruiter}}, \citenamefont
  {{Valsecchi}}, \citenamefont {{Vink}},\ and\ \citenamefont
  {{Hurley}}}]{2010ApJ...714.1217B}%
  \BibitemOpen
  \bibfield  {author} {\bibinfo {author} {\bibfnamefont {K.}~\bibnamefont
  {{Belczynski}}}, \bibinfo {author} {\bibfnamefont {T.}~\bibnamefont
  {{Bulik}}}, \bibinfo {author} {\bibfnamefont {C.~L.}\ \bibnamefont
  {{Fryer}}}, \bibinfo {author} {\bibfnamefont {A.}~\bibnamefont {{Ruiter}}},
  \bibinfo {author} {\bibfnamefont {F.}~\bibnamefont {{Valsecchi}}}, \bibinfo
  {author} {\bibfnamefont {J.~S.}\ \bibnamefont {{Vink}}}, \ and\ \bibinfo
  {author} {\bibfnamefont {J.~R.}\ \bibnamefont {{Hurley}}},\ }\href {\doibase
  10.1088/0004-637X/714/2/1217} {\bibfield  {journal} {\bibinfo  {journal}
  {\apj}\ }\textbf {\bibinfo {volume} {714}},\ \bibinfo {pages} {1217}
  (\bibinfo {year} {2010}{\natexlab{a}})},\ \Eprint
  {http://arxiv.org/abs/0904.2784} {arXiv:0904.2784 [astro-ph.SR]} \BibitemShut
  {NoStop}%
\bibitem [{\citenamefont {{Planck Collaboration}}\ \emph
  {et~al.}(2016)\citenamefont {{Planck Collaboration}}, \citenamefont {{Ade}},
  \citenamefont {{Aghanim}}, \citenamefont {{Arnaud}}, \citenamefont
  {{Ashdown}}, \citenamefont {{Aumont}}, \citenamefont {{Baccigalupi}},
  \citenamefont {{Banday}}, \citenamefont {{Barreiro}}, \citenamefont
  {{Bartlett}},\ and\ \citenamefont {et~al.}}]{2016A&A...594A..13P}%
  \BibitemOpen
  \bibfield  {author} {\bibinfo {author} {\bibnamefont {{Planck
  Collaboration}}}, \bibinfo {author} {\bibfnamefont {P.~A.~R.}\ \bibnamefont
  {{Ade}}}, \bibinfo {author} {\bibfnamefont {N.}~\bibnamefont {{Aghanim}}},
  \bibinfo {author} {\bibfnamefont {M.}~\bibnamefont {{Arnaud}}}, \bibinfo
  {author} {\bibfnamefont {M.}~\bibnamefont {{Ashdown}}}, \bibinfo {author}
  {\bibfnamefont {J.}~\bibnamefont {{Aumont}}}, \bibinfo {author}
  {\bibfnamefont {C.}~\bibnamefont {{Baccigalupi}}}, \bibinfo {author}
  {\bibfnamefont {A.~J.}\ \bibnamefont {{Banday}}}, \bibinfo {author}
  {\bibfnamefont {R.~B.}\ \bibnamefont {{Barreiro}}}, \bibinfo {author}
  {\bibfnamefont {J.~G.}\ \bibnamefont {{Bartlett}}}, \ and\ \bibinfo {author}
  {\bibnamefont {et~al.}},\ }\href {\doibase 10.1051/0004-6361/201525830}
  {\bibfield  {journal} {\bibinfo  {journal} {\aap}\ }\textbf {\bibinfo
  {volume} {594}},\ \bibinfo {eid} {A13} (\bibinfo {year} {2016})},\ \Eprint
  {http://arxiv.org/abs/1502.01589} {arXiv:1502.01589} \BibitemShut {NoStop}%
\bibitem [{\citenamefont {{O'Shaughnessy}}\ \emph {et~al.}(2010)\citenamefont
  {{O'Shaughnessy}}, \citenamefont {{Kalogera}},\ and\ \citenamefont
  {{Belczynski}}}]{2010ApJ...716..615O}%
  \BibitemOpen
  \bibfield  {author} {\bibinfo {author} {\bibfnamefont {R.}~\bibnamefont
  {{O'Shaughnessy}}}, \bibinfo {author} {\bibfnamefont {V.}~\bibnamefont
  {{Kalogera}}}, \ and\ \bibinfo {author} {\bibfnamefont {K.}~\bibnamefont
  {{Belczynski}}},\ }\href {\doibase 10.1088/0004-637X/716/1/615} {\bibfield
  {journal} {\bibinfo  {journal} {\apj}\ }\textbf {\bibinfo {volume} {716}},\
  \bibinfo {pages} {615} (\bibinfo {year} {2010})},\ \Eprint
  {http://arxiv.org/abs/0908.3635} {arXiv:0908.3635} \BibitemShut {NoStop}%
\bibitem [{\citenamefont {{Belczynski}}\ \emph
  {et~al.}(2010{\natexlab{b}})\citenamefont {{Belczynski}}, \citenamefont
  {{Holz}}, \citenamefont {{Fryer}}, \citenamefont {{Berger}}, \citenamefont
  {{Hartmann}},\ and\ \citenamefont {{O'Shea}}}]{2010ApJ...708..117B}%
  \BibitemOpen
  \bibfield  {author} {\bibinfo {author} {\bibfnamefont {K.}~\bibnamefont
  {{Belczynski}}}, \bibinfo {author} {\bibfnamefont {D.~E.}\ \bibnamefont
  {{Holz}}}, \bibinfo {author} {\bibfnamefont {C.~L.}\ \bibnamefont {{Fryer}}},
  \bibinfo {author} {\bibfnamefont {E.}~\bibnamefont {{Berger}}}, \bibinfo
  {author} {\bibfnamefont {D.~H.}\ \bibnamefont {{Hartmann}}}, \ and\ \bibinfo
  {author} {\bibfnamefont {B.}~\bibnamefont {{O'Shea}}},\ }\href {\doibase
  10.1088/0004-637X/708/1/117} {\bibfield  {journal} {\bibinfo  {journal}
  {\apj}\ }\textbf {\bibinfo {volume} {708}},\ \bibinfo {pages} {117} (\bibinfo
  {year} {2010}{\natexlab{b}})},\ \Eprint {http://arxiv.org/abs/0812.2470}
  {arXiv:0812.2470} \BibitemShut {NoStop}%
\bibitem [{\citenamefont {{Chen}}\ \emph {et~al.}(2015)\citenamefont {{Chen}},
  \citenamefont {{Bressan}}, \citenamefont {{Girardi}}, \citenamefont
  {{Marigo}}, \citenamefont {{Kong}},\ and\ \citenamefont
  {{Lanza}}}]{2015MNRAS.452.1068C}%
  \BibitemOpen
  \bibfield  {author} {\bibinfo {author} {\bibfnamefont {Y.}~\bibnamefont
  {{Chen}}}, \bibinfo {author} {\bibfnamefont {A.}~\bibnamefont {{Bressan}}},
  \bibinfo {author} {\bibfnamefont {L.}~\bibnamefont {{Girardi}}}, \bibinfo
  {author} {\bibfnamefont {P.}~\bibnamefont {{Marigo}}}, \bibinfo {author}
  {\bibfnamefont {X.}~\bibnamefont {{Kong}}}, \ and\ \bibinfo {author}
  {\bibfnamefont {A.}~\bibnamefont {{Lanza}}},\ }\href {\doibase
  10.1093/mnras/stv1281} {\bibfield  {journal} {\bibinfo  {journal} {\mnras}\
  }\textbf {\bibinfo {volume} {452}},\ \bibinfo {pages} {1068} (\bibinfo {year}
  {2015})},\ \Eprint {http://arxiv.org/abs/1506.01681} {arXiv:1506.01681
  [astro-ph.SR]} \BibitemShut {NoStop}%
\bibitem [{\citenamefont {{Fryer}}\ \emph {et~al.}(2012)\citenamefont
  {{Fryer}}, \citenamefont {{Belczynski}}, \citenamefont {{Wiktorowicz}},
  \citenamefont {{Dominik}}, \citenamefont {{Kalogera}},\ and\ \citenamefont
  {{Holz}}}]{2012ApJ...749...91F}%
  \BibitemOpen
  \bibfield  {author} {\bibinfo {author} {\bibfnamefont {C.~L.}\ \bibnamefont
  {{Fryer}}}, \bibinfo {author} {\bibfnamefont {K.}~\bibnamefont
  {{Belczynski}}}, \bibinfo {author} {\bibfnamefont {G.}~\bibnamefont
  {{Wiktorowicz}}}, \bibinfo {author} {\bibfnamefont {M.}~\bibnamefont
  {{Dominik}}}, \bibinfo {author} {\bibfnamefont {V.}~\bibnamefont
  {{Kalogera}}}, \ and\ \bibinfo {author} {\bibfnamefont {D.~E.}\ \bibnamefont
  {{Holz}}},\ }\href {\doibase 10.1088/0004-637X/749/1/91} {\bibfield
  {journal} {\bibinfo  {journal} {\apj}\ }\textbf {\bibinfo {volume} {749}},\
  \bibinfo {eid} {91} (\bibinfo {year} {2012})},\ \Eprint
  {http://arxiv.org/abs/1110.1726} {arXiv:1110.1726 [astro-ph.SR]} \BibitemShut
  {NoStop}%
\bibitem [{\citenamefont {{Janka}}(2013)}]{2013MNRAS.434.1355J}%
  \BibitemOpen
  \bibfield  {author} {\bibinfo {author} {\bibfnamefont {H.-T.}\ \bibnamefont
  {{Janka}}},\ }\href {\doibase 10.1093/mnras/stt1106} {\bibfield  {journal}
  {\bibinfo  {journal} {\mnras}\ }\textbf {\bibinfo {volume} {434}},\ \bibinfo
  {pages} {1355} (\bibinfo {year} {2013})},\ \Eprint
  {http://arxiv.org/abs/1306.0007} {arXiv:1306.0007 [astro-ph.SR]} \BibitemShut
  {NoStop}%
\bibitem [{\citenamefont {{Hobbs}}\ \emph {et~al.}(2005)\citenamefont
  {{Hobbs}}, \citenamefont {{Lorimer}}, \citenamefont {{Lyne}},\ and\
  \citenamefont {{Kramer}}}]{2005MNRAS.360..974H}%
  \BibitemOpen
  \bibfield  {author} {\bibinfo {author} {\bibfnamefont {G.}~\bibnamefont
  {{Hobbs}}}, \bibinfo {author} {\bibfnamefont {D.~R.}\ \bibnamefont
  {{Lorimer}}}, \bibinfo {author} {\bibfnamefont {A.~G.}\ \bibnamefont
  {{Lyne}}}, \ and\ \bibinfo {author} {\bibfnamefont {M.}~\bibnamefont
  {{Kramer}}},\ }\href {\doibase 10.1111/j.1365-2966.2005.09087.x} {\bibfield
  {journal} {\bibinfo  {journal} {\mnras}\ }\textbf {\bibinfo {volume} {360}},\
  \bibinfo {pages} {974} (\bibinfo {year} {2005})},\ \Eprint
  {http://arxiv.org/abs/astro-ph/0504584} {astro-ph/0504584} \BibitemShut
  {NoStop}%
\bibitem [{\citenamefont {{Repetto}}\ \emph {et~al.}(2017)\citenamefont
  {{Repetto}}, \citenamefont {{Igoshev}},\ and\ \citenamefont
  {{Nelemans}}}]{2017MNRAS.467..298R}%
  \BibitemOpen
  \bibfield  {author} {\bibinfo {author} {\bibfnamefont {S.}~\bibnamefont
  {{Repetto}}}, \bibinfo {author} {\bibfnamefont {A.~P.}\ \bibnamefont
  {{Igoshev}}}, \ and\ \bibinfo {author} {\bibfnamefont {G.}~\bibnamefont
  {{Nelemans}}},\ }\href {\doibase 10.1093/mnras/stx027} {\bibfield  {journal}
  {\bibinfo  {journal} {\mnras}\ }\textbf {\bibinfo {volume} {467}},\ \bibinfo
  {pages} {298} (\bibinfo {year} {2017})},\ \Eprint
  {http://arxiv.org/abs/1701.01347} {arXiv:1701.01347 [astro-ph.HE]}
  \BibitemShut {NoStop}%
\bibitem [{\citenamefont {{Repetto}}\ \emph {et~al.}(2012)\citenamefont
  {{Repetto}}, \citenamefont {{Davies}},\ and\ \citenamefont
  {{Sigurdsson}}}]{2012MNRAS.425.2799R}%
  \BibitemOpen
  \bibfield  {author} {\bibinfo {author} {\bibfnamefont {S.}~\bibnamefont
  {{Repetto}}}, \bibinfo {author} {\bibfnamefont {M.~B.}\ \bibnamefont
  {{Davies}}}, \ and\ \bibinfo {author} {\bibfnamefont {S.}~\bibnamefont
  {{Sigurdsson}}},\ }\href {\doibase 10.1111/j.1365-2966.2012.21549.x}
  {\bibfield  {journal} {\bibinfo  {journal} {\mnras}\ }\textbf {\bibinfo
  {volume} {425}},\ \bibinfo {pages} {2799} (\bibinfo {year} {2012})},\ \Eprint
  {http://arxiv.org/abs/1203.3077} {arXiv:1203.3077 [astro-ph.GA]} \BibitemShut
  {NoStop}%
\bibitem [{\citenamefont {{O'Shaughnessy}}\ \emph {et~al.}(2017)\citenamefont
  {{O'Shaughnessy}}, \citenamefont {{Gerosa}},\ and\ \citenamefont
  {{Wysocki}}}]{2017PhRvL.119a1101O}%
  \BibitemOpen
  \bibfield  {author} {\bibinfo {author} {\bibfnamefont {R.}~\bibnamefont
  {{O'Shaughnessy}}}, \bibinfo {author} {\bibfnamefont {D.}~\bibnamefont
  {{Gerosa}}}, \ and\ \bibinfo {author} {\bibfnamefont {D.}~\bibnamefont
  {{Wysocki}}},\ }\href {\doibase 10.1103/PhysRevLett.119.011101} {\bibfield
  {journal} {\bibinfo  {journal} {\prl}\ }\textbf {\bibinfo {volume} {119}},\
  \bibinfo {eid} {011101} (\bibinfo {year} {2017})},\ \Eprint
  {http://arxiv.org/abs/1704.03879} {arXiv:1704.03879 [astro-ph.HE]}
  \BibitemShut {NoStop}%
\bibitem [{\citenamefont {{Fryer}}(1999)}]{1999ApJ...522..413F}%
  \BibitemOpen
  \bibfield  {author} {\bibinfo {author} {\bibfnamefont {C.~L.}\ \bibnamefont
  {{Fryer}}},\ }\href {\doibase 10.1086/307647} {\bibfield  {journal} {\bibinfo
   {journal} {\apj}\ }\textbf {\bibinfo {volume} {522}},\ \bibinfo {pages}
  {413} (\bibinfo {year} {1999})},\ \Eprint
  {http://arxiv.org/abs/astro-ph/9902315} {astro-ph/9902315} \BibitemShut
  {NoStop}%
\bibitem [{\citenamefont {{Fryer}}\ and\ \citenamefont
  {{Kalogera}}(2001)}]{2001ApJ...554..548F}%
  \BibitemOpen
  \bibfield  {author} {\bibinfo {author} {\bibfnamefont {C.~L.}\ \bibnamefont
  {{Fryer}}}\ and\ \bibinfo {author} {\bibfnamefont {V.}~\bibnamefont
  {{Kalogera}}},\ }\href {\doibase 10.1086/321359} {\bibfield  {journal}
  {\bibinfo  {journal} {\apj}\ }\textbf {\bibinfo {volume} {554}},\ \bibinfo
  {pages} {548} (\bibinfo {year} {2001})},\ \Eprint
  {http://arxiv.org/abs/astro-ph/9911312} {astro-ph/9911312} \BibitemShut
  {NoStop}%
\bibitem [{\citenamefont {{Paczynski}}(1976)}]{1976IAUS...73...75P}%
  \BibitemOpen
  \bibfield  {author} {\bibinfo {author} {\bibfnamefont {B.}~\bibnamefont
  {{Paczynski}}},\ }in\ \href@noop {} {\emph {\bibinfo {booktitle} {Structure
  and Evolution of Close Binary Systems}}},\ \bibinfo {series} {IAU Symposium},
  Vol.~\bibinfo {volume} {73}\ (\bibinfo {year} {1976})\ p.~\bibinfo {pages}
  {75}\BibitemShut {NoStop}%
\bibitem [{\citenamefont {{Iben}}\ and\ \citenamefont
  {{Livio}}(1993)}]{1993PASP..105.1373I}%
  \BibitemOpen
  \bibfield  {author} {\bibinfo {author} {\bibfnamefont {I.}~\bibnamefont
  {{Iben}}, \bibfnamefont {Jr.}}\ and\ \bibinfo {author} {\bibfnamefont
  {M.}~\bibnamefont {{Livio}}},\ }\href {\doibase 10.1086/133321} {\bibfield
  {journal} {\bibinfo  {journal} {\pasp}\ }\textbf {\bibinfo {volume} {105}},\
  \bibinfo {pages} {1373} (\bibinfo {year} {1993})}\BibitemShut {NoStop}%
\bibitem [{\citenamefont {{Taam}}\ and\ \citenamefont
  {{Sandquist}}(2000)}]{2000ARA&A..38..113T}%
  \BibitemOpen
  \bibfield  {author} {\bibinfo {author} {\bibfnamefont {R.~E.}\ \bibnamefont
  {{Taam}}}\ and\ \bibinfo {author} {\bibfnamefont {E.~L.}\ \bibnamefont
  {{Sandquist}}},\ }\href {\doibase 10.1146/annurev.astro.38.1.113} {\bibfield
  {journal} {\bibinfo  {journal} {\araa}\ }\textbf {\bibinfo {volume} {38}},\
  \bibinfo {pages} {113} (\bibinfo {year} {2000})}\BibitemShut {NoStop}%
\bibitem [{\citenamefont {{Ivanova}}\ \emph {et~al.}(2013)\citenamefont
  {{Ivanova}}, \citenamefont {{Justham}}, \citenamefont {{Chen}}, \citenamefont
  {{De Marco}}, \citenamefont {{Fryer}}, \citenamefont {{Gaburov}},
  \citenamefont {{Ge}}, \citenamefont {{Glebbeek}}, \citenamefont {{Han}},
  \citenamefont {{Li}}, \citenamefont {{Lu}}, \citenamefont {{Marsh}},
  \citenamefont {{Podsiadlowski}}, \citenamefont {{Potter}}, \citenamefont
  {{Soker}}, \citenamefont {{Taam}}, \citenamefont {{Tauris}}, \citenamefont
  {{van den Heuvel}},\ and\ \citenamefont {{Webbink}}}]{2013A&ARv..21...59I}%
  \BibitemOpen
  \bibfield  {author} {\bibinfo {author} {\bibfnamefont {N.}~\bibnamefont
  {{Ivanova}}}, \bibinfo {author} {\bibfnamefont {S.}~\bibnamefont
  {{Justham}}}, \bibinfo {author} {\bibfnamefont {X.}~\bibnamefont {{Chen}}},
  \bibinfo {author} {\bibfnamefont {O.}~\bibnamefont {{De Marco}}}, \bibinfo
  {author} {\bibfnamefont {C.~L.}\ \bibnamefont {{Fryer}}}, \bibinfo {author}
  {\bibfnamefont {E.}~\bibnamefont {{Gaburov}}}, \bibinfo {author}
  {\bibfnamefont {H.}~\bibnamefont {{Ge}}}, \bibinfo {author} {\bibfnamefont
  {E.}~\bibnamefont {{Glebbeek}}}, \bibinfo {author} {\bibfnamefont
  {Z.}~\bibnamefont {{Han}}}, \bibinfo {author} {\bibfnamefont {X.-D.}\
  \bibnamefont {{Li}}}, \bibinfo {author} {\bibfnamefont {G.}~\bibnamefont
  {{Lu}}}, \bibinfo {author} {\bibfnamefont {T.}~\bibnamefont {{Marsh}}},
  \bibinfo {author} {\bibfnamefont {P.}~\bibnamefont {{Podsiadlowski}}},
  \bibinfo {author} {\bibfnamefont {A.}~\bibnamefont {{Potter}}}, \bibinfo
  {author} {\bibfnamefont {N.}~\bibnamefont {{Soker}}}, \bibinfo {author}
  {\bibfnamefont {R.}~\bibnamefont {{Taam}}}, \bibinfo {author} {\bibfnamefont
  {T.~M.}\ \bibnamefont {{Tauris}}}, \bibinfo {author} {\bibfnamefont
  {E.~P.~J.}\ \bibnamefont {{van den Heuvel}}}, \ and\ \bibinfo {author}
  {\bibfnamefont {R.~F.}\ \bibnamefont {{Webbink}}},\ }\href {\doibase
  10.1007/s00159-013-0059-2} {\bibfield  {journal} {\bibinfo  {journal}
  {\aapr}\ }\textbf {\bibinfo {volume} {21}},\ \bibinfo {eid} {59} (\bibinfo
  {year} {2013})},\ \Eprint {http://arxiv.org/abs/1209.4302} {arXiv:1209.4302
  [astro-ph.HE]} \BibitemShut {NoStop}%
\bibitem [{\citenamefont {{Xu}}\ and\ \citenamefont
  {{Li}}(2010)}]{2010ApJ...716..114X}%
  \BibitemOpen
  \bibfield  {author} {\bibinfo {author} {\bibfnamefont {X.-J.}\ \bibnamefont
  {{Xu}}}\ and\ \bibinfo {author} {\bibfnamefont {X.-D.}\ \bibnamefont
  {{Li}}},\ }\href {\doibase 10.1088/0004-637X/716/1/114} {\bibfield  {journal}
  {\bibinfo  {journal} {\apj}\ }\textbf {\bibinfo {volume} {716}},\ \bibinfo
  {pages} {114} (\bibinfo {year} {2010})},\ \Eprint
  {http://arxiv.org/abs/1004.4957} {arXiv:1004.4957 [astro-ph.SR]} \BibitemShut
  {NoStop}%
\bibitem [{\citenamefont {{Loveridge}}\ \emph {et~al.}(2011)\citenamefont
  {{Loveridge}}, \citenamefont {{van der Sluys}},\ and\ \citenamefont
  {{Kalogera}}}]{2011ApJ...743...49L}%
  \BibitemOpen
  \bibfield  {author} {\bibinfo {author} {\bibfnamefont {A.~J.}\ \bibnamefont
  {{Loveridge}}}, \bibinfo {author} {\bibfnamefont {M.~V.}\ \bibnamefont {{van
  der Sluys}}}, \ and\ \bibinfo {author} {\bibfnamefont {V.}~\bibnamefont
  {{Kalogera}}},\ }\href {\doibase 10.1088/0004-637X/743/1/49} {\bibfield
  {journal} {\bibinfo  {journal} {\apj}\ }\textbf {\bibinfo {volume} {743}},\
  \bibinfo {eid} {49} (\bibinfo {year} {2011})},\ \Eprint
  {http://arxiv.org/abs/1009.5400} {arXiv:1009.5400 [astro-ph.SR]} \BibitemShut
  {NoStop}%
\bibitem [{\citenamefont {{Zuo}}\ and\ \citenamefont
  {{Li}}(2014)}]{2014MNRAS.442.1980Z}%
  \BibitemOpen
  \bibfield  {author} {\bibinfo {author} {\bibfnamefont {Z.-Y.}\ \bibnamefont
  {{Zuo}}}\ and\ \bibinfo {author} {\bibfnamefont {X.-D.}\ \bibnamefont
  {{Li}}},\ }\href {\doibase 10.1093/mnras/stu993} {\bibfield  {journal}
  {\bibinfo  {journal} {\mnras}\ }\textbf {\bibinfo {volume} {442}},\ \bibinfo
  {pages} {1980} (\bibinfo {year} {2014})},\ \Eprint
  {http://arxiv.org/abs/1405.4662} {arXiv:1405.4662 [astro-ph.HE]} \BibitemShut
  {NoStop}%
\bibitem [{\citenamefont {{Dominik}}\ \emph {et~al.}(2012)\citenamefont
  {{Dominik}}, \citenamefont {{Belczynski}}, \citenamefont {{Fryer}},
  \citenamefont {{Holz}}, \citenamefont {{Berti}}, \citenamefont {{Bulik}},
  \citenamefont {{Mandel}},\ and\ \citenamefont
  {{O'Shaughnessy}}}]{2012ApJ...759...52D}%
  \BibitemOpen
  \bibfield  {author} {\bibinfo {author} {\bibfnamefont {M.}~\bibnamefont
  {{Dominik}}}, \bibinfo {author} {\bibfnamefont {K.}~\bibnamefont
  {{Belczynski}}}, \bibinfo {author} {\bibfnamefont {C.}~\bibnamefont
  {{Fryer}}}, \bibinfo {author} {\bibfnamefont {D.~E.}\ \bibnamefont {{Holz}}},
  \bibinfo {author} {\bibfnamefont {E.}~\bibnamefont {{Berti}}}, \bibinfo
  {author} {\bibfnamefont {T.}~\bibnamefont {{Bulik}}}, \bibinfo {author}
  {\bibfnamefont {I.}~\bibnamefont {{Mandel}}}, \ and\ \bibinfo {author}
  {\bibfnamefont {R.}~\bibnamefont {{O'Shaughnessy}}},\ }\href {\doibase
  10.1088/0004-637X/759/1/52} {\bibfield  {journal} {\bibinfo  {journal}
  {\apj}\ }\textbf {\bibinfo {volume} {759}},\ \bibinfo {eid} {52} (\bibinfo
  {year} {2012})},\ \Eprint {http://arxiv.org/abs/1202.4901} {arXiv:1202.4901
  [astro-ph.HE]} \BibitemShut {NoStop}%
\bibitem [{\citenamefont {{Belczy{\'n}ski}}\ and\ \citenamefont
  {{Bulik}}(1999)}]{1999A&A...346...91B}%
  \BibitemOpen
  \bibfield  {author} {\bibinfo {author} {\bibfnamefont {K.}~\bibnamefont
  {{Belczy{\'n}ski}}}\ and\ \bibinfo {author} {\bibfnamefont {T.}~\bibnamefont
  {{Bulik}}},\ }\href@noop {} {\bibfield  {journal} {\bibinfo  {journal}
  {\aap}\ }\textbf {\bibinfo {volume} {346}},\ \bibinfo {pages} {91} (\bibinfo
  {year} {1999})},\ \Eprint {http://arxiv.org/abs/astro-ph/9901193}
  {astro-ph/9901193} \BibitemShut {NoStop}%
\bibitem [{\citenamefont {{Scott}}(2015)}]{2015mdet.book.....S}%
  \BibitemOpen
  \bibfield  {author} {\bibinfo {author} {\bibfnamefont {D.~W.}\ \bibnamefont
  {{Scott}}},\ }\href@noop {} {\emph {\bibinfo {title} {{Multivariate Density
  Estimation: Theory, Practice, and Visualization}}}}\ (\bibinfo  {publisher}
  {John Wiley and Sons},\ \bibinfo {year} {2015})\BibitemShut {NoStop}%
\bibitem [{\citenamefont {{Barrett}}\ \emph {et~al.}(2018)\citenamefont
  {{Barrett}}, \citenamefont {{Gaebel}}, \citenamefont {{Neijssel}},
  \citenamefont {{Vigna-G{\'o}mez}}, \citenamefont {{Stevenson}}, \citenamefont
  {{Berry}}, \citenamefont {{Farr}},\ and\ \citenamefont
  {{Mandel}}}]{2018MNRAS.477.4685B}%
  \BibitemOpen
  \bibfield  {author} {\bibinfo {author} {\bibfnamefont {J.~W.}\ \bibnamefont
  {{Barrett}}}, \bibinfo {author} {\bibfnamefont {S.~M.}\ \bibnamefont
  {{Gaebel}}}, \bibinfo {author} {\bibfnamefont {C.~J.}\ \bibnamefont
  {{Neijssel}}}, \bibinfo {author} {\bibfnamefont {A.}~\bibnamefont
  {{Vigna-G{\'o}mez}}}, \bibinfo {author} {\bibfnamefont {S.}~\bibnamefont
  {{Stevenson}}}, \bibinfo {author} {\bibfnamefont {C.~P.~L.}\ \bibnamefont
  {{Berry}}}, \bibinfo {author} {\bibfnamefont {W.~M.}\ \bibnamefont {{Farr}}},
  \ and\ \bibinfo {author} {\bibfnamefont {I.}~\bibnamefont {{Mandel}}},\
  }\href {\doibase 10.1093/mnras/sty908} {\bibfield  {journal} {\bibinfo
  {journal} {\mnras}\ }\textbf {\bibinfo {volume} {477}},\ \bibinfo {pages}
  {4685} (\bibinfo {year} {2018})},\ \Eprint {http://arxiv.org/abs/1711.06287}
  {arXiv:1711.06287 [astro-ph.HE]} \BibitemShut {NoStop}%
\end{thebibliography}%

\end{document}